\documentstyle[preprint,aps,epsf]{revtex}

\def\plb#1{Phys.~Lett.~{\bf B#1}}
\def\npb#1{Nucl.~Phys.~{\bf B#1}}
\def\prl#1{Phys.~Rev.~Lett.~{\bf #1}}
\def\prd#1{Phys.~Rev.~{\bf D#1}}

\def\thptfn{three-point function}
\def\thptfns{three-point functions}
\def\twptfn{two-point function}
\def\twptfns{two-point functions}
\def\iwfn{Isgur-Wise function}
\def\iwfns{Isgur-Wise functions}
\def\hqs{heavy-quark symmetry}

\def\stat{({\rm stat.})}
\def\syst{({\rm syst.})}

\def\QQp{{Q^{(\prime)}}}

\def\e{\epsilon}
\def\g{\gamma}
\def\k{\kappa}
\def\a{\alpha}
\def\b{\beta}
\def\w{\omega}
\def\d{\delta}
\def\gm{\g^\mu}

\def\l{\left}
\def\r{\right}
\def\ord#1{{\cal O}\l(#1\r)}
\def\la{\langle}
\def\ra{\rangle}

\def\PtoPpl{P\to P'\ell\bar\nu\,}
\def\btodl{\bar B\to D\ell\bar\nu\,}
\def\btodlands{\bar B_{(s)}\to D_{(s)}\ell\bar\nu\,}
\def\bstodsl{\bar B_s\to D_s\ell\bar\nu\,}
\def\btodsl{\bar B\to D^*\ell\bar\nu\,}
\def\btodslands{\bar B_{(s)}\to D^*_{(s)}\ell\bar\nu\,}
\def\bstodssl{\bar B_s\to D_s^*\ell\bar\nu\,}
\def\btod{\bar B\to D\,}

\def\hpw{h^+(\w)}
\def\hplatw{h^+_{\rm lat.}(\w)}
\def\hmw{h^-(\w)}
\def\hmlatw{h^-_{\rm lat.}(\w)}
\def\hpQQpw{h^+(\w;m_Q,m_{Q'})}
\def\hplatQQpw{h^+_{\rm lat.}(\w;m_Q,m_{Q'})}

\def\hmQQpw{h^-(\w;m_Q,m_{Q'})}
\def\hmlatQQpw{h^-_{\rm lat.}(\w;m_Q,m_{Q'})}

\def\iw{\xi(\w)}

\def\lqcd{\Lambda_{\rm QCD}}
\def\lbar{\bar\Lambda}
\def\kcrit{\kappa_{\rm crit}}
\def\ks{\kappa_s}

\def\as#1{\a_s\l(#1\r)}

\def\s#1{{\bf #1}}
\def\be{\begin{equation}}
\def\ee{\end{equation}}
\def\bea{\begin{eqnarray}}
\def\eea{\end{eqnarray}}

\def\s#1{{\bf #1}}

\def\bpw{\b^+(\w)}
\def\bmw{\b^-(\w)}
\def\bpQQpw{\b^+(\w;m_Q,m_{Q'})}

\def\gpw{\g^+(\w)}
\def\gpQQpw{\g^+(\w;m_Q,m_{Q'})}

\def\kQ{\k_Q}
\def\kQp{\k_{Q'}}
\def\kq{\k_q}
\def\eQ{\epsilon_Q}
\def\eQp{\epsilon_{Q'}}

\def\pnotp{^{(\prime)}}
\newcommand{\dl}{\stackrel{\leftarrow}{D}}
\newcommand{\dr}{\stackrel{\rightarrow}{D}}
\def\tGm{\tilde\Gamma^\mu}

\def\tab#1{Table~\ref{#1}}

\def\fig#1{Fig.~\ref{#1}}
\def\figs#1#2{Figs.~\ref{#1} and \ref{#2}}
\def\sec#1{Section \ref{#1}}
\def\eq#1{Eq.~(\ref{#1})}
\def\eqs#1#2{Eqs.~(\ref{#1}) and (\ref{#2})}

\newcommand{\plus}{\makebox[15pt][c]{$+$}}
\newcommand{\minus}{\makebox[15pt][c]{$-$}}
\newcommand{\errr}[2]{
\raisebox{0.08em}{\scriptsize {$\hspace{-0.3em}\begin{array}{@{}l@{}}
                          \plus\makebox[0.9em][r]{#1} \\[-0.12em] 
                          \minus\makebox[0.9em][r]{#2} 
                        \end{array}$}}}
\newcommand{\err}[2]{
\raisebox{0.08em}{\scriptsize {$\hspace{-0.3em}\begin{array}{@{}l@{}}
                          \plus\makebox[0.55em][r]{#1} \\[-0.12em] 
                          \minus\makebox[0.55em][r]{#2} 
                        \end{array}$}}}
\newcommand{\er}[2]{
\raisebox{0.08em}{\scriptsize {$\hspace{-0.3em}\begin{array}{@{}l@{}}
                          \plus\makebox[0.15em][r]{#1} \\[-0.12em] 
                          \minus\makebox[0.15em][r]{#2} 
                        \end{array}$}}}
\newcommand{\gev}{{\rm GeV}}
\newcommand{\mev}{{\rm MeV}}

\begin{document}



\preprint{Edinburgh Preprint 93/525, CPT-95/P.3179, Southampton Preprint SHEP
95-06}

\title{ Lattice Study of Semi-Leptonic $B$ Decays:\\
I. $\btodl$ Decays.}

\author{K.C.~Bowler, N.M.~Hazel, D.S.~Henty,
H.~Hoeber\footnote{Present address: HLRZ, 52425 J\"{u}lich, Germany},
R.D.~Kenway, D.G.~Richards, \\
H.P.~Shanahan\footnote{Present address:
Dept. of Physics and Astronomy, The University, Glasgow G12 8QQ, Scotland},
J.N.~Simone\footnote{Present address: FNAL, PO Box 500, Batavia IL 60510, USA}}
\address{Department of Physics \& Astronomy, The University of Edinburgh,
Edinburgh EH9~3JZ, Scotland}

\author{L.~Lellouch}

\address{Centre de Physique Th\'eorique, CNRS Luminy, Case 907,
F-13288 Marseille Cedex 9, France~\footnote{Unit\'e Propre de Recherche
7061.}}

\author{J.~Nieves, C.T.~Sachrajda, H.~Wittig}

\address{Physics Department, The University, Southampton SO7 1BJ, UK}

\author{(UKQCD Collaboration)}
\date{March 1995}
\maketitle



\tighten

\begin{abstract}
We present a study of semi-leptonic $\btodl$ decays in quenched lattice
QCD through a calculation of the matrix element $\la D|\bar c\gm
b|\bar B\ra$ on a $24^3\times 48$ lattice at $\beta=6.2$, using an
$\ord{a}$-improved fermion action.  We perform the calculation for
several values of the initial and final heavy-quark masses around the
charm mass, and three values of the light-(anti)quark mass around the
strange mass.  Because the charm quark has a bare mass which is almost
1/3 the inverse lattice spacing, we study the ensuing mass-dependent
discretization errors, and propose a procedure for subtracting at
least some of them non-perturbatively.

We extract the form factors $h^+$ and $h^-$. After radiative
corrections, we find that $h^+$ displays no dependence on the
heavy-quark mass, enabling us to identify it with an Isgur-Wise
function $\xi$.  Interpolating the light-quark mass to that of the
strange, we obtain an Isgur-Wise function relevant for $\bar B_s\to
D_s^{(*)}\ell\bar\nu\,$ decays which has a slope
$-\xi'_s=1.2\er{2}{2}\stat\er{2}{1}\syst$ at zero recoil.  An
extrapolation to a massless light quark enables us to obtain an
Isgur-Wise function relevant for $\bar B\to D^{(*)}\ell\bar\nu\,$
decays.  This function has a slope
$-\xi_{u,d}'=0.9\er{2}{3}\stat\er{4}{2}\syst$ at zero recoil.  We
observe a slight decrease in the magnitude of the 
central value of the slope as the
mass of the light quark is reduced; given the errors, however, the
significance of this observation is limited.

We then use these functions, in conjunction with heavy-quark effective
theory, to extract $V_{cb}$ with no free parameters from the $\btodsl$
decay rate measured by the ALEPH, ARGUS and CLEO collaborations.
Using the CLEO data, for instance, we obtain
$|V_{cb}|=0.037\er{1}{1}\er{2}{2}\er{4}{1}
\left(\frac{0.99}{1+\beta^{A_1}(1)}\right)\frac{1}{1+\delta_{1/m_c^2}}$,
where $\delta_{1/m_c^2}$ is the power correction inversely
proportional to the square of the charm quark mass, and
$\beta^{A_1}(1)$ is the relevant radiative correction at zero
recoil.  Here, the first set of errors is experimental, the second
represents the statistical error and the third represents the
systematic error in our evaluation of the Isgur-Wise function.  We
also use our Isgur-Wise functions and heavy-quark effective theory
to calculate branching ratios for $\btodlands$ and $\btodslands$
decays. 

\end{abstract}

\bigskip

\begin{tabbing}
\= PACS Numbers: 13.20.He, 12.15.Hh, 12.38.Gc, 12.39.Hg \\
\\
\> Key Words: \= Semi-Leptonic Decays of $B$ Mesons, Determination of 
Kobayashi-Maskawa Matrix \\ \> \> Elements ($V_{cb}$), 
Lattice QCD Calculation, Heavy Quark Effective Theory.
\end{tabbing}

\vfill


%



\section{Introduction}
\label{intro}

Semi-leptonic decays of $B$ mesons have been the focus of much activity
in the last few years.  Experimentally, their rather large branching
ratios have allowed thorough studies of their properties.
Theoretically, they have been a fertile ground for new ideas.
Moreover, the interplay between these experimental studies and new
theoretical ideas has led to a greater understanding of the flavour
sector of the Standard Model and, in particular, to measurements of
the less well-known Cabibbo-Kobayashi-Maskawa (CKM) matrix elements
$V_{cb}$ and $V_{ub}$\cite{bdecays}.

The main theoretical development in the study of hadrons containing a
heavy quark, such as the $b$ or $c$ quarks, is undoubtedly the
discovery of heavy-quark symmetry\cite{russians,iw} and the
development of the heavy-quark effective theory (HQET)\cite{hqet},
which describes the strong interactions of a heavy quark with gluons
and light quarks at low energies.  If one considers the masses of the
$b$ and $c$ quarks to be much larger than the QCD scale, $\lqcd$, one
finds that the dynamics of the light quarks and gluons coupled to a
$b$ or a $c$ quark become independent of this heavy quark's flavour and spin.
In this limit, QCD exhibits a new $SU(4)_{{\rm spin}\times{\rm
flavour}}$ symmetry, known as heavy-quark symmetry, which acts on the
multiplet $(c\uparrow,c\downarrow,b\uparrow,b\downarrow)$.  This
symmetry simplifies considerably the description of the decays of
hadrons containing a heavy quark.  For instance, the 20 form factors
required to describe the semi-leptonic decays $\btodlands$ and
$\btodslands$ as well as the elastic form factors of $B^{(*)}_{(s)}$
and $D^{(*)}_{(s)}$ mesons,\footnote{The subscript $s$ is used to
distinguish mesons in which the light, spectator antiquark is $\bar s$
from those in which it is either $\bar u$ or $\bar d$.} can all be
expressed in terms of two universal form factors, $\xi_{u,d}$ and
$\xi_s$, known as Isgur-Wise functions\cite{iw}, which parametrize the
non-perturbative dynamics of the light degrees of freedom.
$\xi_{u,d}$ describes the decays of mesons containing a heavy quark
and a $\bar u$ or $\bar d$ antiquark, and $\xi_s$ describes the decays
of mesons containing a heavy quark and an $\bar s$ antiquark.
Moreover, heavy-quark symmetry requires these Isgur-Wise functions to
be 1 when $q^2$, the square of the four momentum transfer, is
maximum\cite{iw}.

In an earlier work\cite{ourprl}, we obtained the Isgur-Wise
functions $\xi_{u,d}$ from a lattice study of elastic
$D$ meson scattering.  A similar approach, but with a different
lattice action, was taken by Bernard et al.\cite{bersheso} and led
to very similar results.  In the present paper, we extend our
earlier work to include decays of the form $\PtoPpl$, where
$P\pnotp$ is a heavy-light pseudoscalar meson composed of a heavy
quark, $Q\pnotp$, with a mass around that of the charm quark, and
a light antiquark, $\bar q$.  
These processes are described by
matrix elements of the vector current $\bar Q'\gm Q$.  These
matrix elements can, in turn, be decomposed in terms of two form
factors, $\hpQQpw$ and $\hmQQpw$, given by
\be
\frac{\la P'(\s p')|\bar Q'\gm Q| P(\s p)\ra}{\sqrt{M_PM_{P'}}}
=(v+v')^\mu\hpQQpw+(v-v')^\mu\hmQQpw
\label{matelt}
\ ,
\ee
where $v\pnotp=p\pnotp/M_{P\pnotp}$, $\w=v\cdot v'=(M_P^2+M_{P'}^2-q^2)/
2M_PM_{P'}$ and $m_{Q^{(\prime)}}$ is the mass of $Q^{(\prime)}$.

In the limit of exact \hqs, the two form factors become independent of
the masses of the initial and final heavy quarks and
\bea
&&\hmQQpw \equiv  0\nonumber\\
&&\hpQQpw \equiv \xi(\w)
\ ,
\label{hql}
\eea
where $\xi(\w)$ is an Isgur-Wise function of the type described above,
whose exact functional form only depends on the quantum numbers of the
light spectator antiquark. The only change we make to these quantum numbers 
in the present paper is to vary the light antiquark mass.
For simplicity of notation, this dependence will be left implicit
unless stated otherwise.

For heavy quarks of finite mass, there are two sources of corrections
to the simple results of \eq{hql}.  The first is hard-gluon exchange
between $Q$ and $Q'$ across the vector current vertex. The second
results from the modifications of the vector current and meson states
by higher-dimension operators in HQET. These latter corrections are
proportional to inverse powers of the heavy-quark masses. Thus, we
have
\be
h^i(\w;m_Q,m_{Q'})=(\a^i+\b^i(\w;m_Q,m_{Q'})+\g^i(\w;m_Q,m_{Q'}))\,\iw
\label{corr}
\ ,
\ee
for $i=+,-$, where $\a^+=1$, $\a^-=0$, 
$\b^i$ represents the radiative corrections and
$\g^i$, the power corrections.  It is important to note that these two
corrections incorporate all of the mass dependence of the form factors
$h^i$. As defined in
\eq{corr}, the Isgur-Wise function, $\iw$, is renormalization-group
invariant\cite{short} and normalized to one at zero recoil
as required by \hqs\cite{iw}:
\be
\label{iwnorm} 
\xi(1)=1
\ .
\ee

The radiative corrections can be evaluated analytically in QCD
since they are perturbative.  To quantify them,
we use Neubert's short-distance expansion of heavy-quark
currents\cite{short}.  He considers semi-leptonic $\btodl$ and
$\btodsl$ decays and computes radiative corrections to the
corresponding heavy-quark matrix elements to order $\a_s$ as a
function of $m_c$ and $m_b$.  His calculation improves the previous
leading logarithmic evaluation of these corrections\cite{lla} in two
ways: firstly, he includes next-to-leading logarithms in running the
$\ord{(m_c/m_b)^0}$ heavy-quark operators from $m_b$ down to scales at
which HQET can be safely used, and secondly, he obtains, to order
$\a_s$, the full dependence of the heavy-quark current on the mass
ratio $z=m_c/m_b$.  The sum of these new contributions is as large as
the leading logarithmic term.  Corrections to Neubert's
computation\footnote{Neubert runs the $\ord{m_c/m_b}$ contribution at
one loop.} are of order $\a_s^2(z\ln z)^n$ with $n=0,1,2$ and should
be smaller than 1\%.  The fact that Neubert's result accounts for the
full order $\a_s$ dependence of the heavy-quark current on the mass
ratio $z$ is important for us, because our range of heavy-quark masses
is quite small (see
\tab{Q_masses_6.2}): $z$ ranges from 0.6 to 1. 

The power corrections are proportional to powers of $\e_\QQp=\lbar/
(2m_\QQp)$ where $\lbar$ is the
energy carried by the light degrees of freedom in the mesons. 
$\lbar$ will of course depend on what these light degrees of
freedom are.  In what follows, we will use
$\lbar=\lbar_\chi=500\mev$\cite{bible} when working with light degrees of
freedom with spin $1/2$ and isospin $1/2$. 
Because $\e_\QQp\simeq 1/6$ for the heavy quarks we are
considering, we would naively expect power corrections in $\hpw$
and $\hmw$ to be of order 15 to 30\%.  These corrections are difficult
to quantify because they involve the light degrees of freedom and
are therefore non-perturbative.  Luke's theorem\cite{luke},
however, guarantees that there are no $\ord{\eQ}$ corrections to
$\hpw$ at zero recoil and one may expect that power corrections to $h^+$
remain small away from zero recoil. This is not expected to be
true for $h^-$ which is not protected by Luke's theorem.

For degenerate transitions where $Q=Q'$, conservation of the vector
current $\bar Q\g^\mu Q$ provides further constraints on the
radiative and power corrections:
\bea
\b^+(1;m_Q,m_Q) &=& 0\nonumber\\
\g^+(1;m_Q,m_Q) &=& 0\nonumber\\
\b^-(\w;m_Q,m_Q) &\equiv& 0\label{degencond}
\\
\g^-(\w;m_Q,m_Q) &\equiv& 0\nonumber
\ ,
\eea
where the last two equations hold for all $\w$.

Our results come from a quenched simulation on a $24^3\times 48$
lattice at $\beta=6.2$ on a sample of 60 gauge field
configurations\cite{qlhms}. The lattice has an inverse lattice spacing
of around $2.7~\gev$~\cite{qhldc}. We do not suffer much here from
errors associated with uncertainties in the determination of the
lattice spacing since our main results are dimensionless and depend at
most logarithmically on the scale. Our light quarks have masses which
bracket the strange quark mass. Because our heavy quarks have masses
in the region of the charm-quark mass which are large in lattice units
(up to a half or more), we must contend with discretization errors
proportional to powers of $am_Q$, where $m_Q$ is the mass of the heavy
quark. In order to reduce these discretization errors, we use the
${\cal O}(a)$-improved fermion action originally proposed by
Sheikholeslami and Wohlert\cite{sw} with which discretization errors
in operator matrix elements and hence in our form factors are reduced
from $\ord{am_Q}$ to $\ord{\a_sam_Q}$\cite{improv}.

The remainder of the paper is organized as follows.  In \sec{dotc} we
present the details of our simulation, as well as our strategy for
obtaining the form factors $h^+$ and $h^-$ from the calculated
three-point functions.  In \sec{discerr} we discuss discretization
errors and describe a procedure which enables us to subtract some of
these errors non-perturbatively.  In \sec{ffftsabe6.2} we present our
results for the form factors $h^+$ and $h^-$ for three values of the
light antiquark mass and all available initial and final heavy-quark
combinations.  We also extrapolate $h^+$ in the light-antiquark mass
to the chiral limit, and interpolate it to the strange quark mass.  In
\sec{hqmdohp} we study the dependence of $h^+$ and $h^-$ on heavy-quark
mass and attempt to extract the leading power corrections. We
find that $h^+$ displays no measurable dependence on heavy-quark mass
which enables us to conclude that this form factor is an 
Isgur-Wise function once radiative corrections are subtracted.  
In \sec{dohpolqm}, we
study the dependence of $h^+$ on the light-quark mass and extract
the Isgur-Wise functions $\xi_{u,d}$ and $\xi_s$.  We find that
the slopes of these functions at $\omega = 1$ are
\be
\label{xiudp}
\xi_{u,d}'(1)=-\l[0.9\er{2}{3}\stat\er{4}{2}\syst\r]
\ee
and
\be
\label{xisp}
\xi_s'(1)=-\l[1.2\er{2}{2}\stat\er{2}{1}\syst\r]
\ .\ee
We thus observe a slight decrease in the magnitude of the slope 
light-antiquark mass; given the errors, however, the significance of this
observation is limited.  We compare our
results for these \iwfns\ to other theoretical as well as experimental
determinations.  We find excellent agreement with experiment. In
\sec{eovcb} we use our Isgur-Wise function $\xi_{u,d}$ to extract
the CKM matrix element $V_{cb}$ from different experimental
measurements of the differential decay rate for $\btodsl$ decays. 
Our results for $|V_{cb}|$ are summarized in \tab{vcbres} and are
compared to other determinations of this matrix element.  Our procedure
for extracting $|V_{cb}|$ differs from that proposed by Neubert\cite{neuplb} 
in that we fix the $\w$ dependence of the differential
decay rate with our calculation instead of obtaining it from experiment.
This enables us not only to extract $|V_{cb}|$ with no free parameters, but
also to check the validity of non-perturbative QCD against
experiment. We find that the $\w$ dependence predicted by our calculation
agrees very well with the results of 
the ALEPH\cite{aleph} and CLEO\cite{cleo} collaborations. In
\sec{edr}, we use $\xi_{u,d}$ and $\xi_s$ to compute the branching
ratios for $\btodlands$ and $\btodslands$ decays, and our results
are summarized in \tab{brs}.  We also compute ratios of
semi-leptonic widths and find
\be
\frac{\Gamma(\btodsl)}{\Gamma(\btodl)}=3.2\er{3}{2}({\rm lat.})
\pm 1.0({\rm hqs})
\ee
and
\be
\frac{\Gamma(\bstodssl)}{\Gamma(\bstodsl)}=3.3\er{2}{1}({\rm lat.})
\pm 1.0({\rm hqs})
\ ,\ee
where the first set of errors was obtained by adding our lattice
statistical and systematic errors in quadrature and the second set of
errors, denoted by ``hqs'', quantifies the uncertainty due to neglected
power and radiative corrections. We confront our predictions for these
branching ratios and ratios of widths with experimental measurements
where available and find that they compare quite favourably.  Finally,
in \sec{ccl} we present our conclusions.



\section{Details of the Calculation}
\label{dotc}


\subsection{Lattice Action and Operators}
\label{laao}

Since we are studying the decays of quarks whose masses are large in
lattice units, we must control discretization errors.
In order to reduce these errors, we use an $\ord{a}$-improved
fermion action
originally proposed by Sheikholeslami and Wohlert\cite{sw}, given
by
\be S_F^{SW} = S_F^W - i\frac{\kappa}{2}\sum_{x,\mu,\nu}\bar{q}(x)
         F_{\mu\nu}(x)\sigma_{\mu\nu}q(x),
\label{sclover}
\ee
where $S_F^W$ is the Wilson action:
\be
S_F^W = \sum _x \Biggl\{\bar{q}(x)q(x)
             -\kappa\sum _\mu\Bigl[
            \bar{q}(x)(1- \gamma _\mu )U_\mu (x) q(x+\hat\mu )
+\ \bar{q}(x+\hat\mu )(1 + \gamma _\mu )
   U^\dagger _\mu(x)
            q(x)\Bigr]\Biggr\}
\label{sfw}
\ .\ee
The leading discretization errors in matrix elements for
heavy-quark decays obtained from numerical simulations with the
fermion action \eq{sclover} are reduced from $\ord{am_Q}$ to
$\ord{\a_s am_Q}$ and $\ord{a^2m_Q^2}$, provided one also uses
``improved'' operators obtained by ``rotating'' the field of the
heavy quark, Q:
\be
Q(x)\longrightarrow (1-\frac{1}{2}\gamma\cdot\dr)\,Q(x)
\ .
\ee
Thus, to obtain an $\ord{a}$-improved evaluation of the matrix element
of \eq{matelt}, we use a ``rotated'' vector current
\be
V^\mu_I\equiv\bar Q'(x)\tGm\,Q(x)
\label{improp}
\ ,\ee
where
\be
\tGm=(1+\frac{1}{2}\g\cdot\dl)\,\gm\,
(1-\frac{1}{2}\gamma\cdot\dr)
\label{gamimprov}
\ee
and where the subscript $I$ indicates that $V^\mu_I$ is an improved
lattice current. 



\subsection{Extended Interpolating Operators}
\label{eio}

In order to isolate the ground state in correlation functions
effectively, it is useful to use extended (or ``smeared'')
interpolating operators for the mesons.  In this study we use
gauge-invariant Jacobi smearing on the heavy-quark field
(described in detail in Ref.~\cite{smearing}), in which the
smeared field, $Q^S(\s x,t)$, is defined by
\be
Q^S(\s x,t)\equiv \sum_{\s x^\prime}K(x,x^\prime)Q(\s x^\prime,t),
\label{eq:kernel}\ee
where 
\be
K(x,x^\prime)=\sum_{n=0}^N\,\kappa_S^n\Delta ^n(x, x^\prime)
\ee
and
\be
\Delta(x,x^\prime)=\sum_{i=1}^{3}\{\delta_{\s x^\prime, \s x-\hat\imath}
U^\dagger_i(\s x-\hat\imath,t) + \delta_{\s x^\prime, \s x+\hat\imath}
U_i(\s x,t)\}.
\ee
Following the discussion in Ref.~\cite{smearing}, we choose $\kappa_S=0.25$
and use the parameter $N$ to control the smearing radius, defined by
\be
r^2\equiv \frac{\sum_{\s x}|\s x|^2 |K(x, 0)|^2}
{\sum_{\s x} |K(x, 0)|^2}.
\ee
We use  $N=75$, giving $r\approx 5.2$.

In terms of the operator $Q^S$ of \eq{eq:kernel}, the spatially
extended source $J_P$ we use to create pseudoscalar mesons
composed of a heavy quark $Q$ and a light antiquark $\bar q$ is given by
\be
\label{jp}
J_P(x)=\bar q(x)(1+\frac{1}{2}\g\cdot\dl)\,\g^5\,
(1-\frac{1}{2}\gamma\cdot\dr)Q^S(x)
\ .\ee
%



\subsection{Three-Point Functions and Lattice Form Factors}
\label{tpfalff}

The computation of the matrix element $\la P'(\s p')|\bar Q'\gm Q|
P(\s p)\ra$ proceeds along lines similar to earlier calculations
of the electromagnetic form factor of the pion and to
determinations of the form factors corresponding to semi-leptonic
decays of the $D$ meson into light mesons.  (For
recent reviews of lattice computations of weak matrix elements and
references to the original literature see, for example, the
reviews in Ref.~\cite{reviews}).  Thus, we calculate the
three-point correlator, 
\be 
\label{thpt} 
C^\mu_3(t;\s p',\s
q)_{Q\to Q'}\equiv\sum_{\s x,\s y}\, e^{-i\s q\cdot\s x}\,e^{-i\s
p'\cdot\s y} \la J_{P'}(t_f,\s y)\,V^\mu_I(t,\s
x)\,J^\dagger_P(0,\s 0)\ra \ , 
\ee 
where $J_P$ is the spatially-extended interpolating field for $P$
defined in \eq{jp}, $V^\mu_I$ is the ${\cal
O}(a)$-improved vector current of \eq{improp} and $\s p=\s q+\s p'$. 
To evaluate these
correlators, we use the standard source method reviewed in Ref.~\cite{sm}. 

Provided the three points in the correlator of \eq{thpt} are
sufficiently separated in time, the ground state contribution dominates
and
\be
\label{thptlim}
C^\mu_3(t;\s p',\s q)_{Q\to Q'}\mathop{\longrightarrow}\limits_{t,t_f-t\to
\infty}
\frac{Z_P(\s p^2)Z_{P'}(\s p^{\prime 2})}{4E_PE_{P'}}
\ e^{-E_Pt-E_{P'}(t_f-t)}\,\la P'(\s p')|V^\mu_I(0)|P(\s p)
\ra
\ ,
\ee
where $E_P$
($E_{P'}$) is the energy of the initial (final) meson and $Z_P(\s
p^2)$ is the matrix element $\la 0|J_P(0)|P(\s p)\ra$.
To cancel the above time-dependence, we normalize
the three-point function by two two-point functions and consider the ratio
\be
R^\mu(t;\s p',\s q)_{Q\to Q'}\equiv
\frac{C^\mu_3(t;\s p',\s q)_{Q\to Q'}}{C_2(t,\s p)_Q
C_2(t_f-t,\s p')_{Q'}}
\ ,
\label{fitratio}
\ee
where
\be
C_2(t,\s p)_Q\equiv \sum_{\s x} e^{-i\s p\cdot\s x}\la J_P(t,\s x)
J^\dagger_P(0)\ra
\label{twopt}
\ee
and
\be
C_2(t,\s p)_Q\mathop{\longrightarrow}\limits_{t\to
\infty}\frac{Z_P(\s p^2)^2}{E_P}
\ e^{-E_P\,T/2}\ {\rm cosh}\l(E_P(T/2-t)\r)
\ .
\label{twoptlim}
\ee
Here, $T$ is the temporal extent of the lattice.  
(For $t\ll T/2$, ${\rm exp}(-E_PT/2){\rm cosh}(E_P(T/2-t))$
$\to (1/2){\rm exp}(-E_Pt)$).
Thus, in terms
of the form factors defined in \eq{matelt}
\begin{displaymath}
R^\mu(t;\s p',\s q)_{Q\to Q'}\mathop{\longrightarrow}\limits_{t,t_f-t\to
\infty}
\frac{1}{Z_P(\s p^2)Z_{P'}(\s p^{\prime 2})}\,\sqrt{M_PM_P'}
\end{displaymath}
\be
\label{fitratasym}
\times
\l((v+v')^\mu\hplatQQpw+(v-v')^\mu\hmlatQQpw\r)
\ ,
\ee
where $h^\pm_{\rm lat.}$ are related to the continuum form
factors, $h^\pm$, by a multiplicative renormalization, up to
discretization errors, as discussed in \sec{discerr}.

To obtain the desired form factors, we fit the ratio $R^\mu$ of 
\eq{fitratio} to the asymptotic form of \eq{fitratasym}
by minimizing, with respect to the parameters $h^+_{\rm lat.}$
and $h^-_{\rm lat.}$, a
$\chi^2$ function which takes into account correlations between
the different times (labelled by $t$), but not between the
different equations (labelled by $\mu$).  We neglect correlations
between equations, because spatial and temporal components of
\eq{fitratasym} may be affected differently by discretization
errors, as we discuss at the end of Section~\ref{npsoame}.  The
$\chi^2$ value that we quote indicates not only whether our ratios
$R^\mu$ are asymptotic, but also whether the decomposition of $R^\mu$
in terms of $\hplatw$ and $\hmlatw$ is good.  In fitting the ratio
$R^\mu$, we fix the wavefunction factors $Z_{P^{(\prime)}}$, the
energies, $E_{P^{(\prime)}}$, and masses, $M_{P^{(\prime)}}$, of the
mesons to the values obtained from a fit of the relevant two-point
functions to the asymptotic form of \eq{twoptlim}, taking into account
correlations in time.

We first 
obtain $\hplatw$ from the time component of
\eq{fitratasym} alone, assuming that the contribution proportional
to $\hmlatw$ can be neglected.  This approximation is exact,
up to discretization errors, for degenerate transitions, i.e.\
transitions in which the initial and final heavy-mesons are the same,
and true up to radiative and power corrections for non-degenerate
transitions, i.e.\ transitions between mesons which contain the same
light antiquark, but different heavy quarks (see \eq{degencond}).  For
these non-degenerate transitions we can get {\it a posteriori} some
idea of the size of the contribution of $\hmlatw$ to the time
component of \eq{fitratasym}.  Holding $\hplatw$ fixed to its
time-component value, we use all non-vanishing components of
\eq{fitratasym} to obtain $\hmlatw$.  We find (see
\sec{raflqm}) that $\hmlatw$'s contribution to the
time-component of \eq{fitratasym} is less than about 1\%, thereby justifying
the approximation we make in obtaining $\hpw_{\rm lat.}$.



\subsection{Lattice Parameters and Details of the Analysis}
\label{lpadotaabe6.2}

We compute the \thptfn\ of \eq{thpt} for four values of both the
initial and final heavy-quark hopping parameters, $\kQ$ and $\kQp$
taken from 0.121, 0.125, 0.129, 0.133 (see \tab{Q_masses_6.2}); three
values of the light antiquark hopping parameter, $\kq$ (0.14144,
0.14226, 0.14262); two values of the initial meson momentum ((0,0,0)
and (1,0,0) in lattice units); and ten values of the momentum carried
by the vector current ($\s qa(12/\pi)=(0,0,0)$, (1,0,0), (0,1,0),
(0,0,1), (1,1,0), (1,0,1), (0,1,1), (1,$-1$,0), ($-1$,0,1), (0,1,$-1$)).  To
improve statistics, we average the ratios of
\eq{fitratio} over all equivalent momenta.  Moreover, data with
initial or final momenta greater than $\pi/12a$ are excluded
because they have larger systematic and statistical uncertainties. 
Finally, we choose $t_f$, the time at which the final meson is
destroyed (see \eq{thpt}), to be half-way across the lattice (i.e. 
$t_f=24$) and symmetrize the \thptfns\ about that point using
Euclidean time reversal, also to reduce the statistical errors. 

We observe a plateau in the ratio $R^\mu(t)$ of
\eq{fitratio} around $t=12$, typically extending over 5 time slices.
Therefore, we fit the ratio $R^\mu(t)$ over the range $t=11,12,13$ to
the form given in \eq{fitratasym} for all momentum and heavy-quark
mass combinations. For the purpose of illustration we plot, in
\fig{plat}, the ratio $R^0(t)$ vs.~$t$ for the case where the initial
meson has momentum $(\pi/12a,0,0)$ and the final meson, momentum
$(0,0,0)$. We fit the
\twptfns\ to the asymptotic form of \eq{twoptlim} in the range $t=11$
to $t=22$. The results of these later fits are given in \tab{twoptfits_6.2}.


Statistical errors are obtained from a bootstrap
procedure~\cite{EFRON}.  This involves the creation of 200 bootstrap
samples from the original set of 60 configurations by randomly
selecting 60 configurations per sample (with replacement). Statistical
errors are then obtained from the central 68\% of the corresponding
bootstrap distributions as detailed in Ref.~\cite{qlhms}.


Use of the HQET implies a choice of the expansion parameter, $m_Q$,
and this requires some care~\cite{beneke,bigi}.  We define $m_Q$ as
follows:
\be
m_Q = \frac{a^{-1}}{4}\l(3M_V^\chi+M_P^\chi\r)-\lbar_\chi
\ ,
\label{hq_mass}
\ee
where $M_P^\chi$ and $M_V^\chi$ are the relevant, chirally
extrapolated pseudoscalar meson and vector meson masses in lattice
units (see \tab{Q_masses_6.2}). Since these masses correspond to
heavy-light mesons whose antiquark is massless, the light degrees
of freedom carry an energy $\lbar_\chi=0.50~\gev$ as discussed after
\eq{corr}.

In \tab{rad_corr_6.2p} and \tab{rad_corr_6.2m}, 
we tabulate the results that we obtain for the
radiative corrections, $\b^+(\w;m_Q,m_{Q'})$ and
$\b^-(\w;m_Q,m_{Q'})$, of \eq{corr} for various combinations of the
heavy-quark masses and for a few values of $\w$. As mentioned in
\sec{intro}, we determine these corrections with the help of Neubert's
work\cite{short}.  Since our results for the form factors are obtained
in the quenched approximation, we set the number of quark flavours to
zero and assume no particle thresholds in Neubert's
expressions\footnote{There is, in fact, no rigorous way of running
quenched lattice QCD results since the lattice cutoff $a^{-1}$ is
adjusted to incorporate in part the effects of quenching.}.



\section{$Z_V$, Discretization Errors and how to Subtract them 
Non-Perturbatively}
\label{discerr}

Throughout this study of semi-leptonic weak decays of heavy mesons, we
use an $\ord{a}$-improved fermion action and take for the lattice
vector current, the ``improved" operator $V^\mu_I$ of \eq{improp}, as
discussed in \sec{laao}. In concrete terms, this means that we expect
mass-dependent discretization errors to be of $\ord{\a_s am_Q}\sim
5\%$ and $\ord{(am_Q)^2}\sim 10\%$ at the charm mass
\footnote{For this estimate, we use the boosted value of the coupling
constant $g_{boost}^2=(8\kcrit)^4\,g^2\simeq 1.66$ and the improved
bare mass defined before \protect{\eq{eq:zvfit}} with $\k_Q=0.129$.}
instead of $\ord{am_Q}\sim 40\%$ and 
$\ord{(am_Q)^2}\sim 10\%$ as they would be without $\ord{a}$-improvement.
Thus, despite the improvement expected, discretization errors
in our calculation could be significant.

Discretization errors in the lattice evaluation of the matrix element
of \eq{matelt} can be parametrized as follows:
\bea
&&Z_V(\as{a})
\ \frac{\la P'(\s p')|\bar Q'\tGm Q| P(\s p)\ra_{\rm lat.}}{\sqrt{M_PM_{P'}}}
=\nonumber\\
&&(v+v')^\mu\l(1+d^+(\w)\r)h^+(\w)+
(v-v')^\mu\l(1+d^-(\w)\r)h^-(\w)
+\ord{a^2}\ ,
\label{discerreq}
\eea
where $Z_V(\as{a})$ is the usual renormalization constant which
relates the lattice vector current to the continuum one
\footnote{It is important to note that similar discretization errors
are present for all definitions of the current, even the conserved
current away from the forward direction.}. Because this constant
describes physics that takes place above and around the lattice
cutoff, it is perturbative and independent of the initial and final
states.  

In \eq{discerreq} $d^+$ and $d^-$ are the Euclidean-invariant
discretization errors to all orders in $a$.  At $\ord{a^2}$ the
hypercubic group allows for additional errors which depend on the
Lorentz index of the vector current.  The discretization errors are
non-perturbative and depend on the initial and final states, because
they correspond to matrix elements of higher-dimension operators which
are artefacts of lattice regularization.  In addition, they depend on
the procedure used to cancel all the factors which relate the
three-point function to the matrix element (see
\eq{thptlim}). We adopt the expedient of assuming that we can absorb
the Euclidean-invariant discretization errors into an effective
renormalization constant $Z_V^{eff}$.

In the remainder of this section, we will attempt to quantify the
discretization errors in our calculation more precisely and describe a
procedure which enables us to subtract them, at least partially.

\subsection{Determination of $Z_V^{eff}$}
\label{dozv}
To study discretization errors, we define an effective renormalization
constant, $Z_V^{eff}$, for vector currents composed of degenerate
quark fields (i.e. of the form $\bar q\g^\mu q$) by
\be
Z_V^{eff}=\frac{1}{2}\frac{C_2(t_f;\s p)}{C_3^0(t;\s p,\s 0)}
\label{eq:zvc2c3}
\ee
for $t_f=T/2$, where $T$ is the temporal extent of the lattice, 
$C_3^\mu$ and $t_f$ are defined in \eq{thpt} and $C_2$ in
\eq{twopt}. In the absence of discretization errors, \eq{eq:zvc2c3}
yields a very accurate non-perturbative determination of the renormalization
constant $Z_V$. To see that the ratio of \eq{eq:zvc2c3} is in effect
$Z_V$, one must use the fact that the
forward matrix element of the temporal component of the vector current
is the charge, up to a trivial normalization factor. The factor of
$1/2$ comes from our boundary conditions (see
\eq{twoptlim}). Unless
stated otherwise, we will take $\s p = \s 0$.  In the presence
of the discretization errors described in \eq{discerreq}, however, the ratio
of \eq{eq:zvc2c3} becomes
\be
Z_V^{eff}=Z_V\left(1-d^+(1)+\ord{a^2}\right)
\label{zveff}
\ .
\ee

We start the discussion with a review of the determinations of
$Z_V^{eff}$ for currents composed of degenerate light-quark fields,
between pseudoscalar states composed of degenerate, light quarks and
antiquarks where we expect $Z_V^{eff}$ to be close to $Z_V$.  Using 10
gluon configurations from our simulation at $\beta = 6.2$, we find
\cite{jonivar}
\bea
Z_V^{eff} & = & 0.8314(4) \hspace{0.25in} {\rm at\ }\kappa = 0.14144\nonumber\\
Z_V^{eff} & = & 0.8245(4) \hspace{0.25in} {\rm at\ }\kappa = 0.14226\nonumber\\
Z_V^{eff} & = & 0.8214(6) \hspace{0.25in} {\rm at\ }\kappa = 0.14262
\label{eq:zvjonivar}
\eea
%

These results confirm that discretization
errors are small for light quarks (less than about 2\%), and we take
\be
Z_V=0.82(1)
\label{eq:zvbest}
\ee
as our best estimate for $Z_V$. This value is also consistent with the
expectations from one-loop perturbation theory \cite{borrelli}:
\be
Z_V = 1 - 0.10 g^2 + O(g^4)\simeq 0.83\mbox{   at   }\beta=6.2
\label{eq:zvpert}
\ee
when evaluated using the boosted value of the coupling constant, 
obtained from the mean field resummation of tadpole diagrams\cite{lm}. 

We now turn to the evaluation of $Z_V^{eff}$  using
\eq{eq:zvc2c3} for degenerate heavy-quark currents between 
pseudoscalar mesons consisting of a heavy quark $Q$ and a light
antiquark $\bar q$.  The results and, in particular, the difference
from the value in
\eq{eq:zvbest}, give us a measure of the size of the 
discretization errors, which are of $\ord{\alpha_s
am_Q}$ and $\ord{a^2 m_Q^2}$ here. In \tab{tab:zv14144}, we
present the results for $Z^{eff}_V$, obtained from the simulation at
$\beta = 6.2$ for four values of the heavy-quark mass, and with the
light-quark mass corresponding to $\kappa_q = 0.14144$, and from a simulation
at $\b=6.0$ for three values of the heavy-quark mass and with the light
quark hopping parameter equal to 0.144
\footnote{The simulation at $\b=6.0$ was performed with 36 quenched 
gauge field configurations on a $16^3\times 48$ lattice using the
$\ord{a}$-improved SW action of \eq{sclover}. For details of the simulation,
please see Ref.~\cite{henty}}.
Also tabulated are estimates of the
improved, bare mass of the heavy quark, $m_Q^I$, defined by
$am_Q^I=am_Q^0(1-(1/2)am_Q^0)$, where $am_Q^0=(1/2)(1/\kQ-1/\kcrit)$.

In \fig{fig:zv1} we plot the results for $Z^{eff}_V$ as a function of
$m_Q^Ia$ for the two values of $\b$.  
Fitting this behavior to a quadratic function of $m_Q^Ia$,
\be
Z^{eff}_V(\kappa_Q) = A + B m_Q^Ia + C (m^I_Qa)^2
\label{eq:zvfit}
\ee
we find A = 0.814(2) (A = 0.791(4)), B = 0.342(12) (B = 0.397(18)) and 
C = $-0.072(18)$ (C  = $-0.120(20))$ at $\b=6.2$ ($\b=6.0$). 
These fits are excellent. It is
interesting to note that the results extrapolate to approximately 0.81
(0.79) in the chiral limit and are thus in good agreement with the values
determined using light quarks as can be seen in \fig{fig:zv1} where we
have also plotted the light-quark values for $Z_V^{eff}$ 
given in \eq{eq:zvjonivar}. 
This fact together with the observation
that the size of the mass-dependent effects for a given $am_Q^I$ 
is very similar at the two 
values of $\b$ gives us confidence that the
mass dependence we observe is indeed due to discretization errors.

Further results from the simulation at $\b=6.2$ are presented in
\tab{tab:zvmup} and in \fig{fig:zv2}. 
For $\kappa_Q$ = 0.129 and 0.121 we have evaluated
$Z^{eff}_V(\kappa_Q)$ at three values of the mass of the light
quark. The results can be seen to be practically independent of the
mass of the light quark. We have also evaluated $Z^{eff}_V$ using
\eq{eq:zvc2c3} with $\s p = (\pi/12,0,0)$ and
$\kappa_q=0.14144$\footnote{\label{footseven}The statistical errors in
$Z^{eff}_V(\kappa_Q)$ are tiny, due to a cancellation in the ratio
(\ref{eq:zvc2c3}) of the fluctuations in the numerator and
denominator.  In order to get such a dramatic cancellation of the
fluctuations it is necesary to have precisely the same momentum in the
numerator and denominator. If, for example, we take $\s p =
(\pi/12a,0,0)$ in $C_3$ but average over all 6 equivalent momenta in
$C_2$, ($(\pm\pi/12a,0,0),\ (0,\pm\pi/12a,0),\ (0,0, \pm\pi/12a)$), 
then the statistical error in the ratio increases enormously.}. The
difference between the results obtained with $\s p = (\pi/12a,0,0)$
and with $\s p = \s 0$ is less than 1\%.  Finally, we have determined
$Z^{eff}_V(\kappa_Q)$ using
\be
Z^{eff}_V = \frac{p^1}{2E_P(\s p^2)} \frac{C_2(t_f;\s p)}{C_3^1(t,;\s
p,\s 0)}
\ ,
\label{zvspat}
\ee
for $\s p = (\pi/12a,0,0)$ and $t_f=T/2$ and where $E_P(\s p^2)$ is
the energy of the meson with momentum $\s p$. Now it is no longer the
charge operator which appears in $C_3$, and the statistical errors
increase significantly (see \fig{fig:zv2}). The values of $Z_V^{eff}$
given by \eq{zvspat} are consistent
with those obtained with $\mu = 0$ to within 1.5 standard deviations.

\subsection{Implications of the Results for $Z_V^{eff}$}
The results for $Z^{eff}_V(\kappa_Q)$ with $\s p=\s 0$ presented above
differ from the value of $Z_V$ given in \eq{eq:zvbest} by about
10-20\% for the range of quark masses used in our simulations (for
$\kappa_Q=0.129$, which corresponds approximately to the charm quark
for both values of $\beta$, the difference is about 12\% ). This
difference is a good indication of the size of mass-dependent
discretization errors in our calculation; it is consistent with our
expectation that they should be of $\ord{\alpha_s am_Q }$ and
$\ord{a^2 m_Q^2}$.

Our results for $Z^{eff}_V$ also enable us to quantify the dependence
of discretization errors on momentum as well as on the Lorentz
component of the current used to obtain them. As noted in the previous
subsection, the difference between the results obtained with $\s p =
(\pi/12a,0,0)$ and with $\s p = \s 0$ is less than 1\%. This is a
clear indication that as long as we limit ourselves to momenta $\s p$
such that $|\s p|\le\pi/12a$, discretization errors proportional to
$a\s p$ are small. As for the dependence of $Z^{eff}_V$ on the Lorentz
index of the current, the situation is less clear. The ratio
$Z^{eff}_V(0.121;\mu=1)/Z^{eff}_V(0.121;\mu=0)$ for $\s
p=(\pi/12a,0,0)$ indicates that this dependence could be as large as
11\%. However, given that the statistical errors on
$Z^{eff}_V(0.121;\mu=1)$ are quite large, much of this dependence
could be a statistical fluctuation.


\subsection{Non-Perturbative Subtraction of $am_Q$-Errors}
\label{npsoame}

Having isolated and quantified the different sources of discretization
errors, we now investigate the possibility of subtracting these 
errors.  It is important to remember that these discretization
errors are given by matrix elements of higher dimension operators:
they are non-perturbative and will depend on the initial and final
states between which the current $V^\mu_I$ is sandwiched. This means
that in any attempt to subtract them, one must evaluate the
relevant corrections with states as similar as possible to the ones which
appear in the matrix element of interest.  With this in mind, we have
devised the following subtraction procedure.

Firstly, as mentioned in \sec{discerr}, we assume that the
mass-dependent discretization errors can be absorbed into an overall
effective normalization:
\be
\la P'(\s p')|\bar Q'\tGm Q| P(\s p)\ra_{\rm lat.}=
\frac{\la P'(\s p')|\bar Q'\gm Q| P(\s p)\ra}{Z^{eff}_V(aM_P,aM_{P'};\mu)}
\ ,
\label{overallzv}
\ee
where $\tGm$ is defined in \eq{gamimprov}.  

Secondly we find a normalization condition, i.e.\ a kinematical point
at which we know the physical value of the matrix element. For the
case of degenerate transitions, this normalization condition is
simple; electromagnetic charge conservation requires that
$h^+(1;m_Q,m_Q)=1$. For the case of non-degenerate transitions, the
normalization condition is slightly more complicated. HQET requires,
as we saw earlier, that $h^+(1;m_Q,m_{Q'})=1+\b^+(1;m_Q,m_{Q'})
+\g^+(1;m_Q,m_{Q'})$. The radiative
corrections, $\b^+(1;m_Q,m_{Q'})$, we know from perturbation theory.
The power corrections, $\g^+(1;m_Q,m_{Q'})$, are non-perturbative and
are yet to be determined in a model-independent and reliable way.  We
are, however, helped here by Luke's theorem which guarantees that
$h^+(1;m_Q,m_{Q'})$ is free of corrections proportional to a single
power of the inverse heavy-quark masses. Thus, $\g^+(1;m_Q,m_{Q'})
\sim \epsilon_{Q,Q'}^2+\ord{\epsilon_{Q,Q'}^3}$ 
and is small. In fact, as we shall see
shortly, the exact size of $\g^+(1;m_Q,m_{Q'})$ is not important for
determining the Isgur-Wise function. Thus, we will take our
normalization condition to be 
\be
h^+(1;m_Q,m_{Q'})\equiv 1+\b^+(1;m_Q,m_{Q'})
\label{normcond}
\ee
for both degenerate and non-degenerate transitions.

This condition determines
$Z^{eff}_V$. With $Z^{eff}_V$ defined by \eq{overallzv}
we find
\be
Z^{eff}_V=\frac{1+\b^+(1;m_Q,m_{Q'})}{h^+_{\rm lat.}(1;m_Q,m_{Q'})}
+\ord{a^2}
\ ,
\label{zveffhq}
\ee
where $h^+_{\rm lat.}(1;m_Q,m_{Q'})$ is the zero-recoil form factor obtained
from our lattice calculation and the $\ord{a^2}$ stands for discretization
errors which are not Euclidean invariant. Because, as we mentioned earlier,
discretization errors made in the evaluation of a matrix element 
depend not only on the
initial and final states considered, but also on the procedure used to
obtain the matrix element, it is very important to obtain
$h^+_{\rm lat.}(1;m_Q,m_{Q'})$ with a procedure as similar as possible to
the one used to obtain $\hpQQpw$ for $\w\neq 1$. Thus, we get
$h^+_{\rm lat.}(1;m_Q,m_{Q'})$ from the time-component of the ratio of
\eq{fitratio} with $\s p'=\s q=0$. For degenerate
transitions, there is another zero-recoil channel, which corresponds
to the forward scattering of a meson with one unit of lattice
momentum. We do not use the $h^+_{\rm lat.}(1)$ from this channel
to determine $Z^{eff}_V$ because it is statistically
much noisier than the one at zero momentum, and because it does not
correspond to a zero-recoil transition in the non-degenerate case.

Now, to subtract the discretization errors that $Z_V^{eff}$
incorporates, we simply define the continuum form factors to be
\bea
\hpQQpw & \equiv & (1+\b^+(1;m_Q,m_{Q'})) 
\frac{\hplatQQpw}{h^+_{\rm lat.}(1;m_Q,m_{Q'})} \nonumber \\
\hmQQpw & \equiv & (1+\b^+(1;m_Q,m_{Q'}))
\frac{\hmlatQQpw}{h^+_{\rm lat.}(1;m_Q,m_{Q'})}
\ .
\label{hphmdef} 
\eea
This definition yields
\bea
\hpQQpw \simeq &&[1+\bpQQpw+\gpQQpw-\g^+(1;m_Q,m_{Q'})\nonumber\\
& & +d^+(\w;m_Q,m_{Q'})
-d^+(1;m_Q,m_{Q'})]\iw
\ ,
\label{myhp}
\eea
up to higher-order discretization errors, radiative and power
corrections. It is clear from \eq{myhp} that part of the
discretization errors have been subtracted. The subtraction is only
complete, however, if $d^+(\w)$ is a constant. For the form factor
$h^-(\w)$ it is less clear that we are subtracting the relevant
discretization errors.  Indeed, according to the definition of
\eq{hphmdef} the discretization errors in $h^-$ are
$(d^-(\w)-d^+(1))+\ord{a^2}$. However, the assumption behind
this subtraction is the same as the one made by Lepage, Mackenzie
and Kronfeld \cite{kronmac} in their attempt to remove discretization
errors by modifying the normalization factors which match fermion
fields to their continuum counterparts.

We wish to emphasize here that our subtraction procedure removes
non-perturbatively all discretization errors which do not break
Euclidean invariance and does so to all orders in $a$. Thus, amongst
others, all discretization errors which are removed in mean-field
theory by the procedure of Kronfeld, Lepage and Mackenzie will be
removed non-perturbatively by our procedure.

As \eq{myhp} indicates, in subtracting discretization errors in $h^+$,
we also subtract the zero-recoil power corrections, $\g^+(1)$, thereby
losing the ability to determine them.  This is not a serious concern
in practice because these ought to be small --- they are
proportional to the square of the inverse heavy-quark mass --- and
therefore difficult to isolate reliably.  It does mean, however, that
even if we can reduce all of our errors to the percent level, we will
be unable to obtain the zero-recoil power corrections to the form
factor $h_{A_1}$ relevant for $\btodsl$ decays if we use an analogous
subtraction procedure. This is unfortunate
because these $1/m_c^2$-corrections are one of the dominant
theoretical uncertainties in the extraction of the CKM matrix element
$V_{cb}$ from experimental studies of these decays (see \sec{eovcb}).

For obtaining the Isgur-Wise function, however, the fact that our
normalization procedure subtracts these zero-recoil power corrections,
which are non-perturbative and difficult to quantify, is an advantage.
Our hope is that, once these corrections are subtracted, the resulting
form factor will have smaller power corrections away from zero recoil.

There is one additional issue surrounding normalization that we wish
to address. As indicated in the previous subsection, the
discretization errors on our
\thptfns\ are typically larger for spatial than for temporal channels 
(see \tab{tab:zvmup}).  Thus, we ought to normalize spatial and
temporal channels differently.  For degenerate transitions, this is
possible because there is a zero-recoil \thptfn\ which has a non-zero
spatial component: $C_3^\mu(t;a\s p=(\pi/12,0,0),\s 0)_{Q\to Q}$. As
mentioned above, however, this
\thptfn\ does not correspond to a zero-recoil decay when $Q\ne Q'$.
We have no zero-recoil \thptfn\ with a non-vanishing spatial
component for non-degenerate transitions (momenta are quantized on the
lattice). So, in order to treat degenerate and non-degenerate
transitions in the same way, we will normalize $h^+$ and $h^-$ as
described in \eq{hphmdef}.

It is important to note that because $h^+$ is obtained from the
temporal component of \eq{fitratasym} alone (see end of \sec{tpfalff})
and is correctly normalized, it does not suffer from the possible
discrepancy in normalization between temporal and spatial channels. It
is $h^-$, obtained from both temporal and spatial components, which in
fact will absorb this discrepancy. For degenerate transitions, where $h^-$
is in principle zero, the values of $h^-$ that we obtain are therefore
an indication of how large an error this discrepancy can induce in the
form factors. For non-degenerate transitions, the values of $h^-$ we
obtain, though contaminated to some extent by discretization errors,
can be used to put bounds on the physical $h^-$.




\section{The Form Factors $\hpw$ and $\hmw$}
\label{ffftsabe6.2}


\subsection{Results at Fixed Light-Quark Mass}
\label{raflqm}

In \tab{hp.4144}  ---  \tab{hp.4262} we present the measurements of
$\hpw$, $\hpw/(1+\bpw)$ and $\hmw$ which we obtain for all available
combinations of the initial and final heavy-quark masses for light
antiquarks with $\kq=0.14144$ (\tab{hp.4144}), 0.14226 (\tab{hp.4226})
and 0.14262 (\tab{hp.4262}). In these tables, the first
$\chi^2/d.o.f.$-column corresponds to the fit which yields $h^+_{\rm
lat.}$ from the temporal component of the ratio $R^\mu$ assuming
$h^-_{\rm lat.}(\w)=0$. The second $\chi^2/d.o.f.$-column corresponds
to the fit which gives $h^-_{\rm lat.}(\w)$ from both temporal and
spatial components when holding $h^+_{\rm lat.}$ fixed to its
temporal-component value. The number of degrees of freedom ($d.o.f.$)
that we quote in this second column depends on the momentum channel
because the number of non-vanishing equations for $h^+_{\rm lat.}$
and $h^-_{\rm lat.}$ varies with initial and final meson momenta.

As evidenced by the low values in the first $\chi^2/d.o.f.$ column of
all three tables, the fits which give $h^+_{\rm lat.}$ from the
temporal component of $R^\mu$ are very good. The fact that the values
in the second $\chi^2/d.o.f.$ column of these tables are generally
larger may be due to the fact that spatial and temporal components of our
\thptfns\ may have different discretization errors, as mentioned in
\sec{tpfalff}. When we fit these components simultaneously to the 
asymptotic form of \eq{fitratasym} while holding $h^+_{\rm lat.}$
fixed, we are not fitting to a form which takes into account these
discrepancies and consequently obtain a larger $\chi^2/d.o.f.$. As
discussed in \sec{tpfalff}, however, this fitting strategy is the
only one that guarantees that $h^+$ does not suffer significantly from
discretization errors.

Given the number of different mass combinations and momentum channels
we have, our results for $\hpw/(1+\bpw)$ are remarkably consistent.
Keeping the light-quark mass fixed we find that for recoils $\w$ which
are approximately the same, the values of $\hpw/(1+\bpw)$ 
are equal within errors even when they
are obtained from different momentum and/or heavy-quark mass
combinations.  This supports the validity of our procedure and is
also an indication that the radiative corrections obtained using Neubert's
results\cite{short} are accurate. The fact that $\hpw/(1+\bpw)$ does
not appear to depend strongly on the mass of the heavy quarks is also
an indication that the coefficients of the corrections proportional to
inverse powers of the heavy-quark masses are not very large (see
\sec{hqmdohp}).

There are two momentum combinations on which we wish to comment. The
first is $\s p=(\pi/12a,0,0)$ to $\s p'=(\pi/12a,0,0)$ which, for
degenerate transitions, has zero recoil. For such transitions, current
conservation requires that $h^+((\pi/12a,0,0)\to(\pi/12a,0,0))$ equal
1. We find values of $h^+((\pi/12a,0,0)\to(\pi/12a,0,0))$ which are
just barely consistent with 1 at the level of 1 $\sigma$ for
$\kq=0.14144$. The situation deteriorates when the mass of the light
quark decreases (see \tab{hp.4226} and \tab{hp.4262}). Since for given
quark masses $h^+((\pi/12a,0,0)\to(\pi/12a,0,0))$ is extracted from
a single three-point function (the one with $\s p'=(\pi/12a,0,0)$ and
$\s q=(0,0,0)$), it is much more susceptible to statistical fluctuations
than most other values of $h^+$ which are obtained from averages
of three-point functions over many equivalent momentum combinations. 
To show that this slight discrepancy is statistical, we consider two measures
of $h^+((\pi/12a,0,0)\to(\pi/12a,0,0))$ which use the same \thptfn\
and normalization. The first is
\be
h^+((\pi/12a,0,0)\to(\pi/12a,0,0))=\frac{Z^{eff}_V\l(\kQ;\mu=4;(0,0,0)\to
(0,0,0)\r)}{Z^{eff}_V\l(\kQ;\mu=4;(\pi/12a,0,0)\to(\pi/12a,0,0)\r)}
\ee
with $Z^{eff}_V$ defined in \sec{dozv}. The second is the 
expression above multiplied by the ratio $C_2(t_f;(\pi/12a,0,0))/$
$\bar C_2(t_f;\pi/12a)$ where $\bar C_2(t_f;\pi/12a)$ is
the average of the six $\s p=(\pm \pi/12a,0,0)$,
$(0,\pm \pi/12a,0)$ and $(0,0,\pm \pi/12a)$ \twptfns
\footnote{This ratio
of two-point functions should of course be 1 in the limit of infinite
statistics.}.
Using the values of
$Z^{eff}_V$ given in
\tab{tab:zvmup}, the first procedure gives 
$h^+((\pi/12a,0,0)\to(\pi/12a,0,0))$ equal to 1 to within 1\%, even
when the mass of the light spectator antiquark is reduced, while the
second procedure gives results very much in line with the rather low
results of \tab{hp.4144}, \tab{hp.4226} and \tab{hp.4262}. The
reason why the first procedure is more precise is explained
in footnote~\ref{footseven}. Moreover, that the
results given by the second
procedure agree better with our standard procedure for
obtaining $h^+$ should not be too surprising, as the latter
also makes use of average two-point functions.

The second small inconsistency we wish to comment on is the one
arising from the comparison of $h^+\l((0,0,0)\to(\pi/12a,0,0);m_Q,m_{Q'}\r)$
with $h^+\l((\pi/12a,0,0)\to(0,0,0);m_{Q'},m_Q\r)$, for which
$\w$ is the same. To check the
validity of our results, we have re-analysed our data by fitting our
\thptfns\ directly to the asymptotic form given in \eq{thptlim}, fixing
the energies and wavefunction factors which appear in this asymptotic
form to their \twptfn\ values, and
normalizing the resulting $\hplatw$ according to
\eq{hphmdef}.  This procedure yields values for $h^+$ which are nearly
identical to the ones given in Tables \ref{hp.4144}, \ref{hp.4226} and
\ref{hp.4262}. The only values that change significantly compared to
the size of their error bars are those corresponding to
$h^+\l((0,0,0)\to(\pi/12a,0,0);m_Q,m_{Q'}\r)$.  In this different way
of analyzing the data, the values we find for
$h^+\l((0,0,0)\to(\pi/12a,0,0);m_Q,m_{Q'}\r)$ are lower, making them
nearer the values for $h^+\l((\pi/12a,0,0)\to(0,0,0);m_{Q'},m_Q\r)$.
This partial bridging of the gap, however, comes at the expense of
large $\chi^2/d.o.f.$'s, ranging from 2 to 5. One can fix both
problems --- bridging the gap completely and bringing the $\chi^2/d.o.f.$
down --- by fitting the time component of our \thptfns\ to %
\be
\label{thptlimde}
C^0_3(t;\s p',\s q)_{Q\to Q'}\mathop{\longrightarrow}\limits_{t,t_f-t\to
\infty}
\frac{Z_P(\s p^2)Z_{P'}(\s p^{\prime 2})}{4E_PE_{P'}}
\ e^{-(E_P-E_{P'}+\delta E)t-E_{P'}t_f}\,\la P'(\s p')|V^0_I(0)
|P(\s p)\ra\ ,
\ee
with an extra parameter $\delta E$, instead of to the form given in
\eq{thptlim}.  The parameter $\delta E$ is designed to absorb
slight statistical differences in the time behavior of two- and
\thptfns. One would worry about the consistency of adding this
extra parameter if it were to be large compared to the values of
the various energies which enter the exponential factor in
\eq{thptlimde} since it is inconsistent to allow for changes in the
energies while holding wavefunction factors fixed --- the two quantities
are extremely correlated --- and it is inconsistent to claim that
$E_P-E_{P'}$ is different for two- and \thptfns\ but that
$E_{P'}$ is the same.  However, we find values
of $\delta E$ which are on the order of $10^{-3}$ and consistent with
zero.

In addition to reconciling the values for
$h^+\l((0,0,0)\to(\pi/12a,0,0);m_Q,m_{Q'}\r)$ and
$h^+\l((\pi/12a,0,0)\to(0,0,0);m_{Q'},m_Q\r)$, this method increases
the statistical errors on all values of $\hpw$ because of the
additional freedom introduced by the new parameter. We do not use this
new fitting method as our main one because of the potential
inconsistencies mentioned above and because the introduction of the
extra parameter $\d E$ is difficult to generalize sensibly to
situations where one simultaneously fits more than one four-vector
component of a \thptfn.

The results given by all of these different methods of analyzing the data
are consistent within statistical errors. 
This gives us faith
that the results for $h^+$ in
\tab{hp.4144}, \tab{hp.4226} and \tab{hp.4262} are 
valid representations of our data.  The most likely reason, then, for
the slight discrepancy between
$h^+\l((0,0,0)\to(\pi/12a,0,0);m_Q,m_{Q'}\r)$ and
$h^+\l((\pi/12a,0,0)\to(0,0,0);m_{Q'},m_Q\r)$ is that it arises from
the same statistical fluctuation that yields the low value for
$h^+\l((\pi/12a,0,0)\to(\pi/12a,0,0);m_Q,m_Q\r)$. Like the three-point
function which gives $h^+\l((\pi/12a,0,0)\to(\pi/12a,0,0);m_Q,m_Q\r)$,
the one from which $h^+\l((\pi/12a,0,0)\to(0,0,0);m_{Q'},m_Q\r)$ is
obtained is not averaged with equivalent three-point functions.
$h^+\l((0,0,0)\to(\pi/12a,0,0);m_Q,m_{Q'}\r)$, on the other hand, is
obtained from the average of the six \thptfns\ corresponding to the
transitions $(0,0,0)\to(\pm\pi/12a,0,0)$, $(0,\pm\pi/12a,0)$,
$(0,0,\pm\pi/12a)$.

\bigskip
As mentioned earlier, current conservation requires that $\hmw\equiv
0$ for degenerate transitions. In order to determine whether our
results are consistent with this requirement, we must know how large
the discretization errors on $\hmw$ might be. As suggested by the results
for $Z_V^{eff}$ (see \tab{tab:zvmup}), there may be discretization
errors of the order of 10\% which cause 
the spatial components of our \thptfns\ to be 
low compared to the temporal components. One can easily convince
oneself, by considering the set of equations corresponding to different
components of the vector current in
\eq{fitratasym}, that such discretization errors would cause 
$|\hmw|$ to take on values up to
about $0.1$. This is indeed what we find. Thus, to the level of accuracy
with which we can determine $|\hmw|$, we can conclude that $\hmw$ is
consistent with zero for degenerate transitions.

For non-degenerate transitions, the results we obtain for $\hmw$
resemble very much those found in degenerate transitions. They are
consistent with zero at the level of 2$\sigma$\footnote{ $\hmw$ is
large with large errors when $\s p=\s p'=(\pi/12a,0,0)$, because its
coefficient in the equation which determines it is
$v^1-v^{1\prime}=(\pi/12)(1/am_Q- 1/am_{Q'})$, a small number when
$m_Q\simeq m_{Q'}$.}.  Thus, as far as we can resolve, $\hmw$ is small
for all $\w$, most probably less than about $0.1$ to $0.2$.

Using this information, we can now put a bound on the size of the
error that we are making on $\hpw$ by neglecting the contribution of
$\hmw$ to the temporal component of \eq{fitratasym}. Using the fact
that the ratio of velocity factors, $r\equiv
|v^0-v^{0\prime}|/(v^0+v^{0\prime})$, is at most $0.07$, and that
$\hpw$ is always greater than $0.6$, we find that the error we make on
$\hpw$ is at most $r_{\rm max}\hmw_{\rm max}/
\hpw_{\rm min}\simeq 1\%$ to 2\%. In most situations, if not all,
it will be smaller than that. Thus, neglecting the contribution of
$\hmw$ in obtaining $\hpw$ is a very good approximation indeed.



\subsection{Chiral Extrapolation}
\label{ce}

In the previous section, we determined $\hpw$ for many different
combinations of initial and final heavy quarks and for three light
antiquarks whose masses straddle that of the strange quark. In the
present section, we describe the extrapolation of our results for
$\hpw/(1+\bpw)$ to vanishing light-antiquark mass for which
$\kq=\kcrit=$0.14315(2)\cite{strange}. The chirally extrapolated
results are relevant for the study of semi-leptonic decays of
heavy-light mesons whose light antiquark is a $\bar u$ or a $\bar d$.
These results are summarized in \tab{hp.k_cr}.

The extrapolations are covariant and linear in the improved, bare
quark mass, $am_q^I=am_q(1-(1/2)am_q)$, where $am_q=(1/2\k_q
-1/2\kcrit)$. We fit $h^+/(1+\b^+)$ and $\w$ to the forms $\a_{h^+}
(am_q^I)+\b_{h^+}$ and $\a_\w(am_q^I)+\b_\w$, respectively. Then,
$h^+_{\rm crit}/(1+\b^+)=\b_{h^+}$ and $\w_{\rm crit}=\b_\w$. The
$\chi^2/d.o.f.$ for these extrapolations are given in columns four and
six of \tab{hp.k_cr}. As evidenced by the small values of these
$\chi^2/d.o.f.$'s, the extrapolations are for the most part very
smooth. The only extrapolation which has an anomalously large
$\chi^2/d.o.f.$ is the one for
$h^+\l((\pi/12a,0,0)\to(\pi/12a,0,0);m_Q,m_{Q'}\r)$. As mentioned in
\sec{raflqm}, even though current conservation requires that
$h^+\l((\pi/12a,0,0)\to(\pi/12a,0,0);m_Q,m_{Q'}\r)=1$ when $Q=Q'$, our
results do not quite satisfy this constraint due to a statistical
fluctuation. As Tables \ref{hp.4144},
\ref{hp.4226} and \ref{hp.4262} further indicate, this constraint is less and
less well-satisfied as the mass of the light quark is reduced. The
correlated extrapolation appears to correct for this downward trend in
the data, but does so at the expense of a large $\chi^2/d.o.f$.

We do not extrapolate $\hmw$ because this form factor potentially
suffers from rather large discretization errors as discusssed in
\sec{tpfalff} and is therefore not entirely physical.



\subsection{Interpolation to the Strange Quark}
\label{ittsq}

In the present section, we describe the interpolation of our results
for $\hpw/(1+\bpw)$ in the light-antiquark mass to the mass of the
strange quark ($\ks=$0.1419(1)\cite{strange}). The interpolated
results are relevant for the study of semi-leptonic decays of
heavy-light mesons which contain a strange antiquark. The results are
summarized in
\tab{hp.k_st}. They are obtained from the same covariant, linear fits
as the chirally-extrapolated results of \sec{ce} so that the
$\chi^2/d.o.f.$ are the same as in \tab{hp.k_cr}. The only difference
is that the interpolated results are
$h^+_s/(1+\b^+)=\a_{h^+}(am_s^I)+\b_{h^+}$ and
$\w_s=\a_\w(am_s^I)+\b_\w$, where $m_s^I$ is the improved, bare mass
of the strange quark.




\section{Dependence of $\hpw$ on Heavy-Quark Mass}
\label{hqmdohp}

Having obtained $\hpw$ to good accuracy, we can now attempt
to determine the dependence of this quantity on the masses of the
initial and final heavy quarks. For the purpose of this study, we have
calculated $\hpw/(1+\bpw)$ 
for additional heavy-quark combinations
when $\kq=0.14144$. We concentrate on the results which correspond to
our heaviest, light antiquark ($\kq=0.14144$) because these results
have smaller statistical uncertainties and will therefore enable us to
resolve the dependence of these results on heavy-quark mass more
accurately. We will assume, in the following, that our findings for
$\kq=0.14144$ provide a good description of the
behavior on heavy-quark mass of our results for smaller light-antiquark
masses.  That this assumption may be justified is confirmed by the
mild dependence of $\hpw$ on light-antiquark mass (see
\sec{dohpolqm}).

The first indication that the dependence of $\hpw/(1+\bpw)$ on
heavy-quark mass must be
very weak is shown in \fig{4144.degen}. In this figure, we plot
together the 
form factors $\hpw/(1+\bpw)$ for each of our four $Q\to Q$, degenerate
transitions with $\kq=0.14144$.  It is
natural to begin looking for small heavy-quark mass effects in this 
data because its
normalization is free of uncertainties associated with radiative or power
corrections (see \sec{npsoame}).

The four sets of data lie very much on the same curve. To show this
more precisely, we fit each set individually to the parametrizations
$\xi_{NR}(\w)$ and $s\xi_{NR}(\omega)$\footnote{Since we are only
interested in comparing $\hpw/(1+\bpw)$ for different heavy-quark
mass, any reasonable parametrization will do.}, where
\be
\xi_{NR}(\omega)\equiv{2\over \omega+1}\exp\left(-(2\rho^2-1)
{\omega-1\over\omega+1}\right)
\label{bsw}
\ee
is a parametrization for the Isgur-Wise function suggested by
M.~Neubert and V.~Rieckert in\cite{neurieck}.  In \eq{bsw},
$\rho^2=-\xi'(1)$. We have introduced the supplementary parameter $s$
to absorb possible normalization errors. We summarize our findings in
\tab{rho2.degen.4144} and plot the fit curves in \fig{4144.degen}. 
These results clearly show that the four
different data sets are entirely compatible and suggest that the
dependence of $\hpw/(1+\bpw)$ on heavy-quark mass 
is expected, therefore, to be quite small over most of the range
of experimentally accessible recoils
\footnote{The fact that the values of $\chi^2/d.o.f.$ are relatively high for
all of these fits is explained after \eq{quad} in \sec{dohpolqm}.}.

Before interpreting this observation, let us quantify this
heavy-quark-mass dependence more precisely.  We will do so under the
assumption that this small dependence is due to power corrections. We
have also tested the assumption that it is due to $am_Q$-discretization 
errors but find that this assumption is less well
satisfied by our data (i.e. it leads to higher $\chi^2/d.o.f.$). 
According to \eq{corr}, we have
\be
\frac{\hpw}{1+\bpw}\simeq (1+\gpw)\iw
\label{hpw_bpw}
\ .\ee
Now, to leading order in the heavy-quark expansion,
\bea
\gpQQpw=&&g_Q\l(\w,\as{m_Q},z\r)\ \eQ+g_{Q'}\l(\w,\as{m_{Q'}},z\r)\ \eQp
\nonumber\\&&+\ord{\eQ^2,\eQp^2,\eQ\eQp}
\ ,
\eea
where $\epsilon_{Q\pnotp}=\lbar_{4144}/(2m_{Q\pnotp})$
\footnote{Here, $\lbar_{4144}$ is the energy carried by the
light degrees of freedom when $\kq=0.14144$.  We take it to be
$\lbar_{4144}=\frac{a^{-1}}{4}\l(3(M_V^{4144}-M_V^\chi)
+(M_P^{4144}-M_P^\chi)\r)+\lbar_\chi=0.63~\gev$.}, and $z=m_Q/m_{Q'}$.
The functions $g_Q$ and $g_{Q'}$ correspond to matrix elements of
dimension-five operators in the HQET Lagrangian evaluated at order
$\ord{\epsilon_{Q\pnotp}^0}$. These two functions must be 
equal when $Q=Q'$. They must also be equal in the absence of radiative
corrections as HQET cannot distinguish the flavour of a heavy quark at
order $\ord{\epsilon_{Q\pnotp}^0}$. In the presence of radiative
corrections, however, the two functions will have different values
when $Q\neq Q'$. The amount by which they differ will be partly
governed by logarithms of the heavy-quark masses, as indicated by the
presence of the running coupling constant in the functions' arguments.
The way in which $g_Q$ and $g_{Q'}$ depend on $z$ will also be
different. Nevertheless, since the difference between $g_Q$ and
$g_{Q'}$ is a difference of radiative corrections, it is very small.
We will neglect this difference in what follows and assume that
\be
\label{g_def}
\gpQQpw=g(\w) \l(\eQ+\eQp\r)
+\ord{\eQ^2,\eQp^2,\eQ\eQp}
\ .
\ee
It is worthwhile noting, at this point, that Luke's theorem requires
\be
g(1)=0
\ .
\label{goneeqzero}
\ee

To evaluate $g(\w)$ we need $\hpQQpw$ at a fixed $\w$ for different
$Q$ or $Q'$. Because momenta on the lattice are quantized this is
difficult to achieve. There is one kinematical situation, however,
where we have enough measurements of $\hpw$ at fixed $\w$ for
different heavy quarks to determine $g(\w)$. When the momentum of one
of the mesons vanishes, $\w$ becomes independent of that
meson's mass. There are four values of $\w$ for which this happens, 
corresponding to $|\s p|=\pi/12a$ and $|\s p'|=0$ for $\k_Q=0.121,\,0.129$, 
and $|\s p|=0$ and $|\s p'|=\pi/12a$ for $\k_{Q'}=0.125,\,0.133$. For
each of these four points, we have four measurements of $\hpw$ corresponding
to four different values of the mass of the meson which is at rest. We
pick one of these four measurements and use it to normalize the
remaining three. Thus, we construct the ratio:
\bea
R^+(\w,x) & \equiv & \frac{1}{\epsilon_{Q^1}}\l(1-\frac{\hpQQpw/(1+\bpQQpw)}
{h^+(\w;m_{Q^1},m_{Q'})/(1+\b^+(\w;m_{Q^1},m_{Q'}))}\r)\nonumber \\
& = & g(\w) (1 - x)+\ord{\e_Q,\e_{Q_1}}\ ,
\label{rp_def}
\eea
with $x\equiv m_{Q^1}/m_Q$. Here we have assumed that it is the
initial meson which has vanishing momentum. We then fit the resulting
three data points for $R^+$ at fixed $\w$ to a straight line in
$x$. The slope and intercept of this line is $g(\w)$ (see
\eq{rp_def}). We summarize the details of these fits in
\tab{rp_fit}. In \fig{rp_vs_x}, we show this data with the
corresponding fit (solid line) for each one of the four values of
$\w$. The data for $R^+$ satisfies the parameterization of
\eq{rp_def} surprisingly well. One should remember that all
power corrections are subtracted at $\w=1$ by our normalization
procedure (see \eq{myhp} and ensuing discussion). However,
this is not a problem if one is interested only in $\ord{\e_Q,\e_{Q'}}$
power corrections to $h^+$ since these must vanish at zero
recoil according (\eq{goneeqzero}).

In \fig{g_vs_w} we plot $g$ as a function of $\w$.  $g(\w)$ is
consistent with zero over the range of recoils $\w$ that we can
explore ($1\leq\w\leq1.1$). Since $g(\w)$ shows no trend over that
range and since the functions $\hpw/(1+\bpw)$ plotted in
\fig{4144.degen} exhibit no mass dependence over a range of recoils
from 1 to 1.4, we conclude that $g(\w)$ ought to remain small (less
than about 0.2) over the full range of experimentally interesting
recoils ($1\leq\w\leq 1.55$).  We believe that these results indicate
that the $1/m_Q$-corrections to $\hpw$ and the remaining
$am_Q$-discretization errors in $\hpw$ are genuinely small because we
explore a non-negligible range of heavy-quark masses --- from about 1
to 2 $\gev$.  It seems quite unlikely that discretization errors or
higher order power corrections would cancel the leading power
corrections over such a range.

Because $g(\w)$ appears to be less than about 0.2
over the full range of recoils, we predict that power corrections to the
form factor $h^+$ corresponding to physical $\btodl$ decays must be less
than about $\ord{\e_c^2}\simeq 3\%$ to
$\ord{0.2\times(\e_b+\e_c),\e_c^2}\sim 5\%$ to 10\% over the full range of
recoils for $m_b=4.80~\gev$, $m_c=1.45~\gev$ and
$\lbar=0.50~\gev$\cite{bible}\footnote{We
have included, in this estimate, potential higher order corrections
that may have been subtracted by our normalization procedure. }.
This is significantly smaller than the
$\ord{\e_c}\simeq 15\%$ ($\ord{\e_c^2}$ and $\ord{\e_b}$ may each contribute
an additional 5\%) that one may naively have expected.  It
appears, then, that the protection Luke's theorem
provides at zero recoil extends over the full 
range of recoils and that for the particular combination $\hpw/(1+\bpw)$
the flavour component of the heavy-quark symmetry is well
satisfied in the charmed sector. This is in stark contrast with our
findings for the decay constant, $f_D$, of the pseudoscalar $D$
meson\cite{qhldc}.  In Ref.\cite{qhldc} we find that the $\ord{\e_c}$
corrections to the heavy-quark limit prediction for this decay constant are
of the order of 30\%.

These results for $g(\w)$ also mean
that our results for $\hpw/(1+\bpw)$ are, to a good
approximation, infinite heavy-quark-mass results. Thus, the functions
$\hpw/(1+\bpw)$ that we measure are effectively 
\iwfns\ and we can consistently combine data corresponding to
different initial and final heavy quarks. This is what we do in the
following.

\bigskip
In principle one could also try to quantify power corrections to $h^-(\w)$.
In the absence of radiative corrections, we find from 
the results of Ref. \cite{bible} that these power corrections are given by:
\be
\g^-(w;m_Q,m_{Q'})=(1-2\eta(\w))\,(-\e_Q+\e_{Q'})
\ ,
\label{pcm}
\ee
where, like $g(\w)$ defined in \sec{hqmdohp}, 
$\eta(\w)$ is a subleading, universal form factor
\footnote{Luke's theorem does not constrain $\eta(\w)$ at $\w=1$ as it
did $g(\w)$.}.
\eq{pcm} indicates that power corrections proportional to $\e_Q$
are equal and opposite to those proportional to $\e_{Q'}$. This prediction
is consistent with the mass dependence we observe in our 
results. However, because our normalization procedure is optimized for
determining $h^+(\w)$ and not $h^-(\w)$, it is not clear to what extent
the mass-dependence due to power corrections can be resolved from that
due to discretization errors and to higher-order
power corrections coming from our normalization procedure (see \eq{hphmdef}). 



\section{Dependence of $\hpw$ on Light-Quark Mass:  Isgur-Wise Functions}
\label{dohpolqm}

We established in the previous section that, for fixed light-antiquark
mass, we can combine the results for $\hpw/(1+\bpw)$ corresponding to
different values of the initial and final heavy-quark mass. We further
established that the resulting function is an \iwfn: $\xi_{u,d}(\w)$
when the mass of the light antiquark vanishes; $\xi_s(\w)$
when the light antiquark is given the mass of the strange quark.

We plot $\xi_{u,d}(\w)$ and $\xi_s(\w)$ in \fig{iwud} and \fig{iws},
respectively.  We fit the corresponding data to the parametrizations
$s\,\xi_{NR}(\w)$, $\xi_{NR}(\w)$.  The parameter $s$ is added to
absorb possible uncertainties in the normalization of these form
factors.  Because the parametrization $\xi_{NR}(\w)$ is only one
of many possible parametrizations, we also fit our results to 
$s\xi_{lin}(\w)$, $\xi_{lin}(\w)$, $\xi_{quad}(\w)$ and $s\xi_{quad}(\w)$
where
\be
\label{lin}
\xi_{lin}(\w)=1-\rho^2(\w-1)
\ee
is a simple linear parametrization, and 
\be
\label{quad}
\xi_{quad}(\w)=1-\rho^2(\w-1)+\frac{c}{2}(\w-1)^2
\ee
is a quadratic parametrization. The parameter $c$ in \eq{quad} is, of
course, the curvature of the \iwfn\ at $\w=1$. 
We tabulate the results of these different fits in
\tab{rho2.all}.  In this table, we also present the results of performing
these fits on the data corresponding to $\kq=$0.14144, 0.14226 and
0.14262. 

The fact that the values of $\chi^2/d.o.f.$ are relatively high for
all of these fits should not in itself be taken as an indication that
the parametrizations of \eq{bsw}, (\ref{lin}) and (\ref{quad}) are
poor representations of the Isgur-Wise functions. These large values of
$\chi^2/d.o.f.$ are due to the discrepancy that we mentioned in
\sec{raflqm} between our measurements of
$h^+\l((0,0,0)\to(\pi/12a,0,0);m_Q,m_{Q'}\r)$ and of $h^+
\l((\pi/12a,0,0)\to(0,0,0);m_{Q'},m_Q\r)$. Because of this discrepancy, no
parameterization can fit our data with a good value of
$\chi^2/d.o.f.$. $\chi^2/d.o.f.$'s nevertheless seem 
to favor the use of the extra parameter $s$ but does
not seriously discriminate between $s\,\xi_{NR}(\w)$,
$s\,\xi_{lin}(\w)$ and $s\,\xi_{quad}(\w)$. We have tried fitting our
data to yet other parametrizations and of all the fitting functions,
$s\,\xi_{lin}(\w)$ yields the lowest values for $\rho^2$. The reason
for this is that $s\,\xi_{lin}(\w)$ is the only parametrization which
does not have positive curvature. Since $s\,\xi_{lin}(\w)$ is in that
sense an exception, we will not use it as our standard fitting
function but because it is a valid parametrization for these
Isgur-Wise functions we will make certain that our results have errors
which encompass the values it gives for the slope. Furthermore, since
both $s\,\xi_{NR}(\w)$ and $s\,\xi_{quad}(\w)$ give nearly identical
fits (see
\figs{iwud}{iws}), we will use $s\,\xi_{NR}(\w)$ as our standard 
in the following because it has one less parameter and yields better
$\chi^2/d.o.f.$'s.

Having already argued in \sec{hqmdohp} that, with the normalization
that we have adopted, the remaining mass-dependent discretization
errors are small, we turn to momentum-dependent lattice artefacts. To
quantify these momentum-dependent discretization errors we resort to
the following procedure.  We fit the data for $\xi_{u,d}(\w)$ and
$\xi_s(\w)$ for fixed initial and final meson momentum and all
heavy-quark combinations, to the parametrization
$s\,\xi_{NR}(\w)$. The variation in the results of fits to these
different momentum sets should give us some indication of how large
these momentum-dependent lattice artefacts are. Some of this
variation, of course, may be due to statistical fluctuations of the
sort we mentioned in
\sec{raflqm}.

We summarize the results of the fits to the different momentum sets in
\tab{rho2.syst}. It is reassuring that the value of $s$ for the
case $(0,0,0)\to(\pi/12a,0,0)$ is very close to 1, because the
corresponding data are our best points. They are the points for which
our normalization procedure is optimal because they are obtained from
\thptfns\ which are 
much more correlated with the \thptfn\ which yields the
normalization factor 
$h^+_{\rm lat.}(1)$~\footnote{For readers familiar with the
methods used to calculate \thptfns, the reason why the \thptfns\
corresponding to $(0,0,0)\to(\pi/12a,0,0)$ and $(0,0,0)\to(0,0,0)$ are
strongly correlated is because they are built up from the same
exponentiated propagator. Indeed, the initial momentum in our notation
is the momentum of the exponentiated propagator.}. Furthermore, these
data have the smallest statistical errors and should have the smallest
discretization errors, because the momenta of the incoming and
outgoing mesons are less than or equal to the initial and final momenta of
other momentum sets.

To accommodate the spread in the values in the slope
parameters $\rho^2_{u,d}$ and $\rho^2_s$ corresponding to
$\xi_{u,d}(\w)$ and $\xi_s(\w)$, we assign systematic errors to these
parameters which encompass all the central values given in
\tab{rho2.syst}. The central value and 
statistical errors that we quote are given
by fitting $s\,\xi_{NR}(\w)$ to all momentum sets put together (see
\tab{rho2.all}). Thus, our final results for the slope at $\w=1$ are
\be
\label{rho2ud}
\rho^2_{u,d}=0.9\er{2}{3}{\rm \stat}\er{4}{2}{\rm\syst}
\ee
for $\xi_{u,d}(\w)$ and
\be
\label{rho2s}
\rho^2_s=1.2\er{2}{2}{\rm \stat}\er{2}{1}{\rm\syst}
\ee
for $\xi_s(\w)$. Even though the exact values 
of these slope parameters are slightly different
if different parametrizations for the \iwfns\ are used, these differences
are well within our error bars.

In \tab{rho2comp} we compare our predictions for the slope of the
\iwfns\ $\xi_{u,d}$ and $\xi_s$ with those of 
other authors. We find that our predictions for $\rho^2$ lie safely
above the lower bound of Bjorken\cite{bj} and below the upper bound of
de Rafael and Taron\cite{deraf}. Our results for $\rho^2_s$ also agree
with the lattice result of Bernard et al.\ \cite{bersheso} obtained
with Wilson fermions for a light spectator antiquark with mass
$m_q\sim m_s$, although the details and systematics of the two
calculations are different.  The authors of Ref.~\cite{bersheso} do
not quote a value of $\rho^2_{u,d}$ for vanishing light-quark mass.

Also for comparison, we quote an average experimental value for the
slope of the Isgur-Wise function compiled by Neubert\cite{neubup} from
very recent results of the ALEPH\cite{aleph} and CLEO\cite{cleo}
Collaborations as well as older data from the ARGUS
Collaboration\cite{argus}:
\be
\rho^2_{u,d(expt.)}=0.87(12)(20)
\ ,
\label{rho2udexpt}
\ee
where the second error is theoretical and accounts for the uncertainty
in the size of $1/m_c$ corrections\cite{neubup}.
\eqs{rho2ud}{rho2udexpt} agree remarkably well.

\bigskip

As can be inferred from \eqs{rho2ud}{rho2s} and from \tab{rho2.all}
our results are compatible with the statement that $\rho^2$ is
constant with light-quark mass, possibly decreasing slightly as this
mass decreases.  Such a decrease is consistent with one's intuition
that it is more difficult to make the light degrees of freedom recoil,
the heavier these degees of freedom are. Furthermore, H\o gaasen and
Sadzikowski\cite{hogasad} find a decrease in slope which is very
similar to the trend we observe in the central value of $\rho^2$ when
we include the extra parameter $s$ in our fits.  In fact, our
predictions for $\rho^2$ itself are in excellent agreement with
theirs. Their prediction is based on an improved bag model calculation
and is an extension of earlier work by Sadzikowski and
Zalewski\cite{sadzal}. A similar decrease in slope with spectator
quark mass is observed by Close and Wambach\cite{wambach} though the
values they quote for $\rho^2_{u,d}$ and $\rho^2_s$ are slightly
larger than the ones we find.

\medskip
To test the robustness of our predictions for $\rho^2$, we have
explored many different procedures for obtaining $\hpw$, two of which
we have already described in \sec{raflqm}. To obtain $\hpw$ for
degenerate transitions, we have in addition tried normalizing our lattice
results for $\hplatw$ by
$h^+_{\rm lat.}\l((\pi/12a,0,0)\to(\pi/12a,0,0)\r)$ instead of by
$h^+_{\rm lat.}\l((0,0,0)\to(0,0,0)\r)$ (see
\eq{hphmdef}). When fitted to the $s\xi(\w)$ parametrizations, 
the results obtained using all of these methods give very similar
values for the slope parameter $\rho^2$. They only differ in the value
of $s$ they predict, i.e.\ in their overall normalization. Thus, we
are quite confident that our predictions for the slope are reliable
but believe that it is important to allow for the extra normalization
parameter $s$.

\bigskip



\section{Extraction of \protect{$V_{cb}$}}
\label{eovcb}
In \sec{ffftsabe6.2} we obtained $\hpw$ for a variety of $P\to P'$
transitions where $P$ ($P'$) is a pseudoscalar meson composed of a
heavy quark $Q$ ($Q'$) and a light antiquark $\bar q$. In our study, both
$Q$ and $Q'$ are quarks with masses around that of the charm quark. In
\sec{hqmdohp}, however, we showed that our results for
$\hpw/(1+\bpw)$ are independent of heavy-quark mass for masses
around the charm quark mass or larger. This means, modulo the
issue of power corrections at zero-recoil, that our 
results can be used to describe not only $P\to P'$ transitions 
with $Q(Q')\sim c$ but
$\bar B_q\to D_q$
decays as well, where the subscript $q$ labels the flavour of the light
antiquark. In
\sec{dohpolqm}, we studied the dependence of $\hpw/(1+\bpw)$ on the
mass of the light, spectator quark, $m_q$, and obtained results for
$\xi_{u,d}(\w)$ and $\xi_s(\w)$. All of this means that our result
for $\xi_{u,d}$, once multiplied by $(1+\b^+_{b\to c})$, is in fact
the form factor $h^+$ relevant for $\bar B_{u,d}\to D_{u,d}$
transitions, while $(1+\b^+_{b\to c})\xi_s$ is the form factor $h^+$
relevant for $\bar B_s\to D_s$ transitions.

\bigskip
Now, the differential decay width for
$\btodl$ is, in the limit of zero lepton mass,\cite{bible}
\bea
\label{btodrate}
\frac{d\Gamma(\btodlands)}{d\w}= & {G_F^2|V_{cb}|^2\over 48\pi^3}
(M_{B_{(s)}}+M_{D_{(s)}})^2\ M_{D_{(s)}}^3(\w^2-1)^{3/2}\nonumber\\
& \times\l|
\hpw-\frac{M_{B_{(s)}}-M_{D_{(s)}}}{M_{B_{(s)}}+M_{D_{(s)}}}
\hmw\r|^2
\ .
\eea
So, in principle, we could obtain
$|V_{cb}|$ by comparing our theoretical prediction for
$\frac{d\Gamma(\btodlands)}{d\w}$ to an experimental measurement of
this rate. A major problem with this approach, however, is that the
rate $\frac{d\Gamma(\btodlands)}{d\w}$ is helicity suppressed, as
evidenced by the factor $(\w^2-1)^{3/2}$, so that it is very difficult
to get accurate experimental measurements close to $\w=1$ where the
predictions of HQET are most reliable. Another problem with obtaining
$|V_{cb}|$ from $\btodlands$ decays is that one must know $\hmw$ to
better accuracy than is given by our calculation: an error of 0.1 on
$h^-$ leads to an uncertainty of about 10\% in the rate. We should
mention, however, that Neubert\cite{bible} has estimated $h^-$ using
perturbation theory and sum rules in HQET and has found that its
magnitude does not exceed 0.04 over the whole range of recoils
$\w$. If this is true, its contribution to the rate of \eq{btodrate}
should not exceed 4\%.

We have not exhausted the predictions of heavy-quark
symmetry.  We have yet to exploit the spin component of this
symmetry. Using a combination of the spin and flavour symmetry, we can
relate our predictions for $\xi(\w)\simeq\hpw/(1+\bpw)$ to the form
factors required to describe $\btodslands$ decays. These form factors
are defined by\cite{bible}
\bea
\frac{\la D^*_{(s)}(\s p',\e)|\bar c\gm b|\bar B_{(s)}(\s p)
\ra}{\sqrt{M_{B_{(s)}}M_{D^*_{(s)}}}} 
& = & i h^V(\w)\e^{\mu\nu\a\b}\e^*_\nu v'_\a v_\b \ ,\nonumber\\
\frac{\la D^*_{(s)}(\s p',\e)|\bar c\gm\g^5 b|\bar B_{(s)}(\s p)\ra}
{\sqrt{M_{B_{(s)}}M_{D^*_{(s)}}}} 
& = & (\w+1)\e^{*\mu}h^{A_1}(\w)-\e^*\cdot v\l(v^\mu h^{A_2}(\w)+v^{\prime\mu}
h^{A_3}(\w)\r) \ ,
\eea
where $v=p/M_{B_{(s)}}$ and $v'=p'/M_{D_{(s)}^*}$. In the heavy-quark
limit, these four form factors can be expressed in terms of the
single
\iwfn, $\xi_{u,d(s)}$. 
There are, of course, radiative and power corrections to these
heavy-quark symmetry predictions. Thus, one has
\be
\label{ff1}
h^i(\w)=(\a^i+\b^i(\w)+\g^i(\w))\xi_{u,d(s)}(\w)\ ,
\ee
with $i=V,A_1,A_2,A_3$ and
\bea
\label{ff2}
\a^{A_1} & = & \a^{A_3}=\a^V=1\ ,\nonumber\\
\a^{A_2} & = & 0\ .
\eea
Luke's theorem\cite{luke} further guarantees that, at zero recoil,
$h^{A_1}$ is free of $\ord{\e_b,\e_c}$ corrections, i.e
$\g^{A_1}(1)\sim\ord{\e_b^2,\e_c^2}$. Because $h^{A_1}$ is the only form
factor to contribute to the differential decay rate for $\btodslands$
decays at zero recoil, Luke's theorem implies that the leading
non-perturbative corrections to this rate must be small at
$\w=1$. More precisely, in the limit of zero lepton mass,
\begin{eqnarray}
\label{btodsrate}
{d\Gamma(\btodslands)\over d\omega}
 = &{G^2_F\over 48\pi^3}M_{D^*_{(s)}}^3(M_{B_{(s)}}-M_{D^*_{(s)}})^2
\left[1+\beta^{A_1}(1)\right]^2 \nonumber\\
&\times\sqrt{\omega^2-1}(\omega+1)^2|V_{cb}|^2\xi_{u,d(s)}^2(\omega)
\nonumber\\
&\times\left[1+4\left(\frac{\displaystyle\omega}{\displaystyle\omega+1}
\right)\frac{\displaystyle M_{B_{(s)}}^2-2\omega M_{B_{(s)}}M_{D^*_{(s)}}+
M^2_{D^*_{(s)}}}
{\displaystyle(M_{B_{(s)}}-M_{D^*_{(s)}})^2}\right]K(\omega)
\ ,
\end{eqnarray}
where $\beta^{A_1}(1)=-0.01$\cite{short} and $K(\w)$ incorporates the
radiative corrections, $\b_{A_1}(\w)$, away from $\w=1$, the
non-perturbative power corrections, $\g^{A_1}(\w)$, and the contributions
of the three other form factors to the rate.  From what we have said
above, it should be clear that $K(1)=1+\ord{\e_b^2,\e_c^2}$. One can
also show\cite{bible} that in the limit of exact heavy-quark symmetry,
$K(\w)=1$ for all $\w$. Moreover, since we have factored out
$\xi_{u,d(s)}(\w)$ in the expression for the rate, the contributions
of the three other form factors will be normalized by $\xi_{u,d(s)}$.
This means that $K(\w)$ is a collection of radiative and power
corrections (see \eqs{ff1}{ff2}), many of which are kinematically
suppressed: deviations of $K(\w)$ from 1 ought to remain small.
Neubert estimates\cite{bible}, using perturbation theory and sum rules
in HQET, that $K(\w)$ may reduce the slope parameter, $\rho^2$ by 0.09
which corresponds to an enhancement of the rate by about 10\% at
maximum recoil and by a smaller amount for smaller $\w$. However,  we
cannot estimate yet how the physical
$K(\w)$ deviates from its value in the heavy-quark limit 
from our lattice calculation. For that we
need to study $Q\bar q(0^-)\to Q'\bar q(1^-)$ decays, which we are 
currently analyzing. We also need to determine the $1/m_c^2$-corrections to
$h^{A_1}(1)$, which as discussed in \sec{npsoame}, we cannot get with
a procedure analogous to the one presented in this paper. Hence, we will
assume that $K(\w)=K(1)$ for all $\w$ which is a reasonable assumption
given the size of our errors on the slope parameter $\rho^2$. We can
then use our lattice
determination of $\xi_{u,d}(\w)$ to extract $V_{cb}$ from the
experimentally measured differential decay rate for $\btodsl$
($\bstodssl$ has not yet been measured).  This analysis differs from
Neubert's extraction of $|V_{cb}|$ \cite{neuplb} 
in that we fix the $\w$-dependence of the differential
decay rate using our calculation instead of fitting it from experiment.
This enables us not only to extract $|V_{cb}|$ with no free parameters, but
also to check the validity of non-perturbative QCD against
experiment. We find that the $\w$-dependence predicted by our calculation
agrees very well with the results of 
the ALEPH\cite{aleph} and CLEO\cite{cleo} collaborations. 

In \fig{vcbaleph}, \fig{vcbargus} and \fig{vcbcleo} 
we show least-$\chi^2$-fits to experimental data
for $|V_{cb}|(1+\b^{A_1}(1))K(\w)\xi_{u,d}(\w)$ from ALEPH~\cite{aleph},
ARGUS~\cite{argus} and CLEO~\cite{cleo}, respectively.  The only parameter
is $|V_{cb}|$. The slope of the \iwfn\ is constrained to the value
given by our lattice calculation (see \eq{rho2ud}) and the functional
form for the \iwfn\ that is used is $\xi_{NR}$ of \eq{bsw}
\footnote{As can be seen from \fig{iwud} other parametrizations give very
similar curves when fit to our results for $\xi_{u,d}$. Therefore,
results for $|V_{cb}|$ obtained with these other parametrizations will
be well within our quoted error bars.}. The results of these
fits are summarized in \tab{vcbres}. Our results favor ALEPH and CLEO
data over that of ARGUS. Using the data from CLEO, for instance,
we find
\be
|V_{cb}|=0.037\er{1}{1}\er{2}{2}\er{4}{1}
\left(\frac{0.99}{1+\beta^{A_1}(1)}\right)\frac{1}{1+\delta_{1/m_c^2}}
\ ,
\ee
where $\delta_{1/m_c^2}$ are the power corrections proportional to
$1/m_c^2$ in $K(1)$ which have been the subject of much controversy of
late\cite{neubup,shif}.

For comparison, we present recent experimental predictions for
$|V_{cb}|K(1)$ obtained from a linear fit to the data\footnote{The
ARGUS result has been corrected for the new $D$ branching
fractions \protect\cite{ritch}.}
\begin{equation}
   |\,V_{cb}|\,K(1)\,
\left(\frac{1+\beta^{A_1}(1)}{0.99}\right) = \left\{
   \begin{array}{ll}
   0.0351\pm 0.0019\pm 0.0020 &
    ;\quad \mbox{CLEO~\protect\cite{cleo},} \\
   0.0385\pm 0.0044\pm 0.0035 &
    ;\quad \mbox{ALEPH~\protect\cite{aleph},} \\
   0.0392\pm 0.0043\pm 0.0025 &
    ;\quad \mbox{ARGUS~\protect\cite{argus},}
   \end{array} \right.
\end{equation}
where the first error is statistical and the second systematic.  These
results have been rescaled by Neubert\cite{neubup} using the new
lifetime values $\tau_{B^0}=1.61(8)\,ps$ and
$\tau_{B^+}=1.65(7)\,ps$\cite{jongh}. These new lifetimes reduce
$|\,V_{cb}|\,(1+\b^{A_1}(1))\,K(1)$ 
by approximately 1\%. Our results compare very well
with these experimental measurements, especially in the case of the
CLEO result. This is due to the fact that our Isgur-Wise function has
an $\w$ dependence which agrees very well with that of the CLEO data.



\section{Exclusive Decay Rates}
\label{edr}
Having determined the \iwfns\ $\xi_{u,d}$ and $\xi_s$, we can evaluate
exclusive branching ratios. For $\btodlands$ decays, all we have to do
is integrate \eq{btodrate} and multiply the results by the $\bar
B_{(s)}$ meson lifetime. We approximate $h^+$ in \eq{btodrate} by
$(1+\b^+)\xi_{NR}$ with $\rho^2$ given by \eq{rho2ud} or \eq{rho2s}
depending on whether the light antiquark is a $\bar u$, $\bar d$ or an
$\bar s$ (see \sec{eovcb}). We neglect the $\ord{\e_b,\e_c}$ 
power corrections to $h^+$
since they appear to be small (see
\sec{hqmdohp}) but add a 10\% error to account for possible
higher order power corrections. We further neglect the contribution of
$h^-$ in accordance with Neubert's findings that this form factor is
smaller than 0.04 over the whole range of $\w$ (see discussion after
\eq{btodrate}) but 
add 20\% to our errors since an $|h^-(\w)|\sim 0.2\xi(\w)$
is consistent with radiative corrections and order of magnitude
estimates of power corrections. 

The branching ratios for $\btodslands$ decays are equally simple to
obtain.  Here it is \eq{btodsrate} that we must integrate over the
range
$1\le\w\le(M_{B_{(s)}}^2+M_{D^*_{(s)}}^2)/2M_{B_{(s)}}M_{D_{(s)}^*}$.
The \iwfns\ used are the same as for $\btodlands$ decays. As
discussed after \eq{btodsrate}, we assume $K(\w)=K(1)$. We further
assume $K(\w)=1$ which leads to an uncertainty of the
order of $\ord{2\e_c^2}\sim 5-10\%$ in the branching ratios. 

We summarize our results for ${\cal B}\l(\btodlands\r)$ and ${\cal
B}\l(\btodslands\r)$ in \tab{brs}. The first set of errors is 
obtained by adding our lattice statistical and systematic errors in
quadrature. The second set of errors corresponds to the uncertainty
due to deviations from the heavy-quark limit. For comparison,
we list the experimentally measured values for these branching
ratios. Agreement with our predictions is very satisfactory.

Finally, we give a prediction for the ratio of the rates for
$\btodslands$ and for $\btodlands$. In this ratio, the
factors of $|V_{cb}|$ cancel and lifetimes do not appear. This ratio
is thus a purely theoretical prediction. We find
\be
\label{raterat}
\frac{\Gamma(\btodsl)}{\Gamma(\btodl)}=3.2\er{3}{2}({\rm lat.})
\pm 1.0({\rm hqs})
\ee
and
\be
\frac{\Gamma(\bstodssl)}{\Gamma(\bstodsl)}=3.3\er{2}{1}({\rm lat.})
\pm 1.0({\rm hqs})
\ .\ee
where the first set of errors was obtained by adding our lattice
statistical and systematic errors in quadrature and the second set of
errors, denoted by ``hqs'', quantifies the uncertainty due to
neglected power and radiative corrections.  For comparison, the
experimental result for $\frac{\Gamma(\btodsl)}{\Gamma(\btodl)}$ is
2.1(1)\cite{stone}. Though low compared to our prediction, this result
is consistent with ours within errors.



\section{Conclusions}
\label{ccl}
We have presented an extensive study of semi-leptonic $\btodlands$
decays where we evaluate the matrix element, $\la D|\bar c\gm b|\bar
B\ra$, for many different values of $m_b$ and $m_c$ around the
physical charm mass and three values of the light antiquark mass
around that of the strange.  Because the charm quark has a bare mass
which is almost 1/3 the inverse lattice spacing, mass-dependent
discretization errors are a problem that we must contend with. To
reduce these errors we use an $\ord{a}$-improved quark action in which
the leading such errors are no longer $\ord{am_Q}$ but rather
$\ord{\a_s\,am_Q,(am_Q)^2}$. This reduces discretization errors from
$\ord{40\%}$ to $\ord{5-15\%}$ at the mass of the charm.  To reduce
them even further we describe, in \sec{npsoame}, a procedure for
subtracting them at least partially.  Only those discretization errors
which have the same dependence on $\w$ as $\hpw$ will be fully
subtracted. We believe, however, that the observation in \sec{hqmdohp}
of $h^+(\w)$'s lack of dependence on heavy-quark mass indicates that a
fairly large proportion of discretization errors are eliminated with
our procedure.

The fact that we obtain $\hpw$ and $\hmw$ for many values of the
initial and final heavy-quark masses enables us to study their
heavy-quark-mass dependence. We find that the residual dependence of
$h^+/(1+\bpw)$ on the heavy-quark mass is consistent with zero.  Given
our errors, we conclude that power corrections to the form factor
$h^+$ for physical $\btod$ transitions are less than 10\%.  This is
much smaller than the $25\%$ corrections one is entitled to
expect for form factors not protected by Luke's theorem. It is also in
stark contrast with our findings for the decay constant, $f_D$, of the
pseudoscalar $D$ meson\cite{qhldc}.  In Ref.\cite{qhldc} we find that
the $\ord{\e_c}$ corrections to the heavy-quark limit prediction for
this constant are on the order of 30\%. Thus, it appears that the
protection from $\ord{\lqcd/m_c}$ effects that Luke's theorem provides
at zero recoil extends to some extent over the full range of
experimentally accessible $\w$. Our results for $\hpw/(1+\bpw)$ are
then, to a good approximation, the corresponding Isgur-Wise function.

Having obtained the Isgur-Wise function from $\hpw/(1+\bpw)$ for three
values of the mass of the light, spectator antiquark, we can study its
dependence on light-quark mass.  Interpolating the light antiquark to
the strange, we obtain an Isgur-Wise function relevant for $\bar
B_s\to D_s^{(*)}l\bar\nu\,$ decays which has a slope
$-\xi'_s=1.2\er{2}{2}\stat\er{2}{1}\syst$ at zero recoil when fit to a
parametrization proposed by Neubert and Rieckert\cite{neurieck}.
Extrapolating to a massless light antiquark yields an Isgur-Wise
function relevant for $\bar B\to D^{(*)}l\bar\nu\,$ decays. This
function has a slope $-\xi_{u,d}'=0.9\er{2}{3}\stat\er{4}{2}\syst$ at
zero recoil. We observe a slight decrease in the magnitude of the
central value of the slope as the mass of the light antiquark is
reduced in accordance with one's understanding that more massive
degrees of freedom have more inertia. Given the errors,
however, the significance of this observation is limited.

We also use these functions, in conjunction with heavy-quark effective
theory, to extract $V_{cb}$ from the experimentally measured $\btodsl$
decay rate. Our procedure
for extracting $|V_{cb}|$ differs from that proposed by Neubert\cite{neuplb} 
in that we fix the $\w$-dependence of the differential
decay rate using our calculation instead of fitting it from experiment.
This enables us not only to extract $|V_{cb}|$ with no free parameters, but
also to check the validity of non-perturbative QCD against
experiment. We find that the $\w$-dependence predicted by our calculation
agrees very well with the results of 
the ALEPH\cite{aleph} and CLEO\cite{cleo} collaborations. 
Using the data from CLEO, for instance, we find
\[
|V_{cb}|=0.037\er{1}{1}\er{2}{2}\er{4}{1}
\left(\frac{0.99}{1+\beta^{A_1}(1)}\right)\frac{1}{1+\delta_{1/m_c^2}},
\]
where $\delta_{1/m_c^2}$ is the power correction proportional to
$1/m_c^2$ at zero recoil and $\beta^{A_1}(1)$, the relevant radiative
correction.  Here, the first set of errors is due to experimental
uncertainties, the second due to statistical errors and the third to
systematic errors in our evaluation of the Isgur-Wise function. We
also use our Isgur-Wise functions and heavy-quark effective theory to
calculate branching ratios for $\btodlands$ and $\btodslands$ decays.
Agreement with experiment is very good. Finally, we compute the
following ratios of rates:
\be
\frac{\Gamma(\btodsl)}{\Gamma(\btodl)}=3.2\er{3}{2}({\rm lat.})
\pm 1.0({\rm hqs})
\ee
and
\be
\label{raterats}
\frac{\Gamma(\bstodssl)}{\Gamma(\bstodsl)}=3.3\er{2}{1}({\rm lat.})
\pm 1.0({\rm hqs})
\ ,\ee
where the first set of errors was obtained by adding our lattice
statistical and systematic errors in quadrature and the second set of
errors, denoted by ``hqs'', quantifies the uncertainty due to
neglected power and radiative corrections.  In these ratios, the
factors of $|V_{cb}|$ cancel and $B$-meson lifetimes are absent: they
are purely theoretical predictions.

We are currently extending our study to the matrix elements relevant
for $\btodslands$ decays.  This will enable us not only to check our
predictions for the various Isgur-Wise functions but also to test the
heavy-quark spin symmetry. We are also undertaking a study of
semi-leptonic $\Lambda_b\to\Lambda_c$ and $\Xi_b\to\Xi_c$ decays,
where the $\Lambda_{b(c)}$ is a $J^P=1/2^+$ baryon composed of a
$b(c)$ quark and two light quarks coupled to spin and isospin 0 and
the $\Xi_{b(c)}$, another $J^P=1/2^+$ baryon composed of a $b(c)$
quark and two light quarks but this time with spin 0, isospin 1/2 and
strangeness $-1$. That study should provide many interesting
phenomenological predictions which are at the limit of current
experimental knowledge as well as many tests of the heavy-quark
symmetry. Finally, we are planning to repeat these studies on lattices
with different lattice spacing in order to remove discretization
errors in a more systematic way.

\acknowledgements
L.~L.~would like to thank Southampton University's theory group for
its hospitality and financial support during much of this project. We
would also like to thank G.~Martinelli and members of the APE
Collabortaion for stimulating discussions.  This research was
supported by the UK Science and Engineering Research Council under
grants GR/G~32779 and GR/J~21347, by the European Union under contract
CHRX-CT92-0051, by the University of Edinburgh and by Meiko Limited.
We are grateful to Edinburgh University Computing Service and, in
particular, to Mike Brown for maintaining service on the Meiko i860
Computing Surface.  CTS (Senior Fellow), DGR (Advanced Fellow) and NMH
acknowledge the support of the Particle Physics and
\nopagebreak Astronomy Research Council.  JN acknowledges the European
Union for for their support through the award of a Postdoctoral
Fellowship, contract No. CHBICT920066.





\begin{figure}
\caption{
The ratio $R^0(t)$, up to constant factors, vs. $t$ for the
case where the initial meson has momentum $(0,0,0)$ and the
final meson, momentum $(\pi/12a,0,0)$. Here, the initial and final
heavy-quark hopping parameters are $\kQ=\kQp=$0.129 while the
light-quark hopping parameter is $\kq=$0.14144. The solid line is obtained
from our
fit of $R^0(t)$ to the asymptotic form of \protect{\eq{fitratasym}}.
The dashed lines indicate the errors of this fit.
\label{plat}}
\end{figure}

\begin{figure}
\caption{Values of $Z^{eff}_V$ as functions of $m_Q^Ia$. The solid lines
represent fits to quadratic functions of $m_Q^Ia$ for the data at the 
two different values of $\beta$. We have also plotted the light-quark
values of $Z^{eff}_V$ given in \protect{\eq{eq:zvjonivar}} but have
not included them in the fit.
\label{fig:zv1}}
\end{figure}
\begin{figure}
\caption{Values of $Z^{eff}_V(\kappa_Q)$ obtained from the simulation at 
$\beta = 6.2$ at different momenta and Lorentz indices. The three curves
are quadratic fits to the three sets of data.
\label{fig:zv2}}
\end{figure}

\begin{figure}
\caption{
$\hpw/(1+\bpw)$ vs. $\w$ for all four elastic scattering reactions:
$\kQ=\kQp=0.121$, 0.129, 0.133. The light-quark hopping
parameter is fixed to $\kq=0.14144$. The curves are obtained by fitting
each heavy-quark, data set to $s\xi_{NR}(\w)$. The data points as well
as the curves corresponding to different heavy quarks are really
indistinguisible. The $1/m_Q$ corrections to $\hpw/(1+\bpw)$ cannot
therefore be very large. (See text for details.)
\label{4144.degen}}
\end{figure}

\begin{figure}
\caption{$R^+$ vs. $x=m_{Q'}/m_Q$ at fixed $\w$ for four
values of $\w$. The solid lines are obtained by fitting these
results to the parameterization given in \protect{\eq{rp_def}} and the
dotted lines represent errors. The slope and
intercept of this line is the subleading form factor $g(\w)$
(\protect{\eqs{g_def}{hpw_bpw}}). The light-quark hopping parameter is
$\kq=0.14144$.
\label{rp_vs_x}}
\end{figure}

\begin{figure}
\caption{The subleading form factor $g(\w)$ (\protect{\eqs{g_def}{hpw_bpw}}). 
The light-quark hopping parameter is $\kq=0.14144$.
\label{g_vs_w}}
\end{figure}

\begin{figure}
\caption{$\xi_{u,d}(\w)=\hpw/(1+\bpw)$
vs. $\w$ for $\kq=\kcrit$. The
different symbols correspond to different values of initial and final
heavy-quark mass. The solid curve is obtained by fitting these results to
$s\xi_{NR}(\w)$ while the dashed curve corresponds to a fit to
$s\xi_{lin}(\w)$ and the dotted curve corresponds to a fit to
$s\xi_{quad}(\w)$.  The value of $\rho^2$ shown on the plot is the
one given in \protect{\eq{rho2ud}}.
\label{iwud}}
\end{figure}

\begin{figure}
\caption{$\xi_s(\w)=\hpw/(1+\bpw)$
vs. $\w$ for $\kq=\k_s$. The
different symbols correspond to different values of initial and final
heavy-quark mass. The solid curve is obtained by fitting these results to
$s\xi_{NR}(\w)$ while the dashed curve corresponds to a fit to
$s\xi_{lin}(\w)$ and the dotted curve corresponds to a fit to
$s\xi_{quad}(\w)$.  The value of $\rho^2$ shown on the plot is the
one given in \protect{\eq{rho2s}}.
\label{iws}}
\end{figure}

\begin{figure}
\caption{
Least-$\chi^2$-fit to experimental data
for $|V_{cb}|(1+\b^{A_1}(1))\,K(\w)\,\xi(\w)$ from 
ALEPH~\protect{\cite{aleph}} assuming $K(\w)=K(1)$. 
In this fit, the slope of the \iwfn\ is 
constrained to the value given by our lattice calculation (see 
\protect{\eq{rho2ud}}) 
and the functional form for the \iwfn\ that is used is $\xi_{NR}$ of 
\protect{\eq{bsw}}.  The first set of errors on $|V_{cb}|$ 
is due to experimental uncertainties, the second set of errors results
from the lattice statistical errors on $\rho^2$, and the third, from
the lattice systematic errors on $\rho^2$.  The experimental points
were obtained from a measurement of the rate $dB(\bar B\to
D^*l\bar\nu)/d\w$. Also
shown are our appropriately scaled, chirally-extrapolated results
(octagons). 
\label{vcbaleph}}
\end{figure}

\begin{figure}
\caption{
Same fit as in \protect{\fig{vcbaleph}} but for experimental data from 
the ARGUS Collaboration~\protect{\cite{argus}}.
\label{vcbargus}}
\end{figure}

\begin{figure}
\caption{
Same fit as in \protect{\fig{vcbaleph}} but for experimental data from 
the CLEO Collaboration~\protect{\cite{cleo}}.
\label{vcbcleo}}
\end{figure}



\begin{table}
\caption{
Physical heavy-quark masses corresponding to different values the
heavy-quark hopping parameter, $\kQ$.  They are obtained from the
corresponding chirally-extrapolated pseudoscalar and vector meson
masses, as described in \protect{\eq{hq_mass}}.  For completeness,
we also tabulate the chirally-extrapolated meson masses in lattice
units ($a^{-1}\approx 2.7~\gev$\protect{\cite{qhldc}}).  These
masses were obtained by covariant linear extrapolation of the
masses $M_P$ and $M_V$ obtained at three values of the light antiquark
hopping parameter: $\kq=0.14144$, 0.14226, 0.14262.  The
pseudscalar meson masses were computed as described in
\protect{\sec{lpadotaabe6.2}}, while the vector meson masses were
obtained as in \protect{\cite{qhldc}}, with a fitting range $11\le
t\le 23$. 
\label{Q_masses_6.2}}
 %
\begin{tabular}{cccc}
$\kQ$ & $M_P^\chi$ & $M_V^\chi$ & $m_Q\,(\gev)$\\
\hline
\hline
0.121 & 0.874\er{4}{3} & 0.896\er{5}{4} & 1.90\\
\hline
0.125 & 0.773\er{3}{3} & 0.799\er{4}{3} & 1.64\\
\hline
0.129 & 0.665\er{3}{3} & 0.696\er{4}{4} & 1.36\\
\hline
0.133 & 0.547\er{3}{3} & 0.588\er{4}{5} & 1.06\\
\end{tabular}
\end{table}

\begin{table}
\caption{
Wavefunction factors, $Z^2$, and energies, $E_P$ for
our heavy-light, pseudoscalar mesons and for two values
of momentum, $|\s p|$. The energies are quoted
in lattice units ($a^{-1}\simeq 2.7~\gev$\protect{\cite{qhldc}}). 
The $\chi^2/d.o.f.$ for the fits which give
these results are all on the order of 1. 
\label{twoptfits_6.2}}
\begin{tabular}{cccccccccc}
& $\k_Q$ & \multicolumn{2}{c}{0.121} & \multicolumn{2}{c}{0.125} &
\multicolumn{2}{c}{0.129} & \multicolumn{2}{c}{0.133}\\
\hline
$\kq$ & $|\s p|$ & $Z^2$ & $E_P$ & $Z^2$ & $E_P$ & $Z^2$ & $E_P$ 
& $Z^2$ & $E_P$ \\
\hline
0.14144 & 0 & 17.9\er{6}{5} & 0.924\er{2}{2} & 
16.3\er{5}{5} & 0.823\er{2}{2} & 
14.5\er{4}{4} & 0.716\er{2}{2} & 
12.4\er{4}{4} & 0.600\er{2}{2} \\
 & $\pi/12a$ & 12.3\er{5}{5} & 0.958\er{3}{2} & 
11.4\er{4}{5} & 0.861\er{3}{2} & 
10.3\er{4}{5} & 0.760\er{3}{2} & 
9.0\er{4}{4} & 0.653\er{3}{3} \\
\hline
0.14226 & 0 & 15.5\er{5}{5} & 0.901\er{3}{2} & 
14.2\er{5}{5} & 0.800\er{3}{2} & 
12.7\er{4}{4} & 0.692\er{3}{2} & 
10.8\er{5}{3} & 0.575\er{3}{2} \\
 & $\pi/12a$ & 10.5\er{5}{5} & 0.937\er{3}{4} & 
9.7\er{4}{4} & 0.840\er{3}{3} & 
8.8\er{4}{4} & 0.739\er{3}{3} & 
7.7\er{4}{3} & 0.631\er{4}{3} \\
\hline
0.14262 & 0 & 14.7\er{8}{7} & 0.892\er{4}{4} & 
13.5\er{7}{6} & 0.791\er{3}{3} & 
12.0\er{6}{5} & 0.683\er{3}{3} & 
10.3\er{4}{4} & 0.565\er{3}{2} \\
 & $\pi/12a$ & 9.8\er{7}{5} & 0.928\er{5}{4} & 
9.1\er{6}{4} & 0.832\er{4}{4} & 
8.3\er{5}{4} & 0.730\er{5}{4} & 
7.3\er{5}{4} & 0.623\er{4}{3} \\
\end{tabular}
\end{table}


\begin{table}
\caption{$\bpw$ vs. $\w$ for all combinations of initial and final
heavy-quark mass.\label{rad_corr_6.2p}}
\begin{center}
\begin{tabular}{cccccc}
&\multicolumn{5}{c}{$\w$}\\
\hline
$\kQ\to \kQp$ & 1.0 & 1.1 & 1.2 & 1.3 & 1.4\\
\hline
$0.121\to 0.121$ & 0 & -0.025 & -0.047 & -0.068 & -0.088\\
$0.121\to 0.125$ & 0.017 & -0.008 & -0.030 & -0.051 & -0.071\\
$0.121\to 0.129$ & 0.037 & 0.013 & -0.009 & -0.030 & -0.050\\
$0.121\to 0.133$ & 0.063 & 0.040 & 0.018 & -0.002 & -0.022\\ 
$0.125\to 0.125$ & 0 & -0.023 & -0.045 & -0.065 & -0.085\\
$0.125\to 0.129$ & 0.024 & 0.001 & -0.021 & -0.041 & -0.060\\ 
$0.125\to 0.133$ & 0.055 & 0.033 & 0.012 & -0.008 & -0.027\\
$0.129\to 0.129$ & 0 & -0.022 & -0.042 & -0.061 & -0.079\\
$0.129\to 0.133$ & 0.039 & 0.017 & -0.003 & -0.022 & -0.039\\
$0.133\to 0.133$ & 0& -0.019 & -0.038 & -0.055 & -0.071\\
\end{tabular}
\end{center}
\end{table}

\begin{table}
\caption{$\bmw$ vs. $\w$ for all combinations of initial and final
heavy-quark mass.\label{rad_corr_6.2m}}
\begin{center}
\begin{tabular}{cccccc}
&\multicolumn{5}{c}{$\w$}\\
\hline
$\kQ\to \kQp$ & 1.0 & 1.1 & 1.2 & 1.3 & 1.4\\
\hline
$0.121\to 0.121$ & 0 & 0 & 0 & 0 & 0\\
$0.121\to 0.125$ & 0.001 & 0.000 & -0.001 & -0.001 & -0.002\\
$0.121\to 0.129$ & -0.003 & -0.004 & -0.005 & -0.006 & -0.007\\
$0.121\to 0.133$ & -0.014 & -0.016 & -0.017 & -0.019 & -0.021\\ 
$0.125\to 0.125$ & 0 & 0 & 0 & 0 & 0\\
$0.125\to 0.129$ & 0.000 & -0.001 & -0.001 & -0.002 & -0.003\\ 
$0.125\to 0.133$ & -0.008 & -0.009 & -0.011 & -0.012 & -0.014\\
$0.129\to 0.129$ & 0 & 0 & 0 & 0 & 0\\
$0.129\to 0.133$ & -0.001 & -0.002 & -0.004 & -0.005 & -0.006\\
$0.133\to 0.133$ & 0 & 0 & 0 & 0 & 0\\
%
\end{tabular}
\end{center}
\end{table}



\begin{table}
\caption{Values of the effective normalisation constant $Z^{eff}_V$ as
a function of the improved bare mass of the heavy quark. The value of
$\kappa_q$ is 0.14144 at $\beta=6.2$ and 0.144 at $\beta=6.0$.
\label{tab:zv14144}}
\begin{tabular}{cccccc}
\multicolumn{3}{c}{$\beta = 6.2$} & \multicolumn{3}{c}{$\beta = 6.0$}
\\ \hline
$\kappa_Q$ & $m_Q^I a$ & $Z^{eff}_V(\kappa_Q)$ &
$\kappa_Q$ & $m_Q^I a$ & $Z^{eff}_V(\kappa_Q)$ \\ \hline
0.133 & 0.231 & 0.8913\errr{2}{1} & 0.129 & 0.344 & 0.920\errr{1}{1}\\ 
0.129 & 0.310 & 0.9177\errr{3}{2} & 0.125 & 0.405 & 0.945\errr{1}{1}\\ 
0.125 & 0.379 & 0.9428\errr{4}{2} & 0.120 & 0.464 & 0.973\errr{2}{2}\\ 
0.121 & 0.435 & 0.9659\errr{5}{3} & & & \\ 
\end{tabular}
\end{table}

\begin{table}
\caption{Values of $Z^{eff}_V$ for different choices of the Lorentz
index $\mu$, momenta $\vec p$, and light quark masses (given by $\kappa_q$)
from the simulation at $\b=6.2$.
\label{tab:zvmup}}
\begin{tabular}{ccccc}
$\mu$ and $\vec p$ & $\kappa_Q$ &
\multicolumn{3}{c}{$Z^{eff}_V$} \\ \hline
 & & $\kappa_q$ = 0.14144 &$\kappa_q$ = 0.14226 &$\kappa_q$ = 0.14262 \\ 
\hline 
$\mu = 4$, $\vec p = \vec 0$ & 0.133 &  
0.8913\errr{2}{1} &   & \\ 
$\mu = 4$, $\vec p = \vec 0$ & 0.129 &  
0.9177\errr{3}{2} &  0.9168\errr{4}{4} & 0.9165\errr{5}{6}\\ 
$\mu = 4$, $\vec p = \vec 0$ & 0.125 &  
0.9428\errr{4}{2} &  & \\ 
$\mu = 4$, $\vec p = \vec 0$ & 0.121 &  
0.9659\errr{5}{3} &0.9656\errr{6}{6} & 0.9658\errr{8}{11} \\ \hline
$\mu = 4$, $\vec p = (\pi /12, 0, 0)$ & 0.133 &  
0.8976\errr{10}{6} &  & \\ 
$\mu = 4$, $\vec p = (\pi /12, 0, 0)$ & 0.129 &  
0.9248\errr{9}{7} &0.9242\errr{14}{12}  &0.9240\errr{24}{24} \\ 
$\mu = 4$, $\vec p = (\pi /12, 0, 0)$ & 0.125 &  
0.9498\errr{7}{8} &  & \\ 
$\mu = 4$, $\vec p = (\pi /12, 0, 0)$ & 0.121 &  
0.9729\errr{7}{9} &0.9734\errr{12}{16}  &0.9746\errr{27}{25} \\ \hline
$\mu = 1$, $\vec p = (\pi /12, 0, 0)$ & 0.133 &  
0.949\errr{57}{56} & & \\ 
$\mu = 1$, $\vec p = (\pi /12, 0, 0)$ & 0.129 &  
0.994\errr{57}{63} &0.982\errr{83}{86}  &0.924\errr{134}{118} \\ 
$\mu = 1$, $\vec p = (\pi /12, 0, 0)$ & 0.125 &  
1.042\errr{53}{67} & & \\  
$\mu = 1$, $\vec p = (\pi /12, 0, 0)$ & 0.121 &  
1.084\errr{60}{75} &1.089\errr{94}{116}  &1.059\errr{165}{160} \\ 
\end{tabular}
\end{table}
%



\begin{table}
\caption{Results for $\hpw$, $\hpw/(1+\bpw)$ and $\hmw$ obtained
with the fitting procedure described in \protect{\sec{tpfalff}}.
The light-quark hopping parameter is fixed to $\kq=0.14144$
and all heavy-quark mass combinations are presented. Only transitions with
initial and final meson momenta less or equal to $\pi/(12a)$
are included.\label{hp.4144}}
\begin{tabular}{ccccccrr}
$\s p$ & $\s p'$ & $\omega$ & $\hpw$ & $\hpw/(1+\bpw)$ & $\chi^2/d.o.f.$ & $\hmw$& $\chi^2/d.o.f.$ \\
\hline
\hline
\multicolumn{8}{c}{$\kQ=0.121\longrightarrow\kQp=0.121$,$\quad\kq=0.14144$}\\
\hline
(0,0,0) & (1,0,0) & 1.037\err{1}{1} & 0.95\err{1}{1} & 0.96\err{1}{1} & 3.6/2 & 0.12\err{2}{3} & 23.6/5 \\
(1,0,0) & (1,0,0) & 0.995\err{3}{3} & 0.96\err{4}{4} & 0.96\err{4}{4} & 0.5/2 & 0.00\err{0}{0} & 8.3/6 \\
(1,0,0) & (0,0,0) & 1.037\err{1}{1} & 0.90\err{1}{1} & 0.90\err{1}{1} & 1.0/2 & -0.05\err{3}{2} & 1.3/5 \\
(1,0,0) & (0,1,0) & 1.075\err{3}{3} & 0.86\err{2}{2} & 0.87\err{2}{2} & 0.4/2 & 0.03\err{2}{2} & 16.9/8 \\
(1,0,0) & (-1,0,0) & 1.156\err{3}{3} & 0.78\err{3}{3} & 0.81\err{4}{3} & 1.6/2 & 0.03\err{2}{2} & 2.9/5 \\
\hline
\multicolumn{8}{c}{$\kQ=0.129\longrightarrow\kQp=0.121$,$\quad\kq=0.14144$}\\
\hline
(0,0,0) & (1,0,0) & 1.037\err{1}{1} & 0.98\err{1}{1} & 0.96\err{1}{1} & 2.6/2 & 0.22\err{2}{2} & 22.6/5 \\
(1,0,0) & (1,0,0) & 0.997\err{4}{3} & 0.99\err{4}{5} & 0.96\err{4}{4} & 0.1/2 & -0.74\err{47}{35} & 1.7/5 \\
(1,0,0) & (0,0,0) & 1.062\err{2}{2} & 0.89\err{1}{1} & 0.87\err{1}{1} & 0.7/2 & 0.01\err{3}{2} & 0.9/5 \\
(1,0,0) & (0,1,0) & 1.101\err{3}{3} & 0.84\err{2}{2} & 0.83\err{2}{2} & 0.8/2 & 0.07\err{2}{2} & 15.4/8 \\
(1,0,0) & (-1,0,0) & 1.205\err{3}{3} & 0.75\err{3}{3} & 0.75\err{3}{3} & 5.0/2 & 0.08\err{2}{2} & 6.8/5 \\
\hline
\multicolumn{8}{c}{$\kQ=0.121\longrightarrow\kQp=0.125$,$\quad\kq=0.14144$}\\
\hline
(0,0,0) & (1,0,0) & 1.047\err{2}{1} & 0.95\err{1}{1} & 0.95\err{1}{1} & 3.9/2 & 0.08\err{2}{2} & 23.3/5 \\
(1,0,0) & (1,0,0) & 0.995\err{3}{3} & 0.98\err{4}{4} & 0.96\err{4}{4} & 0.7/2 & 1.23\err{71}{94} & 3.0/5 \\
(1,0,0) & (0,0,0) & 1.037\err{1}{1} & 0.91\err{1}{1} & 0.91\err{2}{1} & 0.7/2 & -0.07\err{3}{3} & 0.9/5 \\
(1,0,0) & (0,1,0) & 1.085\err{3}{3} & 0.86\err{2}{2} & 0.86\err{2}{2} & 0.2/2 & 0.00\err{2}{2} & 13.5/8 \\
(1,0,0) & (-1,0,0) & 1.175\err{3}{3} & 0.77\err{3}{2} & 0.79\err{3}{3} & 1.1/2 & 0.01\err{2}{2} & 2.9/5 \\
\hline
\multicolumn{8}{c}{$\kQ=0.129\longrightarrow\kQp=0.125$,$\quad\kq=0.14144$}\\
\hline
(0,0,0) & (1,0,0) & 1.047\err{2}{1} & 0.95\err{1}{1} & 0.94\err{1}{1} & 2.8/2 & 0.17\err{2}{2} & 23.3/5 \\
(1,0,0) & (1,0,0) & 0.995\err{4}{4} & 0.98\err{5}{5} & 0.96\err{4}{4} & 0.3/2 & -1.10\err{86}{60} & 2.4/5 \\
(1,0,0) & (0,0,0) & 1.062\err{2}{2} & 0.88\err{1}{1} & 0.87\err{1}{1} & 0.6/2 & 0.00\err{3}{2} & 0.7/5 \\
(1,0,0) & (0,1,0) & 1.111\err{3}{3} & 0.82\err{2}{2} & 0.82\err{2}{2} & 0.5/2 & 0.05\err{2}{2} & 12.6/8 \\
(1,0,0) & (-1,0,0) & 1.228\err{3}{3} & 0.72\err{3}{2} & 0.74\err{3}{3} & 4.3/2 & 0.06\err{2}{2} & 6.5/5 \\
\hline
\multicolumn{8}{c}{$\kQ=0.121\longrightarrow\kQp=0.129$,$\quad\kq=0.14144$}\\
\hline
(0,0,0) & (1,0,0) & 1.062\err{2}{2} & 0.95\err{1}{1} & 0.93\err{1}{1} & 4.4/2 & 0.03\err{2}{3} & 22.5/5 \\
(1,0,0) & (1,0,0) & 0.997\err{4}{3} & 0.99\err{5}{4} & 0.96\err{4}{4} & 0.9/2 & 0.30\err{38}{41} & 3.8/5 \\
(1,0,0) & (0,0,0) & 1.037\err{1}{1} & 0.93\err{1}{1} & 0.91\err{1}{1} & 0.4/2 & -0.09\err{3}{3} & 0.5/5 \\
(1,0,0) & (0,1,0) & 1.101\err{3}{3} & 0.85\err{2}{2} & 0.84\err{2}{2} & 0.2/2 & -0.03\err{2}{2} & 10.4/8 \\
(1,0,0) & (-1,0,0) & 1.205\err{3}{3} & 0.77\err{3}{3} & 0.77\err{3}{3} & 0.7/2 & -0.01\err{3}{2} & 3.0/5 \\
\hline
\multicolumn{8}{c}{$\kQ=0.129\longrightarrow\kQp=0.129$,$\quad\kq=0.14144$}\\
\hline
(0,0,0) & (1,0,0) & 1.062\err{2}{2} & 0.91\err{1}{1} & 0.93\err{1}{1} & 3.1/2 & 0.11\err{2}{2} & 23.4/5 \\
(1,0,0) & (1,0,0) & 0.994\err{4}{4} & 0.95\err{5}{4} & 0.95\err{5}{4} & 1.3/2 & 0.00\err{0}{0} & 5.1/6 \\
(1,0,0) & (0,0,0) & 1.062\err{2}{2} & 0.87\err{1}{1} & 0.88\err{1}{1} & 0.4/2 & -0.02\err{3}{2} & 0.5/5 \\
(1,0,0) & (0,1,0) & 1.127\err{4}{4} & 0.78\err{2}{2} & 0.81\err{2}{2} & 0.3/2 & 0.02\err{2}{1} & 9.9/8 \\
(1,0,0) & (-1,0,0) & 1.261\err{3}{4} & 0.68\err{3}{2} & 0.72\err{3}{2} & 3.1/2 & 0.04\err{2}{2} & 6.0/5 \\
\hline
\multicolumn{8}{c}{$\kQ=0.121\longrightarrow\kQp=0.133$,$\quad\kq=0.14144$}\\
\hline
(0,0,0) & (1,0,0) & 1.088\err{3}{2} & 0.94\err{1}{1} & 0.90\err{1}{1} & 5.2/2 & -0.03\err{3}{3} & 21.9/5 \\
(1,0,0) & (1,0,0) & 1.005\err{4}{4} & 1.00\err{5}{5} & 0.94\err{5}{4} & 1.5/2 & -0.06\err{24}{27} & 4.3/5 \\
(1,0,0) & (0,0,0) & 1.037\err{1}{1} & 0.96\err{1}{1} & 0.91\err{1}{1} & 0.2/2 & -0.12\err{3}{3} & 0.3/5 \\
(1,0,0) & (0,1,0) & 1.128\err{4}{4} & 0.84\err{2}{2} & 0.81\err{2}{2} & 0.2/2 & -0.08\err{2}{2} & 7.7/8 \\
(1,0,0) & (-1,0,0) & 1.252\err{4}{4} & 0.75\err{3}{3} & 0.74\err{3}{3} & 0.2/2 & -0.04\err{3}{2} & 3.3/5 \\
\hline
\multicolumn{8}{c}{$\kQ=0.129\longrightarrow\kQp=0.133$,$\quad\kq=0.14144$}\\
\hline
(0,0,0) & (1,0,0) & 1.088\err{3}{2} & 0.92\err{1}{1} & 0.90\err{1}{1} & 3.7/2 & 0.06\err{2}{3} & 23.7/5 \\
(1,0,0) & (1,0,0) & 0.996\err{5}{5} & 0.97\err{4}{4} & 0.94\err{4}{4} & 2.1/2 & 0.06\err{57}{65} & 5.0/5 \\
(1,0,0) & (0,0,0) & 1.062\err{2}{2} & 0.90\err{1}{1} & 0.88\err{1}{1} & 0.2/2 & -0.06\err{3}{2} & 0.4/5 \\
(1,0,0) & (0,1,0) & 1.155\err{5}{5} & 0.78\err{2}{2} & 0.78\err{2}{2} & 0.2/2 & -0.02\err{2}{2} & 7.4/8 \\
(1,0,0) & (-1,0,0) & 1.315\err{4}{5} & 0.67\err{2}{2} & 0.69\err{2}{2} & 1.6/2 & 0.01\err{2}{2} & 5.4/5 \\
\hline
\multicolumn{8}{c}{$\kQ=0.125\longrightarrow\kQp=0.125$,$\quad\kq=0.14144$}\\
\hline
(0,0,0) & (1,0,0) & 1.047\err{2}{1} & 0.93\err{1}{1} & 0.95\err{1}{1} & 3.4/2 & 0.12\err{2}{2} & 24.2/5 \\
(1,0,0) & (1,0,0) & 0.994\err{3}{3} & 0.96\err{4}{4} & 0.96\err{4}{4} & 0.9/2 & 0.00\err{0}{0} & 6.5/6 \\
(1,0,0) & (0,0,0) & 1.047\err{2}{1} & 0.88\err{1}{1} & 0.89\err{1}{1} & 0.7/2 & -0.04\err{3}{2} & 0.8/5 \\
(1,0,0) & (0,1,0) & 1.096\err{3}{3} & 0.82\err{2}{2} & 0.84\err{2}{2} & 0.4/2 & 0.02\err{2}{2} & 13.6/8 \\
(1,0,0) & (-1,0,0) & 1.197\err{3}{3} & 0.74\err{3}{2} & 0.77\err{3}{3} & 2.4/2 & 0.04\err{2}{2} & 4.5/5 \\
\hline
\multicolumn{8}{c}{$\kQ=0.133\longrightarrow\kQp=0.125$,$\quad\kq=0.14144$}\\
\hline
(0,0,0) & (1,0,0) & 1.047\err{2}{1} & 0.98\err{1}{1} & 0.94\err{1}{1} & 2.2/2 & 0.23\err{2}{2} & 19.4/5 \\
(1,0,0) & (1,0,0) & 1.000\err{4}{4} & 1.00\err{5}{5} & 0.95\err{4}{4} & 0.1/2 & -0.46\err{41}{33} & 1.7/5 \\
(1,0,0) & (0,0,0) & 1.088\err{3}{2} & 0.88\err{1}{1} & 0.85\err{1}{1} & 0.6/2 & 0.04\err{3}{2} & 0.7/5 \\
(1,0,0) & (0,1,0) & 1.139\err{4}{4} & 0.81\err{2}{2} & 0.79\err{2}{2} & 0.4/2 & 0.08\err{2}{2} & 10.2/8 \\
(1,0,0) & (-1,0,0) & 1.278\err{4}{4} & 0.69\err{3}{2} & 0.70\err{3}{2} & 6.4/2 & 0.09\err{2}{1} & 8.4/5 \\
\hline
\multicolumn{8}{c}{$\kQ=0.125\longrightarrow\kQp=0.133$,$\quad\kq=0.14144$}\\
\hline
(0,0,0) & (1,0,0) & 1.088\err{3}{2} & 0.93\err{1}{1} & 0.90\err{1}{1} & 4.6/2 & 0.01\err{2}{2} & 23.4/5 \\
(1,0,0) & (1,0,0) & 1.000\err{4}{4} & 0.99\err{5}{5} & 0.94\err{4}{4} & 1.8/2 & -0.02\err{30}{36} & 4.7/5 \\
(1,0,0) & (0,0,0) & 1.047\err{2}{1} & 0.94\err{1}{1} & 0.90\err{1}{1} & 0.2/2 & -0.09\err{3}{3} & 0.3/5 \\
(1,0,0) & (0,1,0) & 1.139\err{4}{4} & 0.82\err{2}{2} & 0.80\err{2}{2} & 0.2/2 & -0.05\err{2}{2} & 7.7/8 \\
(1,0,0) & (-1,0,0) & 1.278\err{4}{4} & 0.72\err{3}{2} & 0.72\err{3}{2} & 0.7/2 & -0.02\err{2}{2} & 4.2/5 \\
\hline
\multicolumn{8}{c}{$\kQ=0.133\longrightarrow\kQp=0.133$,$\quad\kq=0.14144$}\\
\hline
(0,0,0) & (1,0,0) & 1.088\err{3}{2} & 0.88\err{1}{1} & 0.90\err{1}{1} & 2.8/2 & 0.10\err{2}{2} & 21.3/5 \\
(1,0,0) & (1,0,0) & 0.994\err{6}{6} & 0.95\err{5}{5} & 0.95\err{5}{5} & 1.6/2 & 0.00\err{0}{0} & 4.3/6 \\
(1,0,0) & (0,0,0) & 1.088\err{3}{2} & 0.84\err{1}{1} & 0.86\err{1}{1} & 0.3/2 & -0.02\err{3}{2} & 0.6/5 \\
(1,0,0) & (0,1,0) & 1.184\err{5}{5} & 0.73\err{2}{2} & 0.75\err{2}{2} & 0.2/2 & 0.01\err{2}{2} & 6.4/8 \\
(1,0,0) & (-1,0,0) & 1.375\err{5}{5} & 0.60\err{2}{2} & 0.65\err{2}{2} & 2.8/2 & 0.03\err{2}{1} & 6.3/5 \\
\end{tabular}
%
\end{table}


\begin{table}
\caption{Results for $\hpw$, $\hpw/(1+\bpw)$ and $\hmw$ obtained
with the fitting procedure described in \protect{\sec{tpfalff}}.
The light-quark hopping parameter is fixed to $\kq=0.14226$
and all heavy-quark mass combinations are presented. Only transitions with
initial and final meson momenta less or equal to $\pi/(12a)$
are included.\label{hp.4226}}
\begin{tabular}{ccccccrr}
$\s p$ & $\s p'$ & $\omega$ & $\hpw$ & $\hpw/(1+\bpw)$ & $\chi^2/d.o.f.$ & $\hmw$& $\chi^2/d.o.f.$ \\
\hline
\hline
\multicolumn{8}{c}{$\kQ=0.121\longrightarrow\kQp=0.121$,$\quad\kq=0.14226$}\\
\hline
(0,0,0) & (1,0,0) & 1.039\err{2}{2} & 0.95\err{1}{1} & 0.96\err{1}{1} & 3.4/2 & 0.14\err{4}{4} & 20.8/5 \\
(1,0,0) & (1,0,0) & 0.996\err{4}{4} & 0.89\err{7}{7} & 0.89\err{7}{7} & 0.2/2 & 0.00\err{0}{0} & 2.9/6 \\
(1,0,0) & (0,0,0) & 1.039\err{2}{2} & 0.87\err{2}{2} & 0.88\err{2}{2} & 0.7/2 & -0.04\err{6}{5} & 0.7/5 \\
(1,0,0) & (0,1,0) & 1.080\err{4}{4} & 0.84\err{4}{3} & 0.86\err{4}{3} & 0.2/2 & 0.04\err{4}{3} & 8.6/8 \\
(1,0,0) & (-1,0,0) & 1.165\err{4}{4} & 0.81\err{5}{4} & 0.84\err{5}{4} & 1.3/2 & 0.05\err{4}{4} & 3.7/5 \\
\hline
\multicolumn{8}{c}{$\kQ=0.129\longrightarrow\kQp=0.121$,$\quad\kq=0.14226$}\\
\hline
(0,0,0) & (1,0,0) & 1.039\err{2}{2} & 0.99\err{1}{1} & 0.96\err{1}{1} & 3.0/2 & 0.23\err{4}{3} & 18.8/5 \\
(1,0,0) & (1,0,0) & 0.999\err{5}{4} & 0.93\err{8}{8} & 0.89\err{7}{7} & 0.1/2 & -0.72\err{83}{60} & 0.6/5 \\
(1,0,0) & (0,0,0) & 1.067\err{3}{2} & 0.87\err{2}{2} & 0.85\err{2}{2} & 0.3/2 & 0.01\err{5}{4} & 0.6/5 \\
(1,0,0) & (0,1,0) & 1.109\err{5}{4} & 0.83\err{4}{3} & 0.83\err{4}{3} & 0.5/2 & 0.06\err{4}{3} & 8.1/8 \\
(1,0,0) & (-1,0,0) & 1.219\err{5}{4} & 0.78\err{4}{4} & 0.79\err{5}{4} & 3.5/2 & 0.10\err{3}{3} & 6.8/5 \\
\hline
\multicolumn{8}{c}{$\kQ=0.121\longrightarrow\kQp=0.125$,$\quad\kq=0.14226$}\\
\hline
(0,0,0) & (1,0,0) & 1.050\err{2}{2} & 0.95\err{1}{1} & 0.95\err{1}{2} & 3.4/2 & 0.09\err{4}{4} & 19.8/5 \\
(1,0,0) & (1,0,0) & 0.996\err{5}{4} & 0.90\err{7}{7} & 0.89\err{7}{7} & 0.3/2 & 0.81\errr{119}{162} & 1.1/5 \\
(1,0,0) & (0,0,0) & 1.039\err{2}{2} & 0.89\err{2}{2} & 0.88\err{2}{2} & 0.6/2 & -0.04\err{5}{5} & 0.6/5 \\
(1,0,0) & (0,1,0) & 1.091\err{4}{4} & 0.84\err{4}{3} & 0.85\err{4}{3} & 0.1/2 & 0.01\err{4}{3} & 6.3/8 \\
(1,0,0) & (-1,0,0) & 1.187\err{4}{4} & 0.81\err{4}{4} & 0.83\err{4}{4} & 1.0/2 & 0.03\err{4}{3} & 3.8/5 \\
\hline
\multicolumn{8}{c}{$\kQ=0.129\longrightarrow\kQp=0.125$,$\quad\kq=0.14226$}\\
\hline
(0,0,0) & (1,0,0) & 1.050\err{2}{2} & 0.96\err{1}{1} & 0.95\err{1}{1} & 3.2/2 & 0.18\err{4}{3} & 19.1/5 \\
(1,0,0) & (1,0,0) & 0.997\err{5}{5} & 0.91\err{7}{7} & 0.89\err{7}{7} & 0.1/2 & -0.89\errr{143}{107} & 0.8/5 \\
(1,0,0) & (0,0,0) & 1.067\err{3}{2} & 0.86\err{2}{2} & 0.85\err{2}{2} & 0.3/2 & 0.00\err{5}{4} & 0.7/5 \\
(1,0,0) & (0,1,0) & 1.120\err{5}{5} & 0.81\err{4}{3} & 0.81\err{4}{3} & 0.2/2 & 0.04\err{3}{3} & 6.0/8 \\
(1,0,0) & (-1,0,0) & 1.244\err{5}{5} & 0.75\err{4}{4} & 0.78\err{4}{4} & 2.8/2 & 0.08\err{3}{3} & 6.6/5 \\
\hline
\multicolumn{8}{c}{$\kQ=0.121\longrightarrow\kQp=0.129$,$\quad\kq=0.14226$}\\
\hline
(0,0,0) & (1,0,0) & 1.067\err{3}{2} & 0.95\err{1}{1} & 0.93\err{1}{1} & 3.7/2 & 0.04\err{4}{4} & 18.6/5 \\
(1,0,0) & (1,0,0) & 0.999\err{5}{4} & 0.91\err{7}{7} & 0.88\err{7}{7} & 0.7/2 & -0.04\err{57}{69} & 1.9/5 \\
(1,0,0) & (0,0,0) & 1.039\err{2}{2} & 0.91\err{2}{2} & 0.89\err{2}{2} & 0.3/2 & -0.04\err{6}{5} & 0.5/5 \\
(1,0,0) & (0,1,0) & 1.109\err{5}{4} & 0.83\err{4}{3} & 0.82\err{4}{3} & 0.1/2 & -0.02\err{4}{3} & 4.4/8 \\
(1,0,0) & (-1,0,0) & 1.219\err{5}{4} & 0.80\err{4}{4} & 0.81\err{4}{4} & 0.7/2 & 0.01\err{4}{4} & 4.1/5 \\
\hline
\multicolumn{8}{c}{$\kQ=0.129\longrightarrow\kQp=0.129$,$\quad\kq=0.14226$}\\
\hline
(0,0,0) & (1,0,0) & 1.067\err{3}{2} & 0.91\err{1}{1} & 0.93\err{1}{1} & 3.5/2 & 0.12\err{4}{3} & 19.2/5 \\
(1,0,0) & (1,0,0) & 0.995\err{7}{5} & 0.88\err{7}{7} & 0.88\err{7}{7} & 0.5/2 & 0.00\err{0}{0} & 1.6/6 \\
(1,0,0) & (0,0,0) & 1.067\err{3}{2} & 0.84\err{2}{2} & 0.86\err{2}{2} & 0.3/2 & -0.01\err{5}{3} & 0.8/5 \\
(1,0,0) & (0,1,0) & 1.138\err{6}{5} & 0.77\err{3}{3} & 0.79\err{4}{3} & 0.1/2 & 0.02\err{4}{3} & 4.2/8 \\
(1,0,0) & (-1,0,0) & 1.282\err{5}{5} & 0.71\err{4}{4} & 0.76\err{4}{4} & 2.1/2 & 0.05\err{3}{2} & 6.6/5 \\
\hline
\multicolumn{8}{c}{$\kQ=0.121\longrightarrow\kQp=0.133$,$\quad\kq=0.14226$}\\
\hline
(0,0,0) & (1,0,0) & 1.097\err{4}{3} & 0.94\err{2}{1} & 0.90\err{2}{1} & 4.2/2 & -0.02\err{4}{3} & 17.8/5 \\
(1,0,0) & (1,0,0) & 1.009\err{7}{5} & 0.91\err{7}{8} & 0.85\err{7}{8} & 1.7/2 & -0.37\err{39}{43} & 3.0/5 \\
(1,0,0) & (0,0,0) & 1.039\err{2}{2} & 0.94\err{2}{2} & 0.89\err{2}{2} & 0.1/2 & -0.06\err{6}{5} & 0.8/5 \\
(1,0,0) & (0,1,0) & 1.141\err{6}{5} & 0.82\err{3}{3} & 0.80\err{3}{3} & 0.1/2 & -0.06\err{4}{3} & 3.0/8 \\
(1,0,0) & (-1,0,0) & 1.273\err{5}{5} & 0.79\err{4}{3} & 0.79\err{4}{3} & 0.4/2 & -0.01\err{4}{4} & 4.4/5 \\
\hline
\multicolumn{8}{c}{$\kQ=0.129\longrightarrow\kQp=0.133$,$\quad\kq=0.14226$}\\
\hline
(0,0,0) & (1,0,0) & 1.097\err{4}{3} & 0.91\err{1}{1} & 0.90\err{1}{1} & 4.1/2 & 0.05\err{4}{4} & 19.5/5 \\
(1,0,0) & (1,0,0) & 0.999\err{8}{6} & 0.88\err{8}{8} & 0.84\err{7}{8} & 2.1/2 & -0.63\errr{92}{104} & 3.3/5 \\
(1,0,0) & (0,0,0) & 1.067\err{3}{2} & 0.88\err{2}{2} & 0.86\err{2}{2} & 0.2/2 & -0.02\err{5}{3} & 1.1/5 \\
(1,0,0) & (0,1,0) & 1.171\err{6}{6} & 0.77\err{3}{3} & 0.77\err{3}{3} & 0.1/2 & -0.02\err{4}{2} & 2.7/8 \\
(1,0,0) & (-1,0,0) & 1.343\err{6}{6} & 0.71\err{4}{3} & 0.73\err{4}{4} & 1.3/2 & 0.03\err{3}{3} & 6.7/5 \\
\end{tabular}
%
\end{table}


\begin{table}
\caption{Results for $\hpw$, $\hpw/(1+\bpw)$ and $\hmw$ obtained
with the fitting procedure described in \protect{\sec{tpfalff}}.
The light-quark hopping parameter is fixed to $\kq=0.14262$
and all heavy-quark mass combinations are presented. Only transitions with
initial and final meson momenta less or equal to $\pi/(12a)$
are included.\label{hp.4262}}
\begin{tabular}{ccccccrr}
$\s p$ & $\s p'$ & $\omega$ & $\hpw$ & $\hpw/(1+\bpw)$ & $\chi^2/d.o.f.$ & $\hmw$& $\chi^2/d.o.f.$ \\
\hline
\hline
\multicolumn{8}{c}{$\kQ=0.121\longrightarrow\kQp=0.121$,$\quad\kq=0.14262$}\\
\hline
(0,0,0) & (1,0,0) & 1.041\err{2}{2} & 0.95\err{2}{2} & 0.96\err{2}{2} & 2.7/2 & 0.12\err{6}{5} & 12.8/5 \\
(1,0,0) & (1,0,0) & 0.997\err{4}{5} & 0.79\err{11}{11} & 0.79\err{11}{11} & 0.1/2 & 0.00\err{0}{0} & 1.1/6 \\
(1,0,0) & (0,0,0) & 1.041\err{2}{2} & 0.84\err{3}{3} & 0.85\err{3}{3} & 0.8/2 & -0.05\err{8}{7} & 1.0/5 \\
(1,0,0) & (0,1,0) & 1.083\err{4}{4} & 0.82\err{5}{5} & 0.84\err{5}{5} & 0.1/2 & 0.02\err{6}{5} & 4.8/8 \\
(1,0,0) & (-1,0,0) & 1.170\err{4}{4} & 0.84\err{6}{7} & 0.88\err{6}{7} & 0.5/2 & 0.04\err{6}{6} & 2.5/5 \\
\hline
\multicolumn{8}{c}{$\kQ=0.129\longrightarrow\kQp=0.121$,$\quad\kq=0.14262$}\\
\hline
(0,0,0) & (1,0,0) & 1.041\err{2}{2} & 0.99\err{2}{2} & 0.96\err{2}{2} & 3.0/2 & 0.22\err{7}{6} & 11.0/5 \\
(1,0,0) & (1,0,0) & 1.001\err{5}{6} & 0.82\err{12}{11} & 0.79\err{11}{11} & 0.1/2 & -0.36\errr{125}{100} & 0.9/5 \\
(1,0,0) & (0,0,0) & 1.070\err{3}{3} & 0.85\err{4}{3} & 0.83\err{4}{3} & 0.2/2 & -0.01\err{7}{7} & 1.1/5 \\
(1,0,0) & (0,1,0) & 1.114\err{5}{5} & 0.82\err{5}{5} & 0.82\err{5}{5} & 0.2/2 & 0.05\err{5}{5} & 5.0/8 \\
(1,0,0) & (-1,0,0) & 1.226\err{6}{5} & 0.81\err{6}{7} & 0.82\err{6}{7} & 1.6/2 & 0.10\err{5}{4} & 4.5/5 \\
\hline
\multicolumn{8}{c}{$\kQ=0.121\longrightarrow\kQp=0.125$,$\quad\kq=0.14262$}\\
\hline
(0,0,0) & (1,0,0) & 1.052\err{2}{2} & 0.96\err{2}{3} & 0.96\err{2}{3} & 2.5/2 & 0.08\err{6}{5} & 11.5/5 \\
(1,0,0) & (1,0,0) & 0.998\err{5}{5} & 0.79\err{11}{12} & 0.78\err{11}{12} & 0.1/2 & -0.25\errr{187}{251} & 1.0/5 \\
(1,0,0) & (0,0,0) & 1.041\err{2}{2} & 0.86\err{3}{3} & 0.86\err{3}{3} & 0.8/2 & -0.03\err{8}{7} & 1.0/5 \\
(1,0,0) & (0,1,0) & 1.095\err{5}{5} & 0.82\err{5}{5} & 0.83\err{5}{5} & 0.2/2 & 0.00\err{5}{5} & 3.4/8 \\
(1,0,0) & (-1,0,0) & 1.192\err{4}{4} & 0.85\err{6}{7} & 0.88\err{6}{7} & 0.4/2 & 0.03\err{6}{6} & 2.6/5 \\
\hline
\multicolumn{8}{c}{$\kQ=0.129\longrightarrow\kQp=0.125$,$\quad\kq=0.14262$}\\
\hline
(0,0,0) & (1,0,0) & 1.052\err{2}{2} & 0.96\err{2}{1} & 0.94\err{2}{1} & 3.0/2 & 0.16\err{5}{5} & 10.6/5 \\
(1,0,0) & (1,0,0) & 0.998\err{6}{6} & 0.80\err{12}{13} & 0.79\err{11}{12} & 0.2/2 & -0.11\errr{211}{177} & 1.1/5 \\
(1,0,0) & (0,0,0) & 1.070\err{3}{3} & 0.84\err{3}{3} & 0.83\err{3}{3} & 0.3/2 & 0.00\err{7}{6} & 1.3/5 \\
(1,0,0) & (0,1,0) & 1.125\err{6}{6} & 0.79\err{5}{5} & 0.80\err{5}{5} & 0.1/2 & 0.03\err{5}{5} & 3.6/8 \\
(1,0,0) & (-1,0,0) & 1.252\err{6}{6} & 0.78\err{5}{6} & 0.81\err{6}{6} & 1.3/2 & 0.07\err{5}{4} & 4.4/5 \\
\hline
\multicolumn{8}{c}{$\kQ=0.121\longrightarrow\kQp=0.129$,$\quad\kq=0.14262$}\\
\hline
(0,0,0) & (1,0,0) & 1.070\err{3}{3} & 0.96\err{2}{3} & 0.94\err{2}{3} & 2.4/2 & 0.03\err{6}{5} & 10.3/5 \\
(1,0,0) & (1,0,0) & 1.001\err{5}{6} & 0.79\err{11}{13} & 0.76\err{11}{13} & 0.2/2 & -0.70\errr{92}{106} & 1.4/5 \\
(1,0,0) & (0,0,0) & 1.041\err{2}{2} & 0.89\err{3}{3} & 0.87\err{3}{3} & 0.6/2 & -0.02\err{8}{8} & 0.8/5 \\
(1,0,0) & (0,1,0) & 1.114\err{5}{5} & 0.81\err{5}{5} & 0.81\err{5}{5} & 0.3/2 & -0.02\err{6}{5} & 2.3/8 \\
(1,0,0) & (-1,0,0) & 1.226\err{6}{5} & 0.85\err{6}{6} & 0.86\err{6}{6} & 0.3/2 & 0.01\err{6}{6} & 2.8/5 \\
\hline
\multicolumn{8}{c}{$\kQ=0.129\longrightarrow\kQp=0.129$,$\quad\kq=0.14262$}\\
\hline
(0,0,0) & (1,0,0) & 1.070\err{3}{3} & 0.91\err{2}{2} & 0.92\err{2}{2} & 3.2/2 & 0.10\err{5}{5} & 10.4/5 \\
(1,0,0) & (1,0,0) & 0.998\err{7}{7} & 0.77\err{11}{12} & 0.77\err{11}{12} & 0.1/2 & 0.00\err{0}{0} & 1.5/6 \\
(1,0,0) & (0,0,0) & 1.070\err{3}{3} & 0.82\err{3}{3} & 0.83\err{3}{3} & 0.3/2 & 0.01\err{6}{5} & 1.3/5 \\
(1,0,0) & (0,1,0) & 1.145\err{6}{6} & 0.75\err{5}{4} & 0.77\err{5}{4} & 0.2/2 & 0.01\err{4}{4} & 2.5/8 \\
(1,0,0) & (-1,0,0) & 1.292\err{6}{6} & 0.74\err{5}{5} & 0.79\err{6}{6} & 0.9/2 & 0.05\err{5}{4} & 4.5/5 \\
\hline
\multicolumn{8}{c}{$\kQ=0.121\longrightarrow\kQp=0.133$,$\quad\kq=0.14262$}\\
\hline
(0,0,0) & (1,0,0) & 1.102\err{4}{4} & 0.95\err{3}{3} & 0.91\err{3}{3} & 2.6/2 & -0.03\err{6}{5} & 9.5/5 \\
(1,0,0) & (1,0,0) & 1.012\err{6}{6} & 0.77\err{13}{15} & 0.73\err{12}{14} & 0.9/2 & -0.87\err{61}{64} & 2.0/5 \\
(1,0,0) & (0,0,0) & 1.041\err{2}{2} & 0.92\err{2}{3} & 0.88\err{2}{3} & 0.3/2 & -0.01\err{8}{8} & 0.7/5 \\
(1,0,0) & (0,1,0) & 1.147\err{6}{6} & 0.80\err{5}{5} & 0.78\err{4}{5} & 0.2/2 & -0.06\err{6}{4} & 1.6/8 \\
(1,0,0) & (-1,0,0) & 1.283\err{6}{6} & 0.84\err{6}{6} & 0.83\err{6}{6} & 0.2/2 & -0.01\err{6}{6} & 2.9/5 \\
\hline
\multicolumn{8}{c}{$\kQ=0.129\longrightarrow\kQp=0.133$,$\quad\kq=0.14262$}\\
\hline
(0,0,0) & (1,0,0) & 1.102\err{4}{4} & 0.91\err{2}{2} & 0.90\err{2}{2} & 3.6/2 & 0.03\err{5}{5} & 10.6/5 \\
(1,0,0) & (1,0,0) & 1.002\err{8}{8} & 0.75\err{13}{13} & 0.72\err{12}{13} & 1.0/2 & -1.65\errr{149}{155} & 2.0/5 \\
(1,0,0) & (0,0,0) & 1.070\err{3}{3} & 0.86\err{3}{3} & 0.84\err{2}{3} & 0.2/2 & 0.00\err{6}{5} & 1.4/5 \\
(1,0,0) & (0,1,0) & 1.179\err{7}{7} & 0.75\err{4}{4} & 0.75\err{5}{4} & 0.2/2 & -0.02\err{5}{4} & 1.8/8 \\
(1,0,0) & (-1,0,0) & 1.357\err{8}{7} & 0.74\err{6}{5} & 0.76\err{6}{6} & 0.6/2 & 0.03\err{5}{4} & 4.7/5 \\
\end{tabular}
%
\end{table}


\begin{table}
\caption{Results for $\w$ and $\hpw/(1+\bpw)$, for $\kq=\kcrit=$0.14315(2),
obtained from covariant, linear extrapolations of the results for
$\kq=$0.14144, 0.14226, 0.14262.  All heavy-quark mass combinations
are presented. The first $\chi^2/d.o.f.$ column corresponds to the $\w$
extrapolation; the second, to the extrapolation of $\hpw/(1+\bpw)$.
\label{hp.k_cr}}
\begin{tabular}{cccrcr}
$\s p$ & $\s p'$ & $\w$ & $\chi^2/d.o.f.$ & $\hpw/(1+\bpw)$ & $\chi^2/d.o.f.$ \\
\hline
\hline
\multicolumn{6}{c}{$\kQ=0.121\longrightarrow\kQp=0.121$,$\quad\kq=\kcrit$}\\
\hline
 (0,0,0) & (1,0,0) & 1.042\err{ 3}{ 3} & 0.1/1 & 0.97\err{ 2}{ 2} & 0.0/1\\
 (1,0,0) & (1,0,0) & 0.997\err{ 6}{ 5} & 0.2/1 & 0.94\err{ 8}{ 9} & 4.6/1\\
 (1,0,0) & (0,0,0) & 1.042\err{ 3}{ 3} & 0.1/1 & 0.87\err{ 3}{ 2} & 2.3/1\\
 (1,0,0) & (0,1,0) & 1.086\err{ 6}{ 5} & 0.1/1 & 0.86\err{ 5}{ 4} & 0.5/1\\
 (1,0,0) & (-1,0,0) & 1.175\err{ 6}{ 5} & 0.1/1 & 0.85\err{ 6}{ 5} & 1.1/1\\
\hline
\multicolumn{6}{c}{$\kQ=0.129\longrightarrow\kQp=0.121$,$\quad\kq=\kcrit$}\\
\hline
 (0,0,0) & (1,0,0) & 1.042\err{ 3}{ 3} & 0.1/1 & 0.97\err{ 2}{ 1} & 0.0/1\\
 (1,0,0) & (1,0,0) & 1.001\err{ 7}{ 7} & 0.3/1 & 0.96\err{ 8}{ 9} & 4.1/1\\
 (1,0,0) & (0,0,0) & 1.073\err{ 4}{ 4} & 0.3/1 & 0.84\err{ 3}{ 2} & 0.3/1\\
 (1,0,0) & (0,1,0) & 1.118\err{ 6}{ 6} & 0.2/1 & 0.83\err{ 5}{ 4} & 0.1/1\\
 (1,0,0) & (-1,0,0) & 1.235\err{ 7}{ 6} & 0.2/1 & 0.80\err{ 6}{ 5} & 0.8/1\\
\hline
\multicolumn{6}{c}{$\kQ=0.121\longrightarrow\kQp=0.125$,$\quad\kq=\kcrit$}\\
\hline
 (0,0,0) & (1,0,0) & 1.054\err{ 3}{ 3} & 0.2/1 & 0.95\err{ 2}{ 2} & 0.4/1\\
 (1,0,0) & (1,0,0) & 0.998\err{ 6}{ 6} & 0.2/1 & 0.94\err{ 8}{ 9} & 4.3/1\\
 (1,0,0) & (0,0,0) & 1.042\err{ 3}{ 3} & 0.1/1 & 0.87\err{ 3}{ 2} & 1.1/1\\
 (1,0,0) & (0,1,0) & 1.098\err{ 6}{ 5} & 0.2/1 & 0.84\err{ 6}{ 4} & 0.3/1\\
 (1,0,0) & (-1,0,0) & 1.199\err{ 6}{ 5} & 0.1/1 & 0.83\err{ 6}{ 5} & 1.7/1\\
\hline
\multicolumn{6}{c}{$\kQ=0.129\longrightarrow\kQp=0.125$,$\quad\kq=\kcrit$}\\
\hline
 (0,0,0) & (1,0,0) & 1.054\err{ 3}{ 3} & 0.2/1 & 0.95\err{ 1}{ 1} & 0.2/1\\
 (1,0,0) & (1,0,0) & 0.998\err{ 8}{ 7} & 0.3/1 & 0.95\err{ 8}{ 9} & 4.1/1\\
 (1,0,0) & (0,0,0) & 1.073\err{ 4}{ 4} & 0.3/1 & 0.85\err{ 3}{ 2} & 1.3/1\\
 (1,0,0) & (0,1,0) & 1.130\err{ 7}{ 7} & 0.3/1 & 0.81\err{ 5}{ 4} & 0.4/1\\
 (1,0,0) & (-1,0,0) & 1.262\err{ 7}{ 6} & 0.2/1 & 0.79\err{ 5}{ 5} & 0.6/1\\
\hline
\multicolumn{6}{c}{$\kQ=0.121\longrightarrow\kQp=0.129$,$\quad\kq=\kcrit$}\\
\hline
 (0,0,0) & (1,0,0) & 1.073\err{ 4}{ 4} & 0.3/1 & 0.93\err{ 2}{ 2} & 0.6/1\\
 (1,0,0) & (1,0,0) & 1.001\err{ 7}{ 7} & 0.3/1 & 0.93\err{ 9}{ 9} & 4.7/1\\
 (1,0,0) & (0,0,0) & 1.042\err{ 3}{ 3} & 0.1/1 & 0.88\err{ 3}{ 2} & 1.6/1\\
 (1,0,0) & (0,1,0) & 1.118\err{ 6}{ 6} & 0.2/1 & 0.82\err{ 5}{ 3} & 0.4/1\\
 (1,0,0) & (-1,0,0) & 1.235\err{ 7}{ 6} & 0.2/1 & 0.81\err{ 6}{ 5} & 1.7/1\\
\hline
\multicolumn{6}{c}{$\kQ=0.129\longrightarrow\kQp=0.129$,$\quad\kq=\kcrit$}\\
\hline
 (0,0,0) & (1,0,0) & 1.073\err{ 4}{ 4} & 0.3/1 & 0.93\err{ 2}{ 1} & 0.1/1\\
 (1,0,0) & (1,0,0) & 0.997\err{ 9}{ 7} & 0.4/1 & 0.93\err{ 9}{ 9} & 4.0/1\\
 (1,0,0) & (0,0,0) & 1.073\err{ 4}{ 4} & 0.3/1 & 0.85\err{ 3}{ 2} & 1.9/1\\
 (1,0,0) & (0,1,0) & 1.150\err{ 8}{ 8} & 0.3/1 & 0.79\err{ 5}{ 4} & 0.6/1\\
 (1,0,0) & (-1,0,0) & 1.304\err{ 8}{ 8} & 0.2/1 & 0.77\err{ 5}{ 5} & 0.7/1\\
\hline
\multicolumn{6}{c}{$\kQ=0.121\longrightarrow\kQp=0.133$,$\quad\kq=\kcrit$}\\
\hline
 (0,0,0) & (1,0,0) & 1.108\err{ 5}{ 5} & 0.4/1 & 0.90\err{ 2}{ 2} & 0.5/1\\
 (1,0,0) & (1,0,0) & 1.013\err{ 9}{ 7} & 0.4/1 & 0.92\err{ 9}{ 9} & 5.1/1\\
 (1,0,0) & (0,0,0) & 1.042\err{ 3}{ 3} & 0.1/1 & 0.89\err{ 3}{ 2} & 2.2/1\\
 (1,0,0) & (0,1,0) & 1.154\err{ 7}{ 7} & 0.3/1 & 0.80\err{ 4}{ 3} & 0.7/1\\
 (1,0,0) & (-1,0,0) & 1.296\err{ 7}{ 7} & 0.3/1 & 0.79\err{ 5}{ 5} & 1.7/1\\
\hline
\multicolumn{6}{c}{$\kQ=0.129\longrightarrow\kQp=0.133$,$\quad\kq=\kcrit$}\\
\hline
 (0,0,0) & (1,0,0) & 1.108\err{ 5}{ 5} & 0.4/1 & 0.90\err{ 2}{ 1} & 0.0/1\\
 (1,0,0) & (1,0,0) & 1.002\err{11}{ 9} & 0.4/1 & 0.90\err{ 8}{10} & 4.0/1\\
 (1,0,0) & (0,0,0) & 1.073\err{ 4}{ 4} & 0.3/1 & 0.86\err{ 2}{ 2} & 1.6/1\\
 (1,0,0) & (0,1,0) & 1.188\err{ 9}{ 9} & 0.4/1 & 0.77\err{ 5}{ 4} & 0.6/1\\
 (1,0,0) & (-1,0,0) & 1.374\err{ 9}{ 9} & 0.3/1 & 0.74\err{ 5}{ 4} & 0.9/1\\
\end{tabular}
%
\end{table}


\begin{table}
\caption{Results for $\w$ and $\hpw/(1+\bpw)$, for $\kq=\k_s=$0.1419(1),
obtained from covariant, linear interpolations of the results for
$\kq=$0.14144, 0.14226, 0.14262.  All heavy-quark mass combinations
are presented. The $\chi^2/d.o.f.$ are the same as for the chiral
extrapolations (see \protect{\tab{hp.k_cr}}).
\label{hp.k_st}}
\begin{tabular}{cccc}
$\s p$ & $\s p'$ & $\w$ & $\hpw/(1+\bpw)$ \\
\hline
\hline
\multicolumn{4}{c}{$\kQ=0.121\longrightarrow\kQp=0.121$,$\quad\kq=\k_s$}\\
\hline
 (0,0,0) & (1,0,0) & 1.039\er{2}{2} & 0.96\er{1}{1}\\
 (1,0,0) & (1,0,0) & 0.996\er{4}{3} & 0.97\er{5}{5}\\
 (1,0,0) & (0,0,0) & 1.039\er{2}{2} & 0.90\er{2}{1}\\
 (1,0,0) & (0,1,0) & 1.079\er{4}{3} & 0.87\er{3}{2}\\
 (1,0,0) & (-1,0,0) & 1.161\er{4}{4} & 0.82\er{4}{3}\\
\hline
\multicolumn{4}{c}{$\kQ=0.129\longrightarrow\kQp=0.121$,$\quad\kq=\k_s$}\\
\hline
 (0,0,0) & (1,0,0) & 1.039\er{2}{2} & 0.96\er{1}{1}\\
 (1,0,0) & (1,0,0) & 0.999\er{4}{4} & 0.97\er{5}{5}\\
 (1,0,0) & (0,0,0) & 1.065\er{2}{2} & 0.86\er{2}{1}\\
 (1,0,0) & (0,1,0) & 1.106\er{4}{4} & 0.83\er{3}{2}\\
 (1,0,0) & (-1,0,0) & 1.213\er{4}{4} & 0.76\er{4}{3}\\
\hline
\multicolumn{4}{c}{$\kQ=0.121\longrightarrow\kQp=0.125$,$\quad\kq=\k_s$}\\
\hline
 (0,0,0) & (1,0,0) & 1.049\er{2}{2} & 0.95\er{1}{1}\\
 (1,0,0) & (1,0,0) & 0.996\er{4}{4} & 0.97\er{5}{6}\\
 (1,0,0) & (0,0,0) & 1.039\er{2}{2} & 0.90\er{2}{1}\\
 (1,0,0) & (0,1,0) & 1.089\er{4}{3} & 0.85\er{3}{2}\\
 (1,0,0) & (-1,0,0) & 1.182\er{4}{4} & 0.80\er{4}{3}\\
\hline
\multicolumn{4}{c}{$\kQ=0.129\longrightarrow\kQp=0.125$,$\quad\kq=\k_s$}\\
\hline
 (0,0,0) & (1,0,0) & 1.049\er{2}{2} & 0.94\er{1}{1}\\
 (1,0,0) & (1,0,0) & 0.996\er{4}{4} & 0.97\er{5}{5}\\
 (1,0,0) & (0,0,0) & 1.065\er{2}{2} & 0.87\er{2}{1}\\
 (1,0,0) & (0,1,0) & 1.117\er{4}{4} & 0.82\er{3}{2}\\
 (1,0,0) & (-1,0,0) & 1.238\er{4}{5} & 0.75\er{3}{3}\\
\hline
\multicolumn{4}{c}{$\kQ=0.121\longrightarrow\kQp=0.129$,$\quad\kq=\k_s$}\\
\hline
 (0,0,0) & (1,0,0) & 1.065\er{2}{2} & 0.93\er{1}{1}\\
 (1,0,0) & (1,0,0) & 0.999\er{4}{4} & 0.96\er{5}{6}\\
 (1,0,0) & (0,0,0) & 1.039\er{2}{2} & 0.90\er{2}{1}\\
 (1,0,0) & (0,1,0) & 1.106\er{4}{4} & 0.84\er{3}{2}\\
 (1,0,0) & (-1,0,0) & 1.213\er{4}{4} & 0.78\er{4}{3}\\
\hline
\multicolumn{4}{c}{$\kQ=0.129\longrightarrow\kQp=0.129$,$\quad\kq=\k_s$}\\
\hline
 (0,0,0) & (1,0,0) & 1.065\er{2}{2} & 0.93\er{1}{1}\\
 (1,0,0) & (1,0,0) & 0.995\er{5}{4} & 0.96\er{5}{6}\\
 (1,0,0) & (0,0,0) & 1.065\er{2}{2} & 0.87\er{2}{1}\\
 (1,0,0) & (0,1,0) & 1.134\er{5}{5} & 0.80\er{3}{2}\\
 (1,0,0) & (-1,0,0) & 1.273\er{4}{6} & 0.73\er{3}{3}\\
\hline
\multicolumn{4}{c}{$\kQ=0.121\longrightarrow\kQp=0.133$,$\quad\kq=\k_s$}\\
\hline
 (0,0,0) & (1,0,0) & 1.094\er{3}{3} & 0.90\er{1}{1}\\
 (1,0,0) & (1,0,0) & 1.007\er{5}{4} & 0.95\er{6}{6}\\
 (1,0,0) & (0,0,0) & 1.039\er{2}{2} & 0.91\er{2}{1}\\
 (1,0,0) & (0,1,0) & 1.136\er{4}{5} & 0.81\er{3}{2}\\
 (1,0,0) & (-1,0,0) & 1.264\er{4}{6} & 0.75\er{3}{3}\\
\hline
\multicolumn{4}{c}{$\kQ=0.129\longrightarrow\kQp=0.133$,$\quad\kq=\k_s$}\\
\hline
 (0,0,0) & (1,0,0) & 1.094\er{3}{3} & 0.90\er{1}{1}\\
 (1,0,0) & (1,0,0) & 0.998\er{6}{6} & 0.94\er{5}{5}\\
 (1,0,0) & (0,0,0) & 1.065\er{2}{2} & 0.87\er{2}{1}\\
 (1,0,0) & (0,1,0) & 1.165\er{5}{6} & 0.78\er{3}{2}\\
 (1,0,0) & (-1,0,0) & 1.332\er{5}{7} & 0.70\er{3}{2}\\
\end{tabular}
%
\end{table}





\begin{table}
\caption{Results of fits of our data for $\hpw/(1+\bpw)$ to 
the parametrizations $s\,\xi_{NR}(\w)$ and $\xi_{NR}(\w)$ described
in the text. The four $\kQ=\kQ'(=0.121,0.125,0.129,0.133)$ transitions
with $\kq=0.14144$ are considered in turn. Only transitions with
initial and final meson momenta less or equal to $(\pi/12a)$ are
included.\label{rho2.degen.4144}}
\begin{tabular}{cccccc}
& \multicolumn{3}{c}{$s\,\xi_{NR}(\w)$} & \multicolumn{2}{c}{$\xi_{NR}(\w)$} \\
\hline
$\kQ=\kQ'$ & \multicolumn{2}{c}{$(\rho^2,s)$} & 
$\chi^2/d.o.f.$ & $\rho^2$ & $\chi^2/d.o.f.$ \\
\hline
\hline
0.121  & 1.4\er{3}{3} & 0.99\er{1}{1} & 12.5/3 & 1.6\er{2}{3} & 13.0/4\\
\hline
0.125  & 1.4\er{2}{3} & 0.99\er{1}{1} & 13.6/3 & 1.6\er{2}{3} & 14.1/4\\
\hline
0.129  & 1.4\er{1}{2} & 0.99\er{1}{1} & 13.5/3 & 1.5\er{2}{2} & 13.9/4\\
\hline
0.133  & 1.4\er{1}{2} & 0.99\er{1}{1} & 11.1/3 & 1.4\er{2}{1} & 11.5/4\\
\end{tabular}
%
\end{table}




\begin{table}
\caption{Power corrections
to $\hpw$ for four values of $\w$ when $\kq=0.14144$.See text
for definition of $R^+$ and $g(\w)$\label{rp_fit}}
\begin{tabular}{cccccccc}
pt & $\kQ$ & $\kQp$ & $\s p_Q$ & $\s p_{Q'}$ & $\w$ & $x$ & $R^+$\\
\hline
1 & 0.121 & 0.121 & (1,0,0) & (0,0,0) & 1.037\er{1}{1} & 1.00 & 0\\
 & & 0.125 & & & & 1.16 & -0.015\err{17}{15}\\
 & & 0.129 & & & & 1.40 & -0.031\err{35}{23}\\
 & & 0.133 & & & & 1.79 & -0.052\err{51}{35}\\
\hline
&\multicolumn{7}{c}{$g(\w)=0.073$\err{52}{81} with $\chi^2$/dof$=$0.1/2}\\
\hline
2 & 0.121 & 0.125 & (0,0,0) & (1,0,0) & 1.047\er{2}{1} & 1.00 & 0\\
 & 0.125 & & & & & 1.16 & 0.002\er{7}{7}\\
 & 0.129 & & & & & 1.40 & 0.021\err{15}{12}\\
 & 0.133 & & & & & 1.79 & 0.031\err{23}{17}\\
\hline
&\multicolumn{7}{c}{$g(\w)=-0.041$\err{23}{29} with $\chi^2$/dof$=$0.4/2}\\
\hline
3 & 0.129 & 0.121 & (1,0,0) & (0,0,0) & 1.062\er{2}{2} & 1.00 & 0\\
 & & 0.125 & & & & 1.16 & -0.006\err{13}{9}\\
 & & 0.129 & & & & 1.40 & -0.049\err{25}{25}\\
 & & 0.133 & & & & 1.79 & -0.062\err{54}{44}\\
\hline
&\multicolumn{7}{c}{$g(\w)=0.083$\err{51}{65} with $\chi^2$/dof$=$0.8/2}\\
\hline
4 & 0.121 & 0.133 & (0,0,0) & (1,0,0) & 1.088\er{3}{2} & 1.00 & 0\\
 & 0.125 & & & & & 1.16 & 0.010\err{11}{12}\\
 & 0.129 & & & & & 1.40 & 0.013\err{23}{26}\\
 & 0.133 & & & & & 1.79 & 0.003\err{37}{44}\\
\hline
&\multicolumn{7}{c}{$g(\w)=-0.025$\err{53}{50} with $\chi^2$/dof$=$0.5/2}\\
\end{tabular}
\end{table}



\begin{table}
\caption{Results of fits of our data for $\hpw/(1+\bpw)$ to 
the parametrizations $\xi_{NR}(\w)$, $\xi_{lin}(\w)$ and
$\xi_{quad}(\w)$ with and without the additional parameter $s$, as
described in the text. For fixed $\kq$, all heavy-quark mass
combinations are used. Only transitions with initial and final meson
momenta less or equal to $\pi/(12a)$ are included. Here
$\kcrit=$0.14315(2) and $\k_s$=0.1419(1).\label{rho2.all}}
\begin{tabular}{cccccc}
& \multicolumn{3}{c}{$s\,\xi_{NR}(\w)$} & \multicolumn{2}{c}{$\xi_{NR}(\w)$} \\
\hline
$\kq$ & $\rho^2$ & $s$ & 
$\chi^2/d.o.f.$ & $\rho^2$ & $\chi^2/d.o.f.$ \\
\hline
\hline
0.14144  & 1.3\er{2}{1} & 0.98\er{1}{1} & 109/38 & 1.5\er{2}{2} & 121/39 \\
\hline
$\k_s$  & 1.2\er{2}{2} & 0.98\er{1}{1} & 95/38 & 1.4\er{2}{2} & 106/39 \\
\hline
0.14226  & 1.0\er{2}{3} & 0.96\er{2}{1} & 113/38 & 1.4\er{2}{3} & 140/39 \\
\hline
0.14262  & 0.7\er{3}{3} & 0.94\er{2}{2} & 100/38 & 1.4\er{3}{4} & 134/39 \\
\hline
$\kcrit$ & 0.9\er{2}{3} & 0.96\er{2}{2} & 69/38 & 1.3\er{3}{3} & 88/39  \\
\end{tabular}

\begin{tabular}{cccccc}
& \multicolumn{3}{c}{$s\,\xi_{lin}(\w)$} & \multicolumn{2}{c}{$\xi_{lin}(\w)$} \\
\hline
$\kq$ & $\rho^2$ & $s$ & 
$\chi^2/d.o.f.$ & $\rho^2$ & $\chi^2/d.o.f.$ \\
\hline
\hline
0.14144  & 1.0\er{1}{1} & 0.97\er{1}{1} & 111/38 & 1.3\er{1}{1} & 159/39 \\
\hline
$\k_s$  & 0.9\er{1}{1} & 0.97\er{1}{1} & 97/38 & 1.2\er{1}{1} & 139/39 \\
\hline
0.14226  & 0.8\er{1}{2} & 0.96\er{2}{1} & 114/38 & 1.2\er{2}{2} & 170/39 \\
\hline
0.14262  & 0.6\er{2}{2} & 0.93\er{2}{2} & 100/38 & 1.1\er{2}{3} & 155/39 \\
\hline
$\kcrit$ & 0.7\er{1}{2} & 0.95\er{2}{2} & 71/38 & 1.1\er{2}{2} & 111/39  \\
\end{tabular}

\begin{tabular}{cccccccc}
& \multicolumn{4}{c}{$s\,\xi_{quad}(\w)$} & \multicolumn{3}{c}{$\xi_{quad}(\w)$} \\
\hline
$\kq$ & $\rho^2$ & $c$ & $s$ & 
$\chi^2/d.o.f.$ & $\rho^2$ & $c$ & $\chi^2/d.o.f.$ \\
\hline
\hline
0.14144  & 1.2\er{2}{2} & 1.6\errr{1.2}{1.3} & 0.98\er{2}{1} & 108/37 & 1.6\er{2}{2} & 3.9\errr{1.4}{1.4} & 115/38 \\
\hline
$\k_s$ & 1.2\er{2}{2} & 2.0\errr{1.2}{1.5} & 0.98\er{2}{2} & 94/37 & 1.5\er{2}{2} & 4.0\errr{1.4}{1.3} & 99/38 \\
\hline
0.14226  & 1.0\er{2}{3} & 1.4\errr{1.3}{1.7} & 0.96\er{2}{1} & 113/37 & 1.6\er{2}{3} & 5.1\errr{1.6}{1.6} & 125/38 \\
\hline
0.14262  & 0.7\er{4}{4} & 1.1\errr{2.2}{2.6} & 0.94\er{2}{2} & 100/37 & 1.7\er{4}{4} & 6.9\errr{2.1}{2.2} & 114/38 \\
\hline
$\kcrit$ & 1.0\er{3}{3} & 2.3\errr{1.5}{2.0} & 0.97\er{2}{3} & 69/37 & 1.5\er{3}{3} & 5.0\errr{1.5}{1.6} & 74/38  \\
\end{tabular}

\end{table}

\begin{table}
\caption{The different momentum sets for $\xi_{u,d}(\w)$ and $\xi_s(\w)$ 
are fit to the parametrization $s\,\xi_{NR}(\w)$ (\protect{\eq{bsw}}).
A momentum set comprises all combinations of initial and final heavy quarks
with fixed initial and final momenta. These different fits are used
to estimate remaining systematics (see text).\label{rho2.syst}}
\begin{tabular}{cccccccc}
& &\multicolumn{3}{c}{$\xi_{u,d}(\w)$} & \multicolumn{3}{c}{$\xi_s(\w)$} \\
\hline
$\s p$ & $\s p'$ & $\rho^2$ & $s$ & 
$\chi^2/d.o.f.$ & $\rho^2$ & $s$ & $\chi^2/d.o.f.$ \\
\hline
\hline
 (0,0,0) & (1,0,0) & 1.1\er{3}{2} & 1.01\er{2}{2} & 0.3/6 & 1.2\er{2}{1} 
& 1.00\er{2}{1} & 0.1/6\\
 (1,0,0) & (0,0,0) & 1.3\er{5}{4} & 0.93\er{4}{3} & 0.8/6 & 1.4\er{3}{3} 
& 0.95\er{3}{2} & 0.7/6\\
 (1,0,0) & (0,1,0) & 1.2\er{4}{3} & 0.95\er{8}{6} & 0.1/6 & 1.4\er{2}{2} 
& 0.96\er{4}{3} & 0.1/6\\
 (1,0,0) & (-1,0,0)& 0.7\er{4}{2} & 0.95\err{11}{8} & 0.2/6 & 1.1\er{2}{2} 
& 0.95\er{7}{5} & 0.4/6\\
\end{tabular}

\end{table}

\begin{table}
\caption{Comparion of our lattice results for $-\xi'_{u,d}(1)$ and 
$-\xi'_s(1)$ to the theoretical predictions of various authors.
\label{rho2comp}}
\begin{tabular}{ccc}
Reference & $-\xi'_{u,d}(1)$ & $-\xi'_s(1)$\\
\hline
UKQCD & $0.9\er{2}{3}\stat\er{4}{2}\syst$ & $1.2\er{2}{2}\stat\er{2}{1}\syst$\\
\hline
Bernard, Shen and Soni\cite{bersheso} & & 1.24(26)\stat(26)\syst\\
\hline
de Rafael and Taron\cite{deraf} & $\rho^2<1.42$ &\\
\hline
Close and Wambach\cite{wambach} & 1.40 & 1.64\\
\hline
Narison\cite{narison} & 1.00(2) &\\
\hline
Neubert\cite{bible} & 0.66(5) &\\
\hline
Voloshin\cite{volosh} & 1.4(3) &\\
\hline
Bjorken\cite{bj} & \multicolumn{2}{c}{$\rho^2>0.25$}\\
\hline
Blok and Shifman\cite{bs} & $0.35<\rho^2<1.15$ &\\
\hline
H\o gaasen and Sadzikowski\cite{hogasad} & 0.98 & 1.135\\
\hline
Rosner\cite{ros} & 1.59(43) &\\
\hline
Burdman\cite{burd} & 1.08(10) &\\
\hline
Dai, Huang and Jin\cite{dai} & 1.05(20)\\
\end{tabular}
\end{table}




\begin{table}
\caption{Results for $|V_{cb}|$ from a fit of 
$|V_{cb}|(1+\b^{A_1}(1))K(\w)\xi_{NR}(\w)$ to
experimental data with $\xi_{NR}(\w)$ fixed by our lattice computation
(i.e.  $\rho^2$ is given by \protect{\eq{rho2ud}}) and $K(\w)=1$. 
The experimental
data are obtained from the differential branching ratio for
$\btodsl$ decays. In the $|V_{cb}|$ column, 
the first set of errors is due to experimental uncertainties, the
second set of errors results from the lattice statistical errors on
$\rho^2$, and the third, from the lattice systematic errors on
$\rho^2$.
\label{vcbres}}
\begin{tabular}{ccc}
Experiment & $|V_{cb}|
\left(\frac{1+\beta^{A_1}(1)}{0.99}\right)(1+\delta_{1/m_c^2})$ & $\chi^2/d.o.f.$\\
\hline
ALEPH & 0.042(2)\er{2}{3}\er{4}{1} & 3.0/5\\
ARGUS & 0.033(2)\er{1}{2}\er{3}{1} & 9.9/7\\
CLEO II & 0.037(1)\er{2}{2}\er{4}{1} & 4.5/6\\
\end{tabular}

\end{table}


\begin{table}
\caption{Our predictions for various branching ratios compared to the
experimentally measured values for these ratios. Our results are
obtained assuming $|V_{cb}|=$0.038\protect{\cite{stone}}, $\tau_{\bar
B^0}=$1.53ps, $\tau_{\bar B^0_s}=$1.54ps\protect{\cite{wilbur}},
$M_{\bar B^0}=5.28~\gev$, $M_{\bar B^0_s}=5.38~\gev$, $M_{\bar
D^+}=1.87~\gev$, $M_{\bar D^+_s}=1.97~\gev$, $M_{\bar
D^{*+}}=2.01~\gev$ and $M_{\bar D^{*+}_s}=2.11~\gev$
\protect{\cite{pdg}}. Our errors are explained in the
text. We only
consider here semi-leptonic $\bar B^0$ and $\bar B^0_s$ decays because
the experimental data for charged $B$ meson decays are much less
precise.  The quoted experimental numbers were taken from
Ref.\protect\cite{stone}.
\label{brs}}
\begin{tabular}{ccccc}
& $\btodl$ & $\bstodsl$ & $\btodsl$ & $\bstodssl$ \\
\hline
UKQCD & 1.5\er{4}{4}$\pm$0.3 & 1.3\er{2}{2}$\pm$0.3
& 4.8\er{8}{9}$\pm$0.5 & 4.4\er{4}{5}$\pm$0.4\\
\hline
ARGUS & 2.1$\pm$0.7$\pm$0.6 & & 4.7$\pm$0.6$\pm$0.6 &\\
CLEO I & 1.8$\pm$0.6$\pm$0.3 & & 4.1$\pm$0.5$\pm$0.7 &\\
CLEO II & & & 4.50$\pm$0.44$\pm$0.44 &\\
\end{tabular}

\end{table}


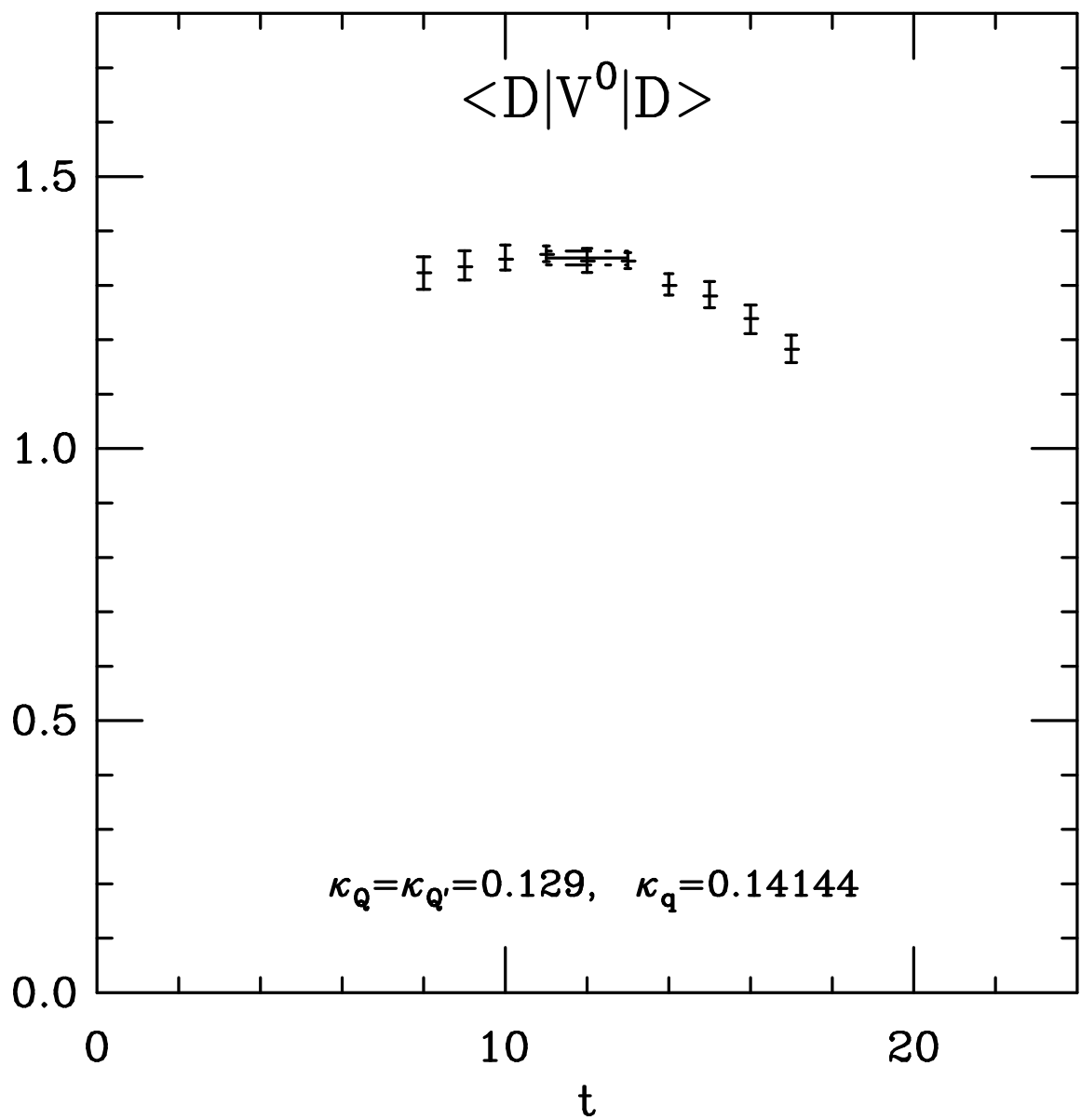
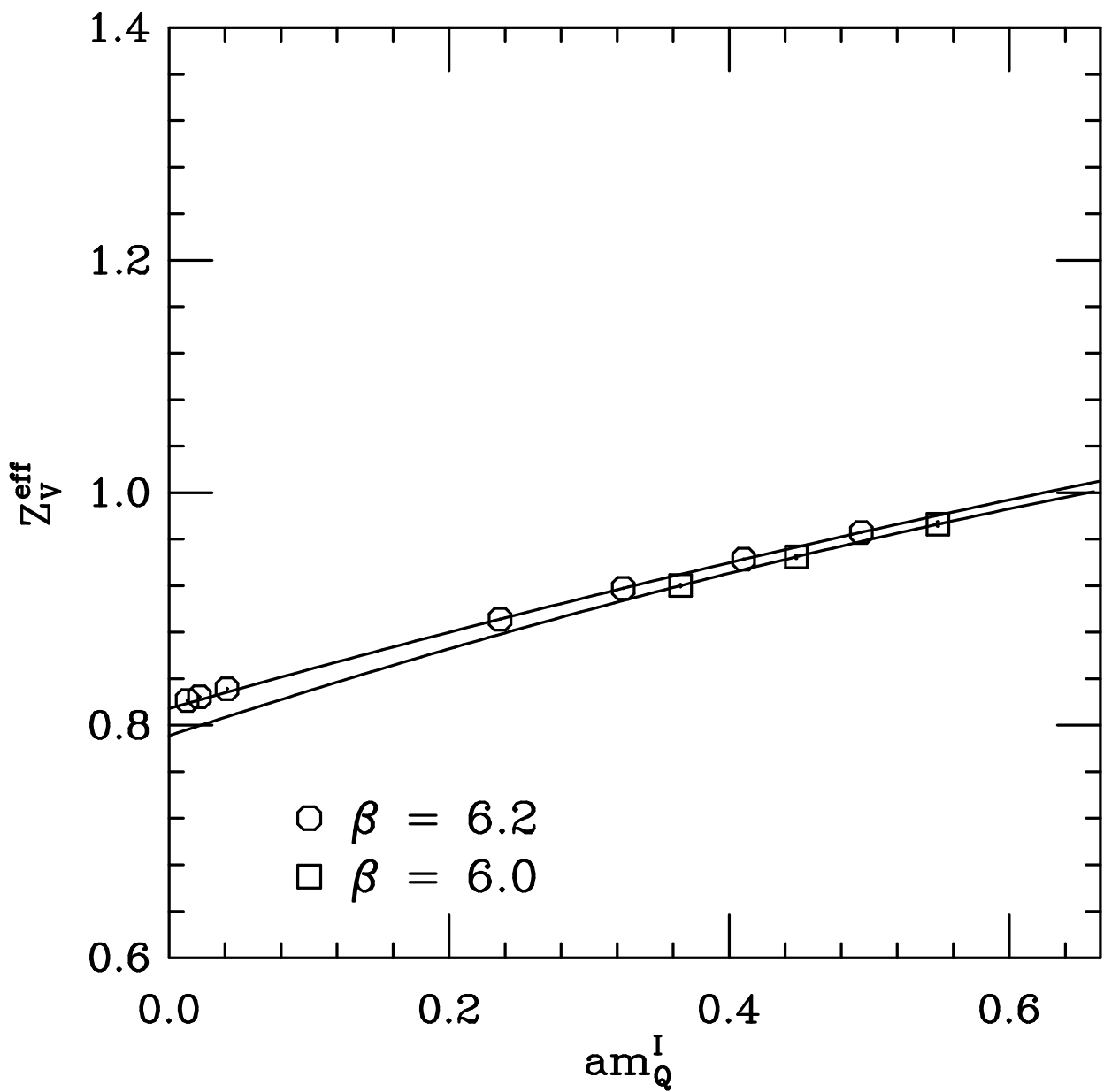
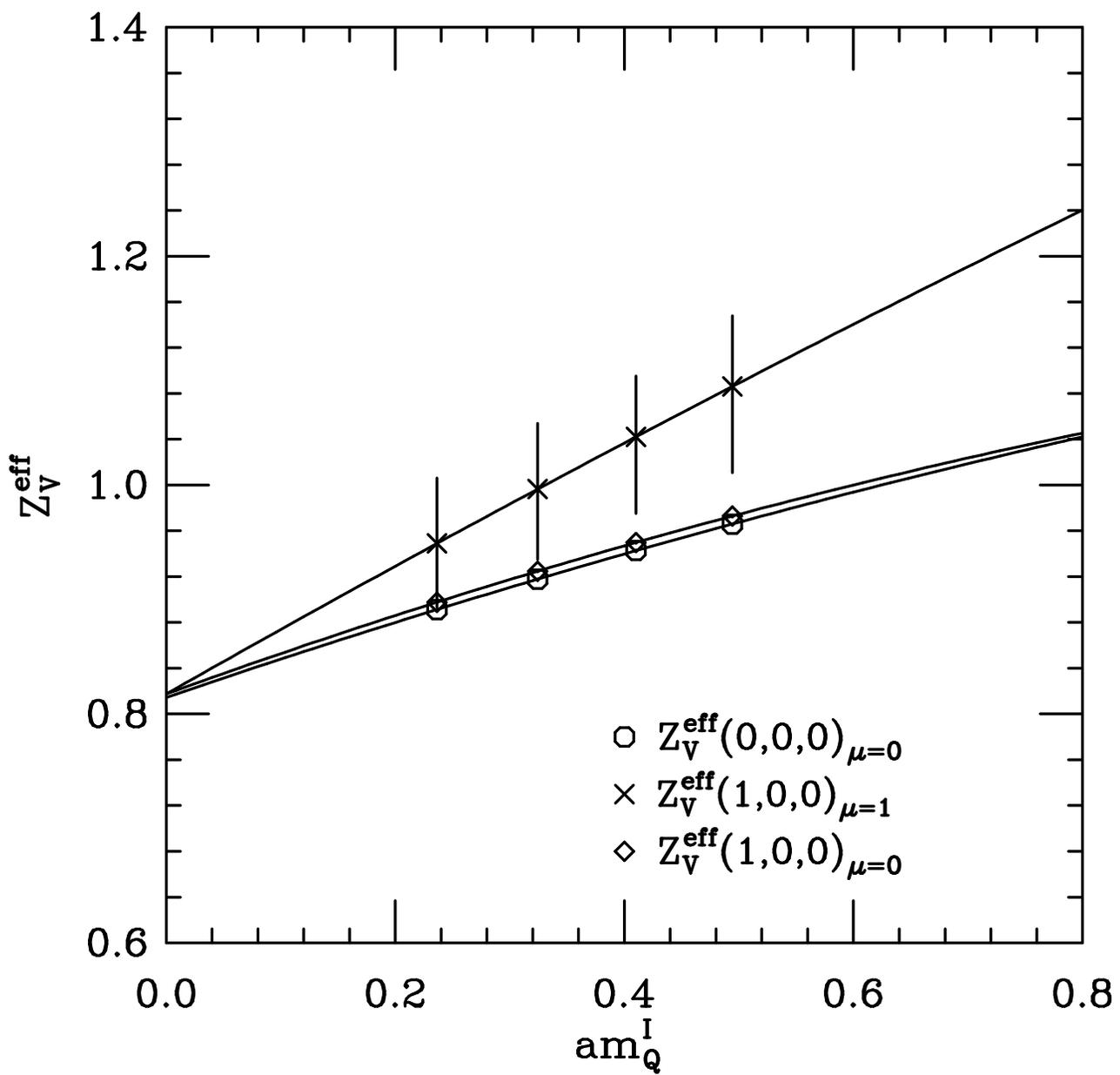
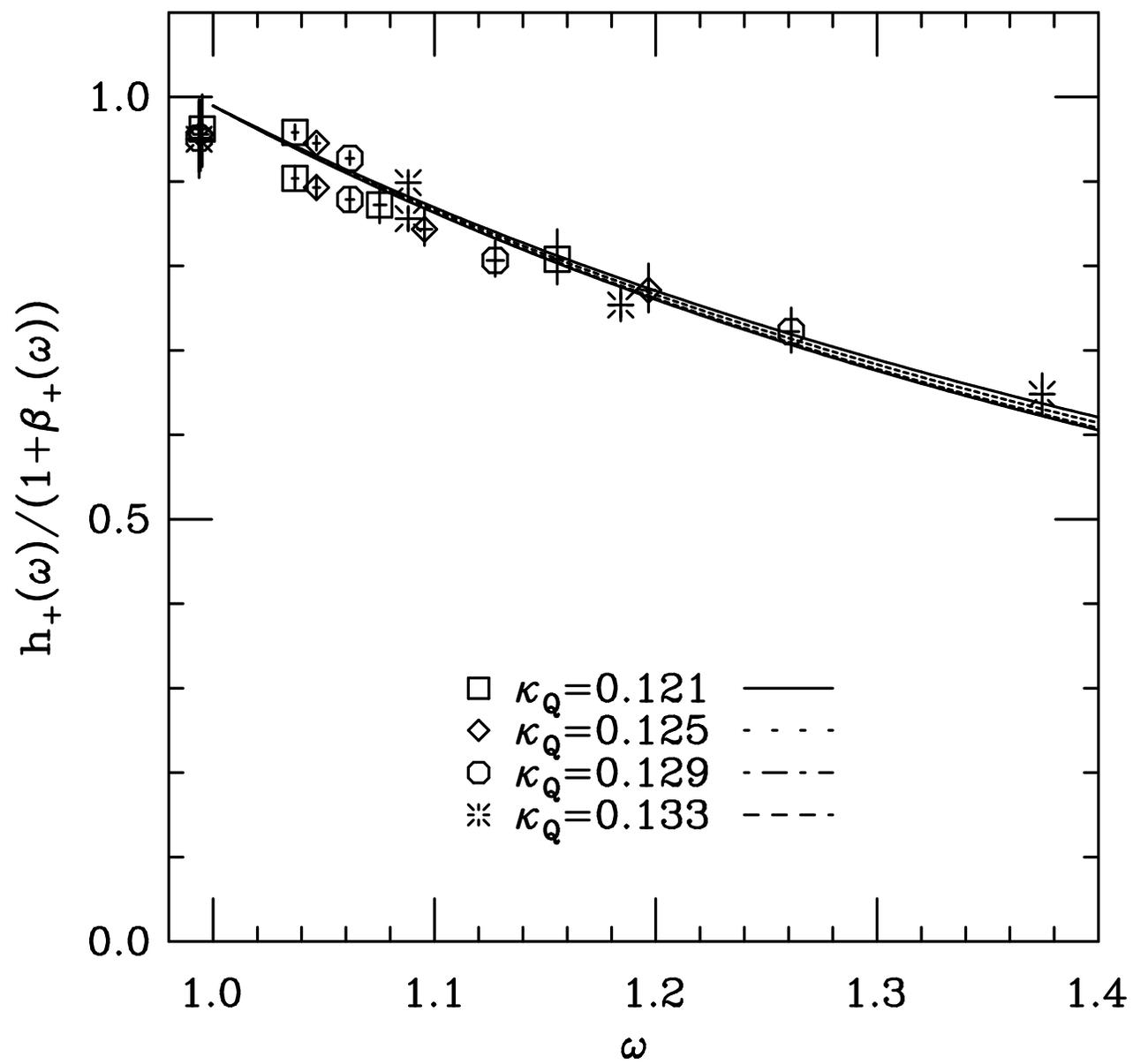
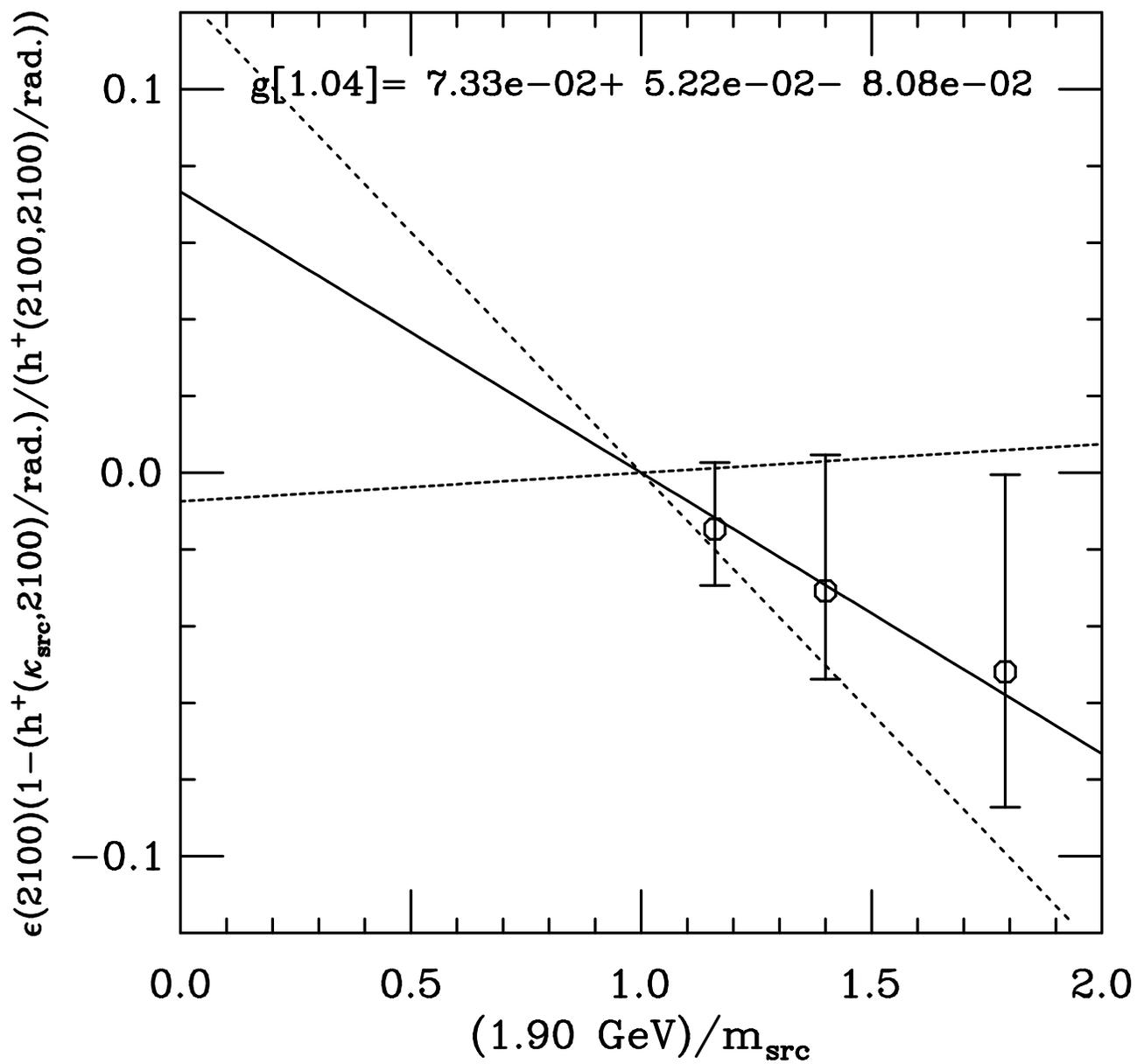
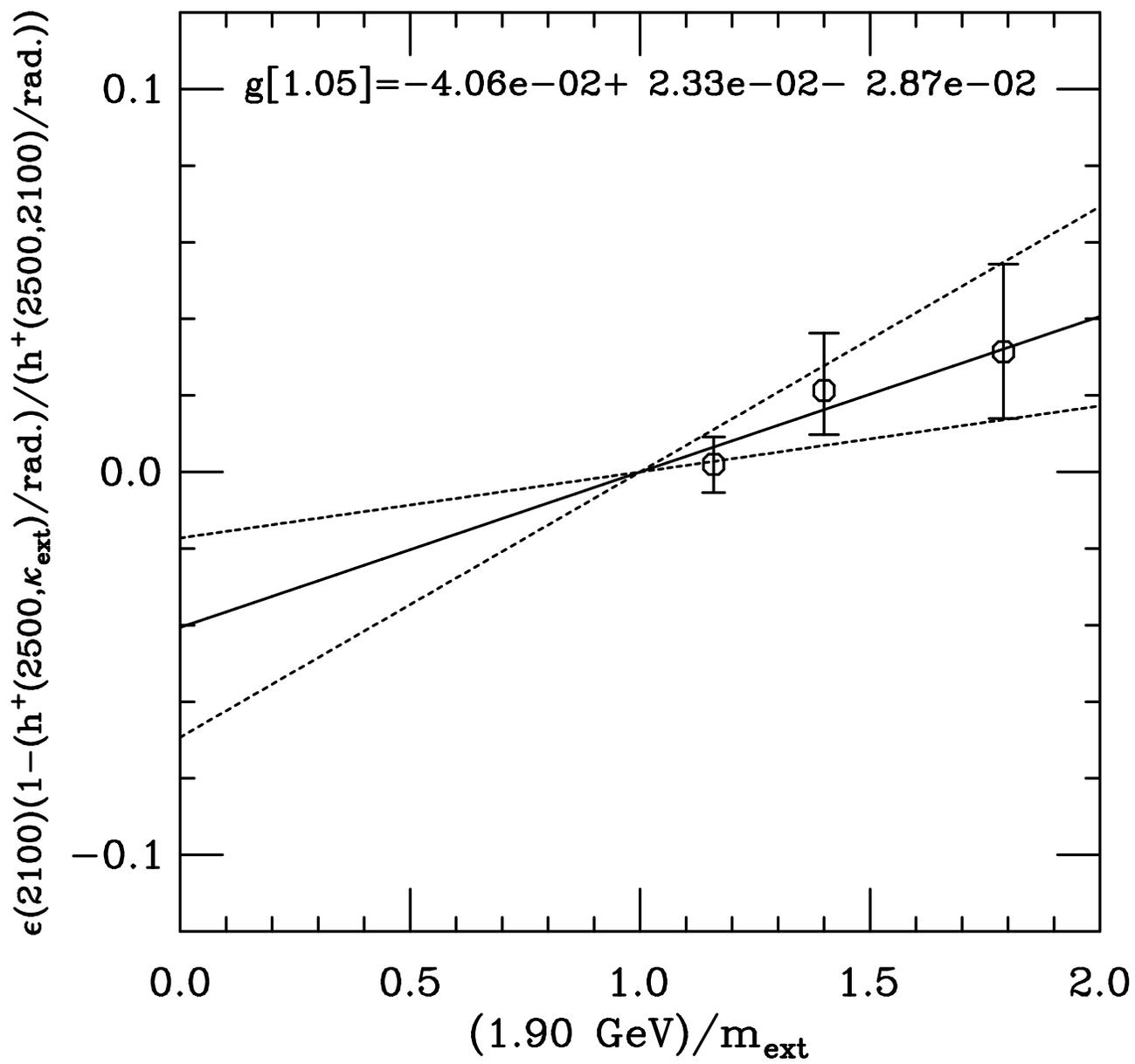
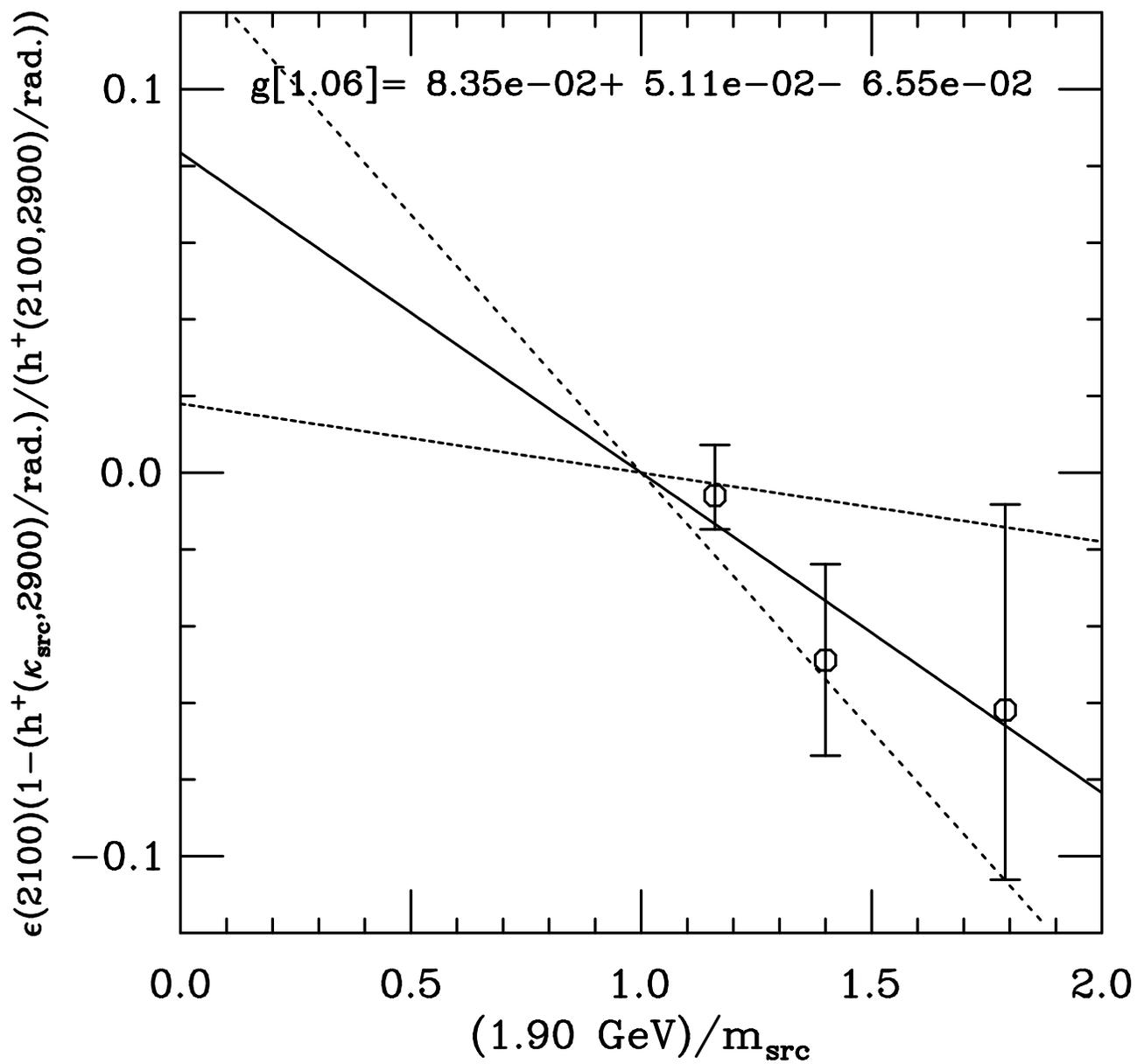
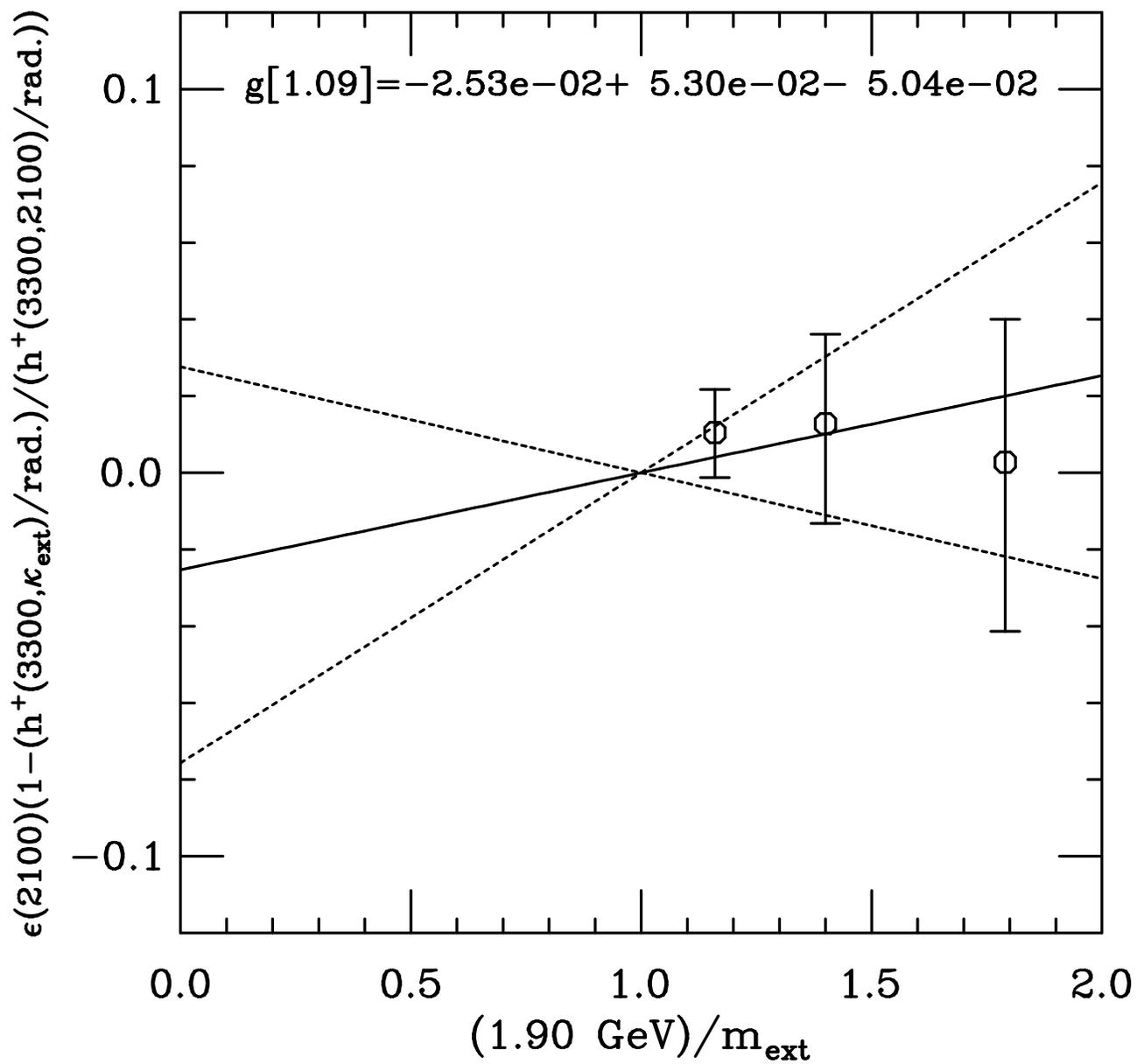
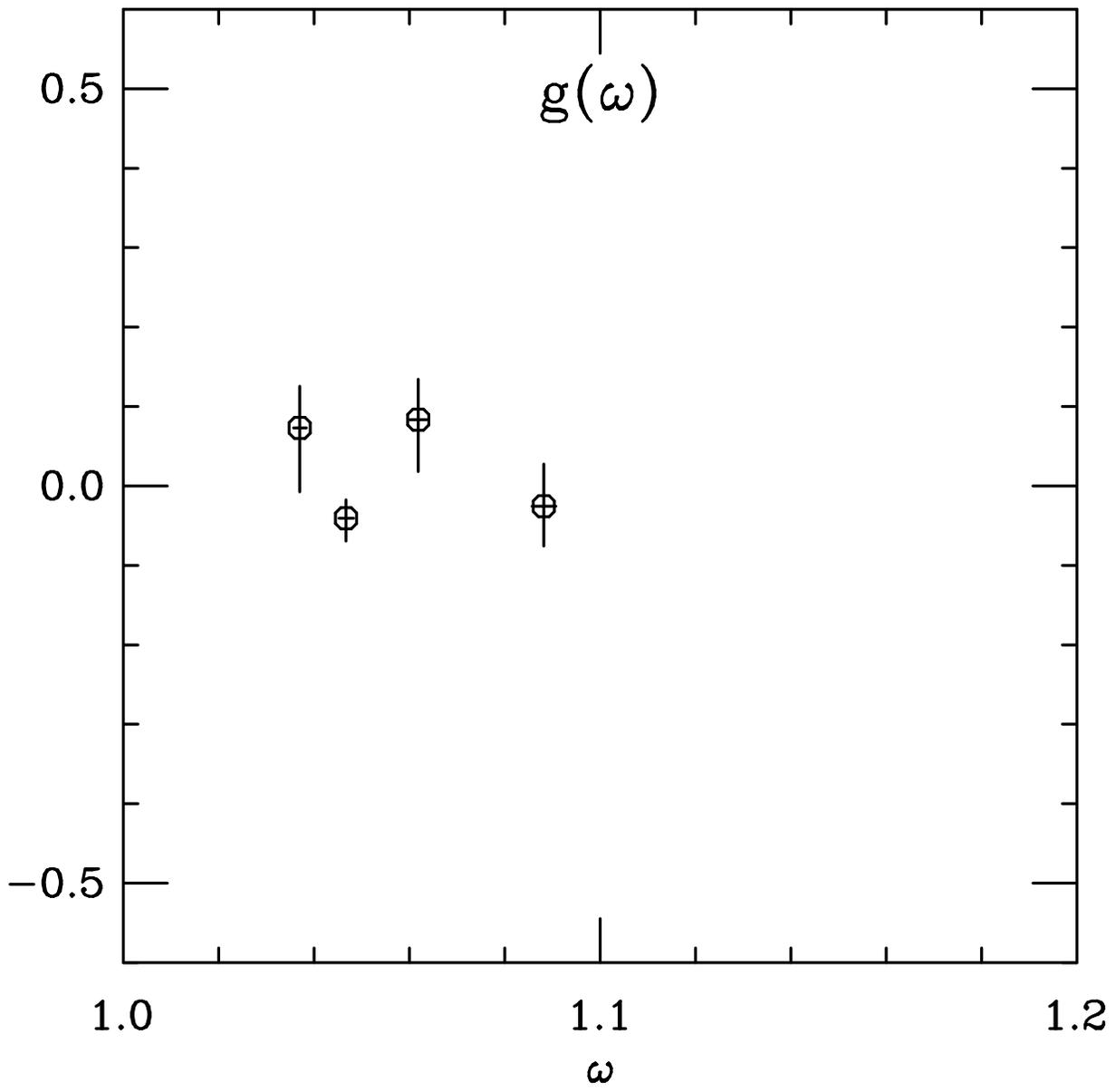
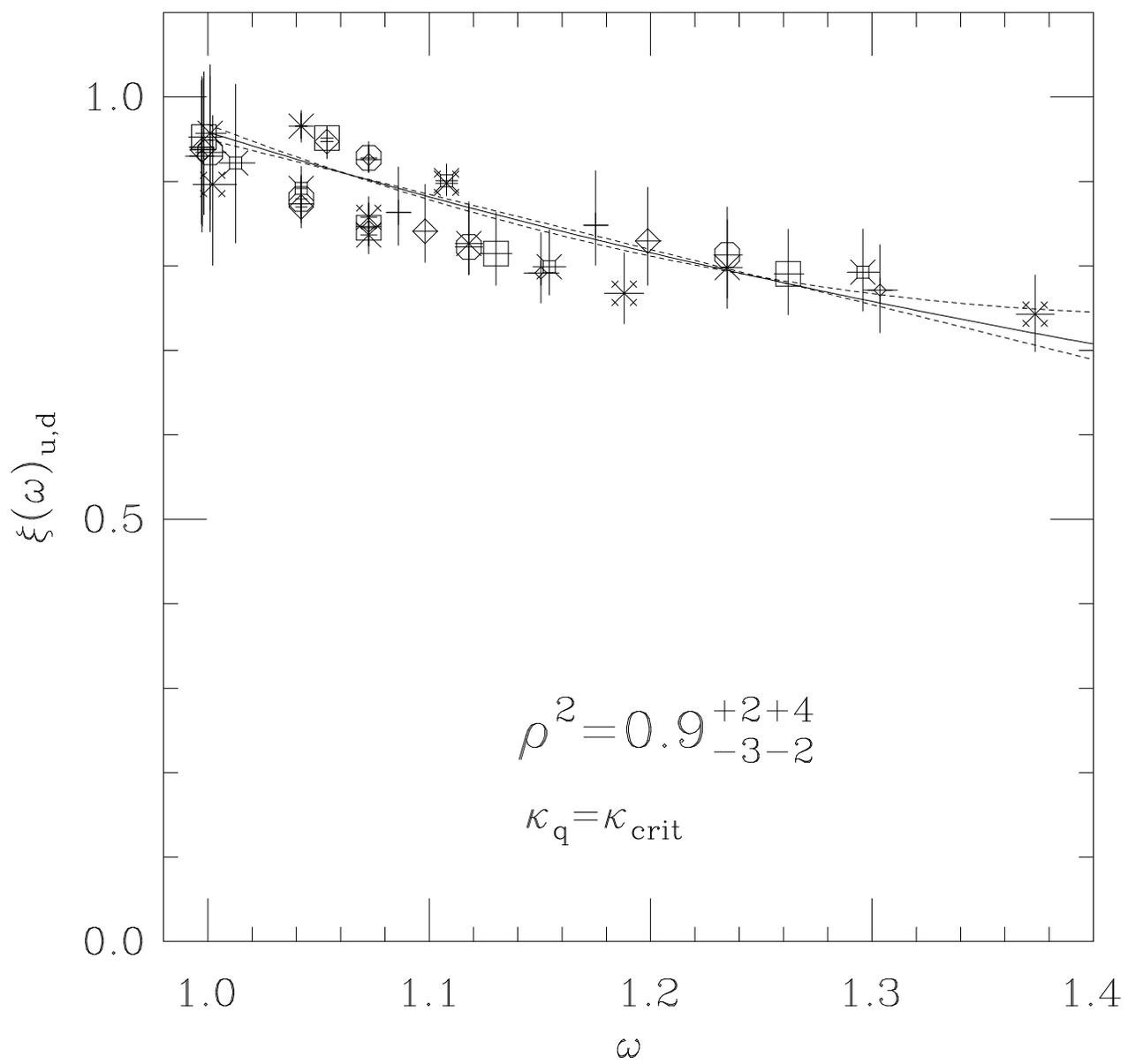
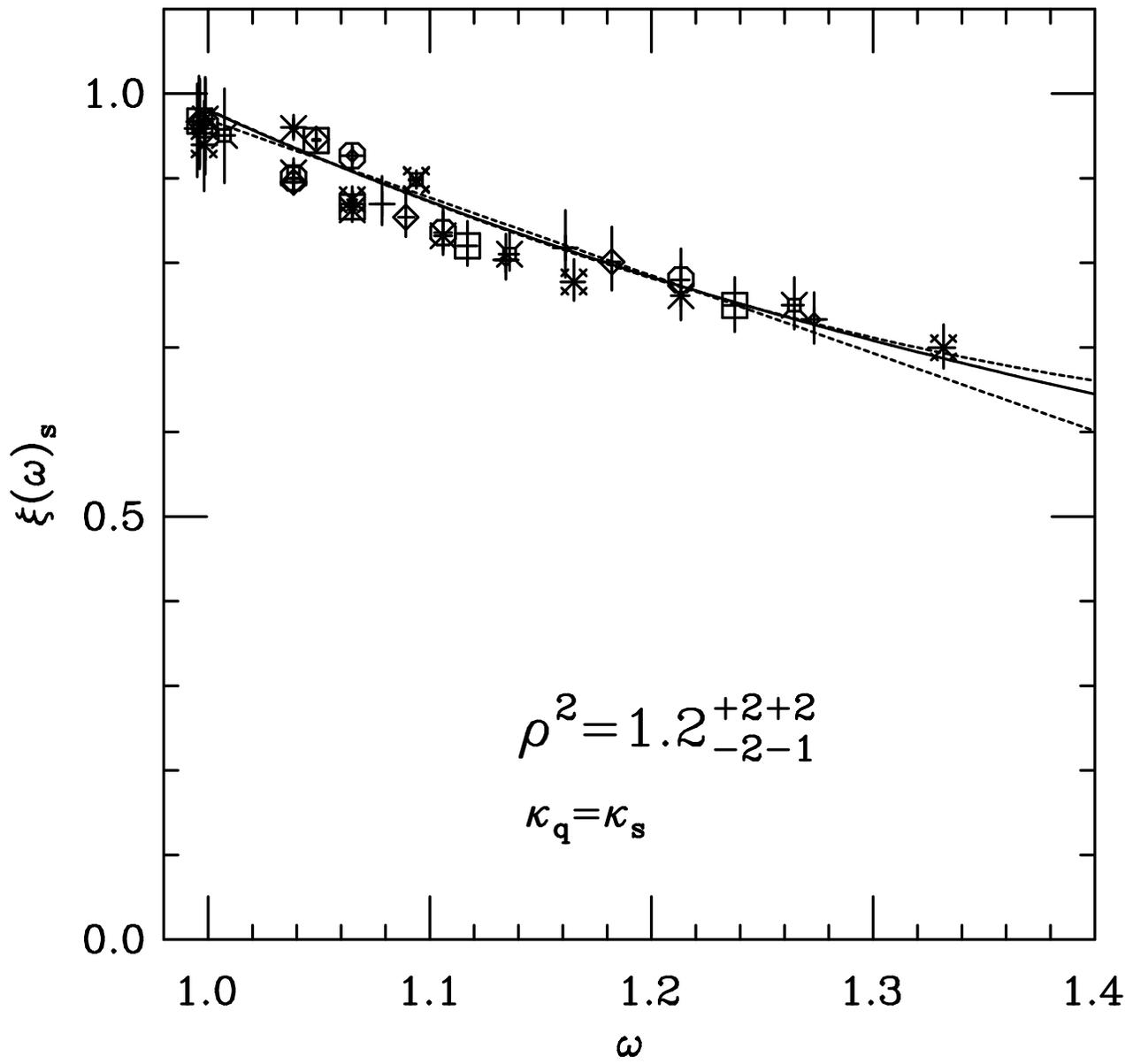
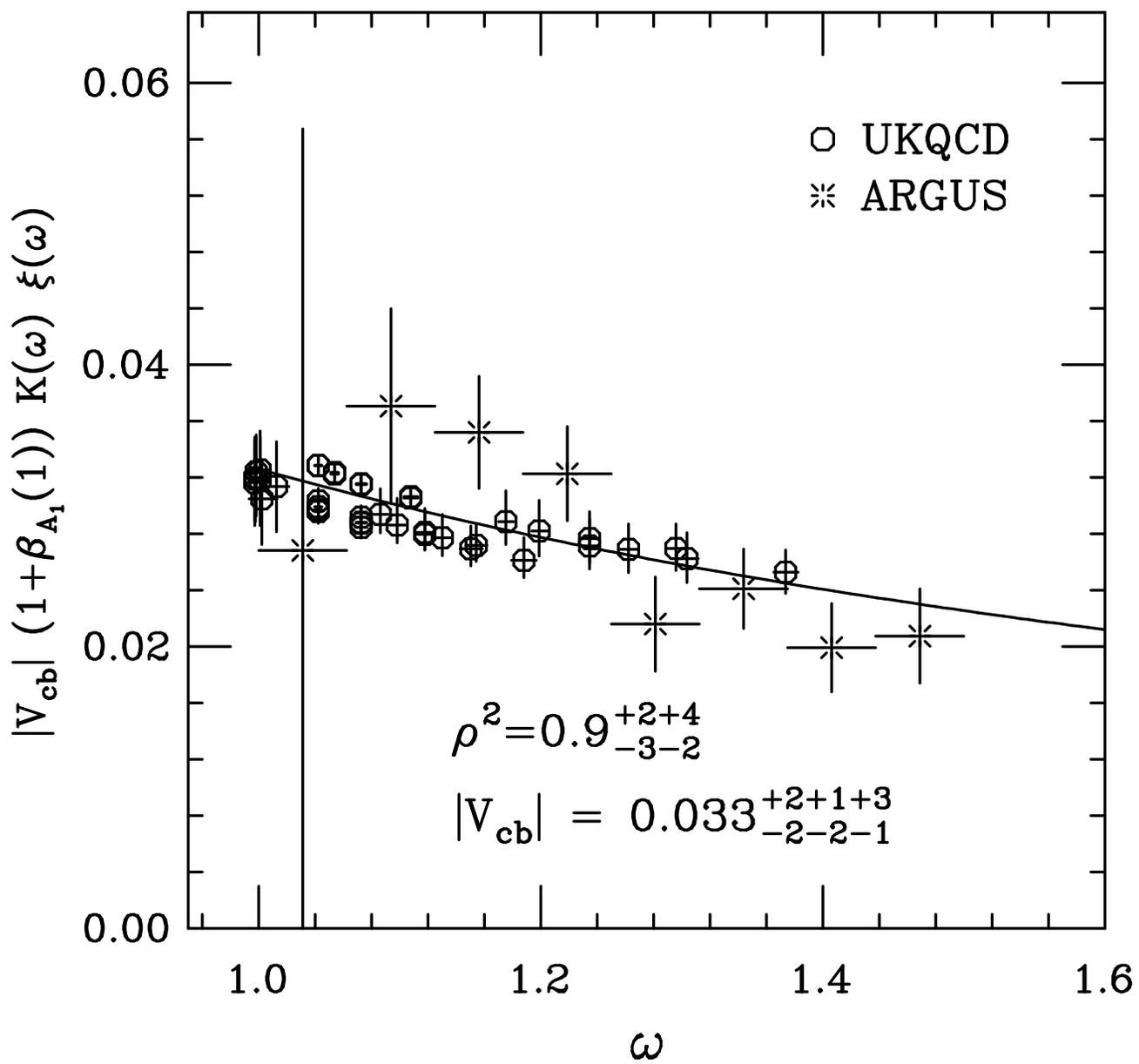
\end{document}